\documentclass[twoside,twocolumn,9pt]{article}
\usepackage{extsizes}
\usepackage[super,sort&compress,comma]{natbib} 
\usepackage[version=3]{mhchem}
\usepackage[left=1.5cm, right=1.5cm, top=1.785cm, bottom=2.0cm]{geometry}
\usepackage{balance}
\usepackage{mathptmx}
\usepackage{sectsty}
\usepackage{graphicx} 
\usepackage{lastpage}
\usepackage[format=plain,justification=justified,singlelinecheck=false,font={stretch=1.125,small,sf},labelfont=bf,labelsep=space]{caption}
\usepackage{float}
\usepackage{fancyhdr}
\usepackage{fnpos}
\usepackage[english]{babel}
\addto{\captionsenglish}{%
  
}
\usepackage{array}
\usepackage{droidsans}
\usepackage{charter}
\usepackage[T1]{fontenc}
\usepackage[usenames,dvipsnames]{xcolor}
\usepackage{setspace}
\usepackage[compact]{titlesec}
\usepackage{hyperref}

\usepackage{epstopdf}
\usepackage{soul}
\usepackage{bm}
\usepackage[normalem]{ulem}
\usepackage{amsbsy,amsmath}
\usepackage{amssymb}
\usepackage{comment}
\usepackage{cuted}
\usepackage{dsfont}

\definecolor{cream}{RGB}{222,217,201}

\begin{document}

\pagestyle{fancy}
\thispagestyle{plain}
\fancypagestyle{plain}{
\renewcommand{\headrulewidth}{0pt}
}

\makeFNbottom
\makeatletter
\renewcommand\LARGE{\@setfontsize\LARGE{15pt}{17}}
\renewcommand\Large{\@setfontsize\Large{12pt}{14}}
\renewcommand\large{\@setfontsize\large{10pt}{12}}
\renewcommand\footnotesize{\@setfontsize\footnotesize{7pt}{10}}
\makeatother

\renewcommand{\thefootnote}{\fnsymbol{footnote}}
\renewcommand\footnoterule{\vspace*{1pt}%
\color{cream}\hrule width 3.5in height 0.4pt \color{black}\vspace*{5pt}} 
\setcounter{secnumdepth}{5}

\makeatletter 
\renewcommand\@biblabel[1]{#1}            
\renewcommand\@makefntext[1]%
{\noindent\makebox[0pt][r]{\@thefnmark\,}#1}
\makeatother 
\renewcommand{\figurename}{\small{Fig.}~}
\sectionfont{\sffamily\Large}
\subsectionfont{\normalsize}
\subsubsectionfont{\bf}
\setstretch{1.125} 
\setlength{\skip\footins}{0.8cm}
\setlength{\footnotesep}{0.25cm}
\setlength{\jot}{10pt}
\titlespacing*{\section}{0pt}{4pt}{4pt}
\titlespacing*{\subsection}{0pt}{15pt}{1pt}

\fancyfoot{}
\fancyfoot[LO,RE]{\vspace{-7.1pt}\includegraphics[height=9pt]{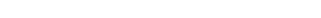}}
\fancyfoot[CO]{\vspace{-7.1pt}\hspace{11.9cm}\includegraphics{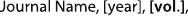}}
\fancyfoot[CE]{\vspace{-7.2pt}\hspace{-13.2cm}\includegraphics{head_foot/RF}}
\fancyfoot[RO]{\footnotesize{\sffamily{1--\pageref{LastPage} ~\textbar  \hspace{2pt}\thepage}}}
\fancyfoot[LE]{\footnotesize{\sffamily{\thepage~\textbar\hspace{4.65cm} 1--\pageref{LastPage}}}}
\fancyhead{}
\renewcommand{\headrulewidth}{0pt} 
\renewcommand{\footrulewidth}{0pt}
\setlength{\arrayrulewidth}{1pt}
\setlength{\columnsep}{6.5mm}
\setlength\bibsep{1pt}

\makeatletter 
\newlength{\figrulesep} 
\setlength{\figrulesep}{0.5\textfloatsep} 

\newcommand{\topfigrule}{\vspace*{-1pt}%
\noindent{\color{cream}\rule[-\figrulesep]{\columnwidth}{1.5pt}} }

\newcommand{\botfigrule}{\vspace*{-2pt}%
\noindent{\color{cream}\rule[\figrulesep]{\columnwidth}{1.5pt}} }

\newcommand{\dblfigrule}{\vspace*{-1pt}%
\noindent{\color{cream}\rule[-\figrulesep]{\textwidth}{1.5pt}} }

\makeatother

\newcommand{\ben}{\begin{equation}}
\newcommand{\een}{\end{equation}}
\newcommand{\bea}{\begin{eqnarray}}
\newcommand{\eea}{\end{eqnarray}}
\def\bra#1{\langle#1\vert}
\def\ket#1{\vert#1\rangle}
\def\braket#1#2{\langle #1 \vert #2 \rangle}

\def\sss{\scriptscriptstyle\rm}

\def\subks{_{\sss KS}}
\def\subext{_{\sss ext}}
\def\subx{_{\sss X}}
\def\subc{_{\sss C}}
\def\subh{_{\sss H}}
\def\subs{_{\sss S}}
\def\subxc{_{\sss XC}}
\def\supS{^{\rm S}}

\def\1s{_{1,\sss S}}
\def\2s{_{2,\sss S}}
\def\x{_{\sss X}}
\def\c{_{\sss C}}
\def\s{_{\sss S}}
\def\xc{_{\sss XC}}
\def\Hx{_{\sss HX}}
\def\Hxc{_{\sss HXC}}
\def\xj{_{{\sss X},j}}
\def\xcj{_{{\sss XC},j}}
\def\N{_{\sss N}}
\def\H{_{\sss H}}
\def\sH{^{\sss H}}
\def\ext{_{\rm ext}}
\def\S{^{\rm S}}
\def\sss{\scriptstyle\rm} 
\def\a{_{\sss a}}

\def\BLA{\rm BLA}

\def\Hxcqq{_{{\sss HXC}\,qq}}
\def\Hxcqqp{_{{\sss HXC}\,qq'}}

\def\Hxcnm{_{{\sss HXC}\,q_n q_m}}
\def\HxcQQ{_{{\sss HXC}\,QQ}}

\def\xcqq{_{{\sss XC}\,qq}}
\def\xcqqp{_{{\sss XC}\,qq'}}
\def\xcnm{_{{\sss XC}\,q_n q_m}}
\def\xcQQ{_{{\sss XC}\,QQ}}
\def\xcQQp{_{{\sss XC}\,QQ'}}
\def\bigsss{\displaystyle\rm}  
\def\a{_{\bigsss a}}


\def\cF{{\cal F }}

\def\trho{\tilde{\rho}}
\def\tk{\tilde{k}}
\def\tn{\tilde{n}}
\def\tv{\tilde{v}}

\def\br{{\bf r}}
\def\bR{{\bf R}}
\def\bj{{\bf j}}
\def\bb{{\bf b}}
\def\bA{{\bf A}}
\def\bJ{{\bf J}}

\def\dulr{{\underline{\underline{\bf r}}}}
\def\dulR{{\underline{\underline{\bf R}}}}
\def\dulq{{\underline{\underline{q}}}}

\twocolumn[
  \begin{@twocolumnfalse}
{\includegraphics[height=30pt]{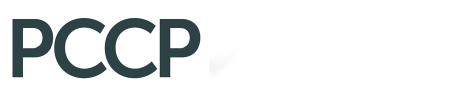}\hfill\raisebox{0pt}[0pt][0pt]{\includegraphics[height=55pt]{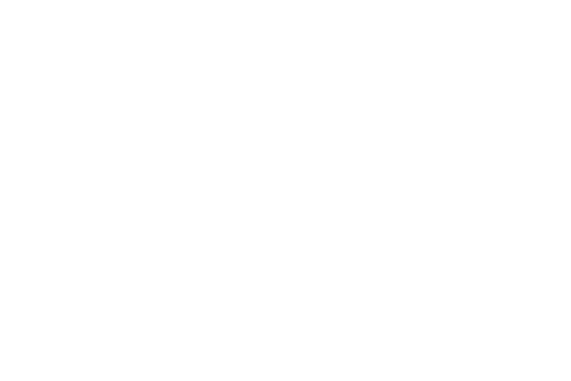}}\\[1ex]
\includegraphics[width=18.5cm]{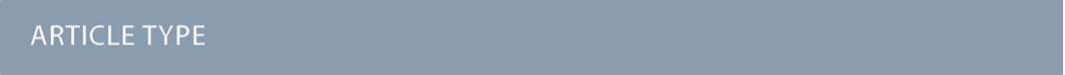}}\par
\vspace{1em}
\sffamily
\begin{tabular}{m{4.5cm} p{13.5cm} }

\includegraphics{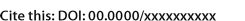} & \noindent\LARGE{\textbf{Dressed Time-Dependent Density Functional Theory: Opening and Closing a Pandora's Box}} \\
\vspace{0.3cm} & \vspace{0.3cm} \\

 & \noindent\large{Jack T. Taylor and Neepa T. Maitra$^{\ast}$} \\

\includegraphics{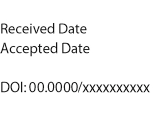} & \noindent\normalsize{
Dressed time-dependent density functional theory (DTDDFT) has been shown to overcome the inability of adiabatic TDDFT to describe excitation energies and oscillator strengths of double-excitation character, however, it is far from being a black-box method.
Stepping towards this goal, here we outline and address several operational issues while benchmarking the approach on a range of molecules. 
We define a protocol to determine whether an excitation resulting from adiabatic TDDFT needs a double-excitation correction, without requiring a priori knowledge of the state, and which Kohn-Sham states to build the frequency-dependent DTDDFT matrix from. 
We present an `undownfolding' of this matrix that replaces this frequency-dependent matrix by a linear frequency-independent one in a larger space, eliminating ambiguities in the iterative procedure required to diagonalize the former. 
We compare the performance of DTDDFT for both excitation energies and oscillator strengths, as measured by an approximate metric for their single-excitation character, for different choices of the underlying adiabatic density functional approximation it is built upon, with and without employing the Tamm-Dancoff approximation, for molecular excitations predicted by wavefunction methods to have either partial or genuine double-excitation character. 
We propose a correction to DTDDFT to account for missing dynamical correlation that is removed in the truncation procedure inherent in the method, and discuss the possibility of unphysical imaginary frequencies reminiscent of the `triplet instability' that may arise. 
We provide benchmark data on a range of molecules: butadiene and hexatriene (for which we present potential energy curves), octatetraene, cyclopentadiene, acrolein, naphthalene, ethylene, borole, and cyclopentadienone. 
In some cases, but not all, DTDDFT predicts frequencies with about as much accuracy as the same underlying adiabatic TDDFT predicts for the constituent and near-lying single excitations.  
}

\end{tabular}

 \end{@twocolumnfalse} \vspace{0.6cm}

  ]

\renewcommand*\rmdefault{bch}\normalfont\upshape
\rmfamily
\section*{}
\vspace{-1cm}


\footnotetext{Department of Physics, Rutgers University, Newark 07102, New Jersey USA\\
E-mail: neepa.maitra@rutgers.edu, jack.t.taylor@rutgers.edu}



\section{Introduction}
\label{sec:Intro}
Time-dependent density functional theory (TDDFT)~\cite{RG84,C95,PGG96} remains one of the most widely used electronic-structure methods for optical spectra and response properties of medium- to large-sized molecules~\cite{LJ13,SBLJ21,SGDLJ25}, stemming from its attractive trade-off between computational affordability, chemical accuracy and fundamental rigor~\cite{CH12,M16,U25}.
The growth of TDDFT applications over the past decades owes much to the seminal contributions from Martin Head-Gordon, whom this Special Issue honors, and his research group, beyond their significant contributions to the quantum chemistry codes, QChem~\cite{qchem} and Gaussian\cite{gaussian16}.
For example, the introduction of the Tamm-Dancoff approximation (TDA)~\cite{T91,D50} to TDDFT by Head-Gordon~\cite{HH99} not only simplifies the implementation and reduces the cost of calculations, but also eliminates the possibility of unphysical imaginary excitation frequencies that can emerge in full linear-response TDDFT (i.e., triplet~\cite{CGGG00,PWT11,peach2012overcoming,peach2013triplet} and near-singlet~\cite{CDICFV07,TTRF08} instabilities).
Equally, the success of TDDFT would not have been possible were it not for increasingly accurate density functional approximations (DFAs)\cite{Mardirossian2017,goerigk2017look} developed for ground-state DFT~\cite{HK64,KS65}.
After all, the very first step in any TDDFT calculation is a ground-state DFT calculation, from which occupied and virtual Kohn-Sham (KS) orbitals are obtained, whose energy differences serve as zeroth-order approximations to the TDDFT excitation frequencies.
The development of the $\omega$B97 series of DFAs~\cite{CH08,CH09} within the Head-Gordon research group has played a key role in this effort.
However, even the exact ground-state functional, were it known, would not be able to describe excited states of double-excitation character -- which is the subject of the present paper --  when using the corresponding adiabatic exchange-correlation (xc) kernel, as is usually done in practice.
The spin-flip TDDFT (SF-TDDFT) approach of Head-Gordon and collaborators~\cite{SHK03} offers a solution: one can access doubly-excited configurations (with respect to a closed-shell singlet ground state) indirectly via spin-flip single excitations out of a high-spin triplet reference~\cite{wZ04,RVA10,LL12,CK20,HM22}, whilst remaining in the adiabatic approximation.
While the spin-flip approach depends on the chosen ground-state DFA being accurate for the reference triplet state and on the xc kernel being suitably adapted for spin-flip excitations out of the triplet reference, the present paper instead considers the dressed TDDFT (DTDDFT)~\cite{MZCB04,CZMB04,MW09,MMWA11,C05,HIRC11,M22,DM23,DM25,dar2026correction} approach, which directly addresses the deficiency of adiabatic TDDFT (ATDDFT) through introducing a frequency-dependent kernel.

In recent years, DTDDFT has exhibited reasonable accuracy for a range of quantities where a reliable description of states of double-excitation character is crucial: excitation energies and oscillator strengths~\cite{DM23,DM25,dar2026correction}, real-space excited-state densities~\cite{BM25}, and excited-to-excited state dipole couplings~\cite{RBDM26}.
The phenomena studied in these works considered both pure response properties, as well as their performance in fully non-equilibrium dynamics within the recently-introduced response-reformulated TDDFT approach~\cite{DBM24}.
While these studies highlight the promise and practical feasibility of DTDDFT for real molecules, capturing for example the elusive 1B$_u$/2A$_g$ curve crossing in trans-butadiene~\cite{DM25,dar2026correction}, the limited number of examples warrants a broader exploration of its overall performance and operational procedures, before the method is ready for routine use in photochemistry.

The present article progresses DTDDFT towards this purpose: we discuss the practical choices that must be made when running DTDDFT, the operational subtleties that arise when dealing with the non-linear eigenvalue problem inherent in the approach, and provide benchmark results spanning 15 excited states across nine molecules, hand-picked from the QUEST database (QUESTDB) double-excitation benchmark set~\cite{VSCL21,KBLJ24,loos2025quest}. 
The paper is organised as follows.
First, in Section~\ref{sec:DTDDFT_theory}, we recap the DTDDFT equations given in Ref.~\citenum{DM25}, and clarify the formulae in the case of hybrid adiabatic DFAs.
This leads us, in Section~\ref{sec:Operational}, to delineate the operational procedures and practical choices that the implementation and use of these equations necessitate.
We present an equivalent `un-downfolded' formulation of DTDDFT in Section~\ref{sec:undownfolding}, which unravels the non-linearity of the original prescription into a higher-dimensional but linear matrix. 
We present our results for each of the nine molecules in turn in Section~\ref{sec:Results}, beginning with general computational details of our numerical calculations in Section~\ref{subsec:Comp_det}. 
Finally Sec.~\ref{sec:Concs} provides a summary and outlook. 

\section{DTDDFT in Theory}
\label{sec:DTDDFT_theory}

Formally, excitation energies and transition densities are obtained in TDDFT from the linear response of the ground-state density to a weak time-dependent external perturbation, as the poles and residues, respectively, of the associated density-density response function~\cite{PGG96,Carstenbook,GM12chap,M16}. 
For systems with discrete spectra, the problem can be equivalently expressed as
matrix equations~\cite{C95,C96,BA96a,GPG00,Carstenbook},
\begin{equation}
\label{eq:ATDDFT_Ahlrichs}
    \begin{pmatrix}
        \mathbf{A} & \mathbf{B} \\
        \mathbf{B^*} & \mathbf{A^*} \\
    \end{pmatrix}
    \begin{pmatrix}
        \mathbf{X}_I \\
        \mathbf{Y}_I \\
    \end{pmatrix}
    =
    \omega_I 
    \begin{pmatrix}
        \mathds{1} & \mathbf{0} \\
        \mathbf{0} & - \mathds{1} \\
    \end{pmatrix}
    \begin{pmatrix}
        \mathbf{X}_I \\
        \mathbf{Y}_I \\
    \end{pmatrix} \, ,
\end{equation}

\noindent whose dimension spans twice the number of KS single excitations.  
Here, $A_{qq'} = \delta_{qq'} \nu_q + 2 f_{{\sss HXC},qq'}$ and $B_{qq'} = 2 f_{{\sss HXC},q-q'}$, where ($q = i \rightarrow a$) denotes a single excitation from occupied KS orbital $i$ to unoccupied $a$, of frequency $\nu_q =  \epsilon_a - \epsilon_i$, and   ($-q = a \rightarrow i$) the corresponding de-excitation.
The Hartree-xc kernel matrix elements are $f_{{\sss HXC}, qq'}= \int d\mathbf{r} \int d\mathbf{r'} \Phi_q^*(\mathbf{r}) f_{\sss HXC}(\mathbf{r}, \mathbf{r'},\omega) \Phi_{q'}(\mathbf{r'})$, where $f_{\sss HXC}(\mathbf{r}, \mathbf{r'},\omega) = \frac{1}{|\mathbf{r} - \mathbf{r'}|} + f\xc[n_0](\mathbf{r}, \mathbf{r'},\omega)$ with the second term being the xc kernel, a functional of the ground-state density $n_0(\br)$; and $\Phi_q^*(\mathbf{r}) = \phi_i(\mathbf{r}) \phi_a(\mathbf{r})$ is the KS transition density.
Assuming real KS orbitals, Eq.~(\ref{eq:ATDDFT_Ahlrichs}) can be recast as an eigenvalue problem whose dimension now equals the number of KS single excitations; it is this form that appears most widely in quantum chemistry program packages:
\begin{equation}
\label{eq:ATDDFT_Casida}
    \boldsymbol{\Omega}(\omega) \mathbf{G}_I = \omega_I^2 \mathbf{G}_I \, ,
\end{equation}
where
\begin{eqnarray}
    \Omega_{qq'}(\omega) &= \left[{\left(\mathbf{A} - \mathbf{B}\right)}^{1/2} {\left(\mathbf{A} + \mathbf{B}\right)} {\left(\mathbf{A} - \mathbf{B}\right)}^{1/2}\right]_{qq'}  \;\;\; {\rm and}\label{eq:ATDDFT_Casida_mat_ele} \\
    G_{I,q} &= {\mathcal N} \left[{\left(\mathbf{A} - \mathbf{B}\right)}^{1/2} \left(\mathbf{X}_I - \mathbf{Y}_I\right)\right]_q  \label{eq:ATDDFT_Casida_G_ele} \; ,
\end{eqnarray}
where ${\mathcal N}$ is a normalization (see shortly). 
The eigenvalues of $\boldsymbol\Omega(\omega)$ supply the square of the excitation frequencies, $\omega_I = E_I - E_0$, where $E_I$ ($E_0$) is the energy of the $I$th excited state (ground state).
The transition densities, and any one-body transition property, such as oscillator strengths $f_I$, are related to the eigenvectors, $\mathbf{G}_I$. 
When these are normalized such that
\begin{equation}
\label{eq:Casida_G_renorm}
    \mathbf{G}_I^\dagger \left( \mathds{1} - \left[ \frac{\partial \boldsymbol{\Omega}(\omega)}{\partial \omega^2} \right]_{\omega = \omega_I} \right) \mathbf{G}_I = 1 \, ,
\end{equation}
then, 
\ben
f_I \equiv \frac{2}{3}\omega_I \vert\bra{\Psi_0}\br\ket{\Psi_I}\vert^2
=\frac{2}{3} \left\vert \sum_{q,q'} \br^{\dagger}_q S_{q,q'}^{-1/2}{G}_{I,q'}\right\vert^2 \, ,
\label{eq:osc_str}
\een
where, on the right, $\br = (x, y, z)$ with $x$  being a vector in single-excitation space with matrix elements $x_q = \bra{\phi_i}x\ket{\phi_a}$ and $S$ is a diagonal matrix with elements $S_{qq} = \frac{1}{\nu_q} = \frac{1}{\epsilon_a -\epsilon_i}$. 

While the $\boldsymbol{\Omega}$ (and $\bf A$ and $\bf B$) matrices are formally frequency-dependent through $f\Hxc(\br,\br',\omega)$, quantum chemistry codes usually implement an adiabatic approximation to the kernel, where there is no frequency-dependence, $f\xc^\text{A}(\mathbf{r}, \mathbf{r'})$. 
In this case, Eq.~(\ref{eq:ATDDFT_Casida}) conveniently becomes an ordinary linear Hermitian eigenvalue problem. However, it is well known that an adiabatic approximation yields fewer excitations than that of the true interacting system~\cite{TAHR99,TH00,MZCB04,M22}: $\Omega_{qq'}$ can only mix KS single excitations, precluding access to double (or higher-multiple) excitations.
Excited states of such electronic character are therefore usually either poorly described or completely absent from the ATDDFT spectrum.
DTDDFT provides a practical means of going beyond the adiabatic approximation, introducing an effective frequency-dependent $f\xc(\mathbf{r}, \mathbf{r'}, \omega)$ based on first-principle arguments.
Refs.~\citenum{DM23,DM25} derived matrix elements of the dressed xc kernel,
\begin{equation}
\label{eq:freq-dep_kern}
    f_{{\sss XC}, qq'}(\omega) = f_{{\sss XC}, qq'}^\text{A} + X_{qq'}(\omega) \, ,
\end{equation}
 by considering the diagonalization of the matrix representation of the operator $(\hat{H}~-~E_0)^2$ in the truncated subspace of $N_q$ `relevant' KS single excitations and (so far) one KS double excitation, $D = (s_1 = i \rightarrow a, \; s_2 = j \rightarrow b)$, in the limit that couplings of the double excitation to all other KS single excitations of the system are negligible (e.g., far away in energy). 
 From mapping the eigenvalues onto the corresponding truncated $\boldsymbol{\Omega}$ matrix, and requiring reduction to ATDDFT in the limit that the singles-double Hamiltonian coupling matrix element goes to zero, a frequency-dependent kernel emerges. 
 This kernel can be written as a sum of the usual adiabatic xc kernel and a frequency-dependent correction [Eq.~(\ref{eq:freq-dep_kern})], so that within this truncated subspace of $N_q$ `relevant' KS single excitations, the DTDDFT matrix has the form
\ben
\Omega_{qq'}^\text{DTDDFT}(\omega) = \Omega^{\text A}_{qq'} + 4\sqrt{\nu_q \nu_{q'}}X_{qq'}(\omega)\,.
\een
From the derivation, $X_{qq'}(\omega)$ consists of Hamiltonian matrix elements, but reducing the computational cost by limiting the number of these needed (and possibly also improving the accuracy), different variants were proposed in Refs~\citenum{DM23,DM25} that replaced these matrix elements with various energies obtained from ATDDFT. 
The two main variants that have been explored are:
\begin{equation}
\label{eq:DTDDFTs}
    X_{qq'}^\text{DTDDFTs}(\omega) = \frac{H_{qD}H_{Dq'}}{4\sqrt{\nu_q\nu_{q'}}}\left[1 + \frac{(\nu_q + \nu_D)(\nu_{q'} + \nu_D)}{\omega^2 - {\nu_D}^2}\right]
\end{equation}
and
\begin{equation}
\label{eq:DTDDFTa}
    X_{qq'}^\text{DTDDFTa}(\omega) = \frac{H_{qD}H_{Dq'}}{4\sqrt{\nu_q\nu_{q'}}}\left[1 + \frac{(\Omega_q^\text{A} + \Omega_D^\text{A})(\Omega_{q'}^\text{A} + \Omega_D^\text{A})}{\omega^2 - {\Omega_D^\text{A}}^2}\right] \, .
\end{equation}
 Here, $\nu_D \equiv \nu_{s_1} + \nu_{s_2}$ and $\Omega_D^\text{A} \equiv \Omega_{s_1}^\text{A} + \Omega_{s_2}^\text{A}$ are the frequencies of the KS double excitation and its ATDDFT counterpart, respectively, with $\Omega_{s_{1(2)}}^\text{A}$ being the ATDDFT frequencies that correct the bare KS frequencies, $\nu_{s_{1(2)}}$.
Likewise, $\nu_q$ and $\Omega_q^\text{A}$ are the KS excitation frequency and its ATDDFT counterpart for the single excitations $q$ that mix with the KS double excitation.  
These ingredients are directly outputted from standard ATDDFT codes.
The $H_{qD}$ are the matrix elements of the interacting Hamiltonian, which couple a given KS single excitation to the chosen KS double excitation (see also Sec.~\ref{subsec:Comp_det}). 

Now we clarify an important point regarding the underlying adiabatic DFA upon which the dressing is applied. In the earlier work~\cite{DM23,DM25}, it appeared that  $X_{qq'}(\omega)$ was derived  assuming a pure underlying adiabatic DFA, for which $ \Omega^\text{A}_{qq'} = \delta_{qq'} \nu_q^2 + 4\sqrt{\nu_q\nu_{q'}}f_{{\sss HXC},qq'}^\text{A}$. 
Indeed, this expression only holds for a pure DFA, and yet several of the examples given in Refs.~\citenum{DM23,DM25} used DTDDFT built upon a hybrid adiabatic functional. 
When using a hybrid DFA, the result for  $X_{qq'}(\omega)$  nevertheless remains the same, provided we view this {\it correction} as arising from a pure functional of the density.  
We may write 
\begin{equation}
\label{eq:DTDDFT_gen}
    \Omega^\text{DTDDFT}_{qq'}(\omega) = \Omega^\text{A}_{qq'} + \left[{\left(\mathbf{A} - \mathbf{B}\right)}^{1/2} {\left(4\mathbf{X}(\omega)\right)} {\left(\mathbf{A} - \mathbf{B}\right)}^{1/2}\right]_{qq'}
\end{equation}
where $\Omega^\text{A}_{qq'}$  can come from either a pure or orbital-dependent adiabatic approximation.  

Due to its frequency-dependence, after $\Omega^\text{DTDDFT}_{qq'}(\omega)$ is constructed, it needs to be iteratively diagonalized in the truncated subspace until eigenvalue convergence is reached, after which the corresponding eigenvectors are appropriately normalized according to Eq.~(\ref{eq:Casida_G_renorm}) and oscillator strengths can be obtained. 
In a loose sense, the magnitude of the eigenvector obtained from Eq.~(\ref{eq:Casida_G_renorm}) can be viewed as an indication of single-excitation character~\cite{DM23}, at least in the weakly correlated limit. 
However, given that such a concept is heavily reference dependent, too much quantitative meaning should not be attached to it. 

The key advantage of the DTDDFT approach of Refs~\citenum{DM23,DM25} compared to the earlier dressed TDDFT of Refs~\citenum{MZCB04,CZMB04}, is that the oscillator strength sum-rule is satisfied. 
The latter work, since it operates in the framework of the TDA, where the matrix $\mathbf{B} = \mathbf{0}$, cannot reliably predict oscillator strengths or transition densities~\cite{DM23}. 
We will compare energies and state characters between DTDDFT of Eqs.~(\ref{eq:DTDDFTs})-(\ref{eq:DTDDFTa}), with this earlier approach, denoted dressed TDA (DTDA).
We recall the relevant equations for DTDA here. 
First the adiabatic TDA (ATDA),
\begin{equation}
\label{eq:ATDA}
    \mathbf{A} \mathbf{X}_I = \omega_I \mathbf{X}_I \, ,
\end{equation}
is an ordinary Hermitian eigenvalue problem which yields $\omega_I$ directly as its eigenvalues (rather than $\omega_I^2$).
 The matrix elements of the corresponding DTDA matrix are then given as~\cite{MZCB04,DM23,M22}
\begin{equation}
\label{eq:DTDA_mat_ele}
   A_{qq'}^\text{DTDA}(\omega) = A_{qq'} + 2X_{qq'}(\omega) \, ,
\end{equation}
where
\begin{equation}
\label{eq:DTDAs}
    X_{qq'}^\text{DTDAs}(\omega) = \frac{H_{qD}H_{Dq'}}{2\left(\omega - \nu_D\right)}
\end{equation}
 and
\begin{equation}
\label{eq:DTDAa}
    X_{qq'}^\text{DTDAa}(\omega) = \frac{H_{qD}H_{Dq'}}{2\left(\omega - \Omega_D^\text{A}\right)} \, .
\end{equation}
As in the full DTDDFT case, the DTDA matrix is diagonalized self-consistently due to its frequency dependence, with the resulting converged eigenvectors, $\mathbf{X}_I$, normalized according to a modified version of Eq.~(\ref{eq:Casida_G_renorm}),
\begin{equation}
\label{eq:Casida_G_renorm_TDA}
    \mathbf{X}_I^\dagger \left( \mathds{1} - \left[ \frac{1}{2\omega} \frac{\partial \mathbf{A}(\omega)}{\partial \omega} \right]_{\omega = \omega_I} \right) \mathbf{X}_I = 1 \, .
\end{equation}

\section{DTDDFT in Practice}
\label{sec:DTDDFT_practice}

\subsection{Operational Considerations}
\label{sec:Operational}

We now discuss a series of operational considerations that are important in the implementation and practical use of the DTDDFT equations presented in Section ~\ref{sec:DTDDFT_theory} above.
They range from the more obvious to the more technical, and can be summarized via the following set of questions, each of which we expand on in turn and loosely connect to our benchmark results reported later in Section~\ref{sec:Results}.

We begin with a key question, which, in answering, defines our protocol in the present study.

\begin{enumerate}
 \item  \textit{How do we choose which ATDDFT state(s) to correct, and which KS excitations to include in the DTDDFT problem we set out to solve?} 
    Sometimes, a priori knowledge, or `common wisdom', informs us which states are expected to have double-excitation character. For example, for polyene chains, the dark 2A$_g$ state is typically flagged, and trans-butadiene has become a sort of fruit-fly for this problem, but a truly black-box method for a general molecule would be independent of  this prior knowledge. 
The DTDDFT matrix is expressed in the space of KS single excitations, so equally, we need to understand which KS single excitations to include in the truncated basis, as well as the double excitation to couple. Note that in the present work, we consider only one double-excitation as this is the most common situation, but the framework in the previous section could be generalized to more than one double.

One way to proceed would be to first consider the lowest KS double excitation in the system, that is, twice the HOMO-LUMO transition (assuming a closed-shell ground state). Symmetry considerations could then help select the ATDDFT states that should be corrected by the DTDDFT procedure: this double excitation transforms as the totally symmetric irreducible representation of the molecular point group, and  the Hamiltonian, given its scalar nature, can only couple to states which have that same symmetry. 
In the present work, the KS double excitations we consider are all HOMO$^2 \to$ LUMO$^2$, except for one case of cyclopentadienone where also the case of (HOMO $\to$ LUMO, HOMO $-1 \to$ LUMO) was considered. 

Energetic considerations  would then suggest to consider the lowest few, say $N$, ATDDFT states of this symmetry (whether or not the molecule possesses point group symmetry). We note that energetic considerations alone are not always good predictors of coupling: for example, we find in some cases that the ATDDFT root that gets the largest shift is not always the one energetically closest to the energy of the KS double excitation included in DTDDFT, due to the size and phase of the $H_{qD}$ elements involved, see also question 9 shortly.
Ultimately how many ATDDFT roots are initially considered is up to the user, but for a chosen KS double excitation, the size of $H_{qD}$ for the KS single excitations that contribute to a given ATDDFT root is a key factor as to which states are significantly affected, as we discuss next.

Once $N$ relevant ATDDFT state(s) and the KS double excitation have been selected, then we use the following protocol to choose the KS single excitations in which to form the DTDDFT matrix: 
\newline
(i) include all the KS single excitations that contribute 10\% or more to any of these ATDDFT states,
\newline
(ii) out of the remaining KS single excitations contributing to the $N$ ATDDFT excitations, consider those that contribute to more than half of the ATDDFT roots of interest,
\newline
(iii) include any of these remaining KS transitions for which $\vert H_{qD}\vert$ is larger than the smallest $\vert H_{qD}\vert$ for the transitions accepted in (i); such KS transitions may contribute only weakly to the ATDDFT energies but couple relatively strongly to the KS double excitations being considered, so can have a finite correction to the DTDDFT result. 

Note that the $H_{qD}$ matrix elements are constructed from two-electron integrals according to the Slater-Condon rules, accounting for spin-adaptation; one-body integrals typically vanish.  Both the symmetry (if present) and the Slater-Condon rules, filter out many KS single excitations that will not couple with the chosen KS double excitation. 
Table I in the Supplementary Information (SI) reports the two-electron orbital patterns and prefactors for constructing the $H_{qD}$ matrix elements for the two types of KS double excitation considered in this work.

The fractions in this protocol are up to the user, however in our experience with the molecules in this study, we found they were a good guide and that including further KS transitions had a negligible effect on the results.  

Table II in the SI reports the KS double and single excitations considered in the DTDDFT problem for each molecule, along with which ATDDFT states are corrected.
  
    \item  \textit{Which variant should we use?} DTDDFT, as presented in Refs~\citenum{DM23} and \citenum{DM25}, focussed on two variants, DTDDFTs and DTDDFTa, but, as noted in the previous section, the particular ingredients in these choices are not derived from first-principles but rather motivated from physical and practical considerations.
    There could be other choices; for example, replacing certain excitation energies with Hamiltonian matrix elements~\cite{DM23,BM25}, using the KS frequencies for the single excitations and ATDDFT for the double or vice-versa,
    in a kind of mixed s- and a-variant, or replacing the adiabatic frequencies in the dressing with matrix elements of the adiabatic $\Omega$ in the truncated basis. This last possibility is however discouraged due to violation of an exact condition,  explained in Sec.~\ref{sec:undownfolding}. 
    For practical reasons, we rule out any variant that would increase the number of two-electron integrals needed. 
    
    In this paper, we explore the originally proposed s- and a-variants, with an aim to determine which gives more accurate results. We might expect, on the one hand, that DTDDFTa should, because the frequency-dependent part of the dressing incorporates adiabatic-TDDFT correlation in its ingredients; on the other hand, one might argue that this is already accounted for in the native $f\Hxc^{\rm A}$ term. In fact, we will find that DTDDFTa is more accurate. A reason for this could be understood from considering the original derivation~\cite{DM23}:  the diagonalization yields matrix elements of the Hamiltonian in the truncated basis extended by the KS double excitation, with the derivation arising out of the assumption that the coupling between these states is much stronger than to any of the other KS states in the system. To replace these with adiabatic counterparts, DTDDFTa would allow us to reintroduce dynamic correlation in the dressing ingredients, while DTDDFTs reintroduces it only at the zeroth order stage [see Eq. (9) of Ref.~\citenum{DM23} and subsequent discussion]. 
    We show the results for both DTDDFTs and DTDDFTa for (trans-)butadiene and (all-trans-)hexatriene in Section~\ref{subsec:polyenes}; for the remaining molecules we show results for only DTDDFTa. 

The DTDDFTa variant comes with another choice to be made: how to associate the $\Omega_q^{\rm A}$ in the dressing numerator to the particular KS transition $q$, and likewise the $\Omega_{s_i}$ that compose the double excitation $\Omega_D$? When one KS transition dominates the ATDDFT root, then the choice is naturally to associate that ATDDFT frequency with the index associated with KS transition. However, when either multiple ATDDFT roots have the same $q$ as their most dominant KS single-excitation contribution, or when one ATDDFT root has equally large KS contributions from multiple $q$, the choice is less clear. In our experience with the molecules examined here, the assignment of which $\Omega_q$ goes with which, has only a secondary effect, provided all the relevant KS single excitations are included.

    \item \textit{Which ground-state DFA should we use for the adiabatic kernel on top of which the dressing is performed?} There exists a ‘zoo’ of density functional approximations  for the xc potential in ground-state DFT~\cite{goerigk2017look}.
    Codes typically take the same approximation for the xc  kernel as was used to obtain the KS orbitals, which would lead to consistent results with a perturbative response approach based on the ground-state functional. The choice of ground-state xc affects the quality of the orbitals, the KS excitation frequencies, and adiabatic TDDFT frequencies, all of which are ingredients in DTDDFT. 
    Ref.~\citenum{HIRC11} had found that with the earlier dressed TDDFT, i.e. DTDA, using a base GGA functional tended to significantly underestimate the double excitation energies, and they recommended using a local hybrid. We will find that observation persists for DTDDFT also, but we further find that range-separated hybrids (not considered in Ref.~\citenum{HIRC11}) often tend to give improved results over local hybrids as the underlying adiabatic functional; this turned out to be the case in about half of the excitations studied here.
    
    \item \textit{It is expected that DTDDFT affords better transition densities than the DTDA, but how do their energies compare?} DTDDFT naturally satisfies the Thomas-Reiche-Kuhn sum rule for oscillator strengths~\cite{thomas1925zahl, reiche1925zahl, kuhn1925gesamtstarke, DM23} whereas DTDA does not because it is based on the Tamm-Dancoff approximation. The studies so far suggest that both the character of the state (fraction of single and double) and the excitation energy are not predicted as accurately with DTDA as with DTDDFT, and we find that this holds somewhat true in general for the wider range of molecules studied here. In some cases, we found the DTDA excitation energies were very close to DTDDFT ones for the same underlying functional, even if their character was not predicted as well.   
    
    \item \textit{Can DTDDFT be expected to describe excited states of ‘partial’ and ‘genuine’ double-excitation character on the same footing?} 
   The characterization of  ‘partial’ or ‘genuine’  double-excitation character has been adopted to analyze the performance of high-level wavefunction methods for these states~\cite{KBLJ24},  based on the value of the $\%T_1$ diagnostic\cite{lee1989theoretical,lee1989diagnostic} of coupled-cluster theory: genuine double ($\%T_1$ $\lesssim$ 20$\%$), partial double (20$\%$ $<$ $\%T_1$ $\lesssim$ 85$\%$), single ($\%T_1$ $>$ 85$\%$). While $|G_I|^2$ gives an indication of the single-excitation character of our DTDDFT excitation, comparing this measure to $\%T_1$ is like comparing apples to pears: not only is the reference out of which the single or double character of the state is defined different, but also $|G_I|^2$ is at best an approximate measure, valid when the true ground state is dominated by a single determinant~\cite{DM23}. Nevertheless, with this in mind, we will still use $\%T_1$ as a benchmark to compare with $\vert G_I\vert^2$, and investigate whether the errors of DTDDFT for excitation energies are significantly different for the different categories of double-excitation character above. It is useful to bear in mind the limit of zero coupling: without any mixing, the frequencies resulting from DTDDFTa reduce to the sum of the ATDDFT-corrected frequencies of the two KS transitions making up the double ($\Omega_D$), and the truncated ATDDFT frequencies. 
We will find  that for the genuine doubles in ethylene (Sec.~\ref{subsec:ethylene}), borole (Sec.~\ref{subsec:borole}) and cyclopentadienone (Sec.~\ref{subsec:cyclopentadienone}) the energies predicted from DTDDFTa are indeed very close to the sum of the ATDDFT-corrected frequencies. 
      
    \item \textit{Does DTDDFT always yield real excitation energies?}
    While the adiabatic TDDFT matrix in its Hermitian form [Eq.~(\ref{eq:ATDDFT_Casida})] always yields real eigenvalues, this does not guarantee they are strictly positive. Negative eigenvalues would mean imaginary frequencies, which has been termed an `instability' in the literature. 
    ATDDFT has been shown to exhibit both triplet instabilities~\cite{PWT11,peach2012overcoming,peach2013triplet}, and singlet near-instabilities~\cite{CDICFV07,TTRF08}  (where the frequency becomes alarmingly close to zero), typically  at geometries away from the Franck-Condon region. A fresh perspective on this unphysical behavior has recently been provided in terms of CT-symmetry breaking~\cite{FC25}. Now with the non-linearity associated with frequency-dependence of DTDDFT, the problem appears less clear: will the eigenvalues of $\Omega^{\rm DTDDFT}$ even be real? In Sec.~\ref{sec:undownfolding}, we will recast the equation into a linear Hermitian matrix form, guaranteeing real eigenvalues for $\omega^2$. 
    Still, we will find that  these are not guaranteed to be non-negative and indeed imaginary frequencies may arise from DTDDFT in some of our examples, as exemplified by (trans-)butadiene and (all-trans-)hexatriene (Section~\ref{subsec:polyenes}).

    \item \textit{How do we know which dressed solution to iterate on in the non-linear matrix?}
 The DTDDFT equations represent a non-linear eigenvalue problem in the space of $N_q$ KS single excitations.
 When $N_q = 1$, this reduces to the dressed small matrix approximation (DSMA), where  one can simply solve the quadratic equation for $\omega^2$, and find two roots, of mixed single and double character. For $N_q>1$, the procedure has been to iterate on a particular excitation, typically starting with its ATDDFT value, finding a correction to that excitation~\cite{MW09,DM23}.  This can cause some ambiguity:  each pass through the iteration generates $N_q$ dressed eigenvalues, and especially in dense manifolds of states, it may be unclear which one should be iterated on further.
    Picking one of these solutions would likely give $N_q$ new solutions different to those if another solution was chosen instead, but these other solutions are ignored anyway. Further, one would expect to obtain $N_q +1$ states if we mix $N_q$ singles with 1 double (consistent with the outcome of DSMA in the case of $N_q = 1$). This implies we should do another iteration starting  from the KS double excitation ($\nu_D$ or $\Omega_D$) but this state has not been explored at all in previous work (apart from the $N_q = 1$ DSMA case~\cite{DM23}). 
  To circumvent this problem and the need to choose which dressed solution to iterate on, we can lift the non-linear $N_q \times N_q$ matrix problem to a larger linear matrix problem that can be solved in one diagonalization. One obtains $N_q + 1$ solutions, $N_q$ of which, by construction, are the same as the $N_q$ solutions obtained from the previously outlined downfolded DTDDFT approach~\cite{DM25} upon appropriate convergence. These can be viewed as representing doubles-corrected ATDDFT energies, with an additional $N_q +1)$th solution arising from a double-excitation corrected by mixing with single excitations (which, depending on the amount of mixing, could retain a primary double character, or a more even mix of double and singles). 
Section~\ref{sec:undownfolding} outlines the derivation and further details of this alternative ``un-downfolded'' form of the DTDDFT equations, and we use it for all the molecules considered in the present work.
    
    \item \textit{How should we proceed if the adiabatic states generated within a chosen truncated subspace differ in energy significantly from the ‘full’ states obtained from the original ATDDFT calculation?} DTDDFT is certainly not intended to reduce to a diagonalization in the basis of KS excitations; this would destroy its favorable system-size scaling. Instead, the DTDDFT procedure should be performed in a truncated  single-excitation  subspace: that which is determined by the working protocol outlined in Question 1. In some cases, however, the ATDDFT excitation may involve small contributions from many high-lying KS states, in addition to the relevant KS states included in the DTDDFT matrix. In this case, we consider an \textit{a posteriori} correction, dubbed $\Delta$c-DTDDFT, which accounts for the difference between the full adiabatic states that are obtained from ATDDFT calculations in the full space of KS single excitations, and their truncated equivalents that the dressing term in DTDDFT directly acts upon, i.e., the difference arises due to the extra dynamical correlation in the former. The correction takes the form of an energetic shift, such that $\Omega_I^{\Delta \text{c-DTDDFT}} = \Omega_I^\text{DTDDFT} + \Delta_I$, where $\Delta_I = \Omega_I^\text{full ATDDFT} - \Omega_I^\text{trunc. ATDDFT}$, and analogously for $\Delta $c-DTDA. Since truncation occurs within the adiabatic single-excitation subspace, $\Delta_I$ is only applied to dressed solutions that have a significant proportion of single-excitation character, that is, to the ``$N_q$ solutions'' referred to in Question 7 above. 
    (Trans-) acrolein and naphthalene serve as two striking examples (Sections~\ref{subsec:acrolein} and \ref{subsec:naphthalene}, respectively), where the effect of truncation is significant. 

    \item\textit{For a given selection of ATDDFT states to correct, can we predict which will be significantly corrected by the double excitation before applying DTDDFT?}
As evident from the procedure itself, a key determining aspect for how much a single excitation mixes with a specified double excitation is the size of the coupling matrix elements $H_{qD}$ for the KS single excitations that constitute the ATDDFT root. If several ATDDFT roots involve a KS single excitation (possibly the same one) with a large $H_{qD}$, a question arises as to whether we can predict which root is most affected. From resonance arguments, one might expect that the closer the ATDDFT excitation energy  to the double excitation (i.e. $\nu_D$ for the case of DTDDFTs, and $\Omega_D$ in DTDDFTa), the more they mix. However, while this argument might have some merit when mixing only one single excitation with one double excitation, it does not typically hold true when several single KS excitations are involved. 
This is because the signs with which these excitations contribute to the ATDDFT root can lead to cancellations within the DTDDFT solution. Typically the lowest ATDDFT root will have the larger number of same signs, and so tends to get shifted the most by DTDDFT.
Illustrative examples are (all-trans-)octatetraene and cyclopentadiene (Section~\ref{subsec:octaetraene_cyclopentadiene}), for which two single KS excitations mix strongly with a double-excitation for which $\nu_D$ and $\Omega_D$ are both far closer to the upper ATDDFT root than the lower one. The correction from DTDDFT is far more significant for the lower root than the upper root. The same is also true for the smaller chain polyenes, butadiene and hexatriene. 
\end{enumerate}

\subsection{From Non-linear to Linear via `Un-downfolding'}
\label{sec:undownfolding}

The iterative solution for the eigenvalues of the non-linear $\Omega^{\rm DTDDFT}$ matrix requires a choice to be made as to which solution to iterate upon, as discussed in Question 7 above. 
To avoid this, we may recast the equation into a linear form. 
The original derivations of DTDDFT (and DTDA) (Refs.~\citenum{MZCB04,DM23}) began with a linear problem of dimension $N_q + 1$ (the truncated TISE), which had been `downfolded' into a matrix of dimension $N_q$ spanning the KS single-excitations, in order to be able to map onto a TDDFT form, identify the $f\Hxc$ matrix elements, and tweak them to restore some dynamic correlation with states outside the truncation.
We now `un-downfold' that procedure to get a linear $(N_q + 1)$-dimensional matrix again whose eigenvalues and eigenvectors are equivalent to those of DTDDFT. 

To avoid notational clutter, we explicitly write the case where $N_q = 2$ KS single excitations couple strongly to one KS double, and illustrate the procedure using DTDDFTa but it is straightforward to see how it works for the other variants, and we will generalize at the end to a larger $N_q$. 

We separate out the frequency-dependence in the DTDDFTa matrix, writing it as:
\begin{equation}
\Omega_{qq'}^\text{DTDDFTa}(\omega) = \tilde{\Omega}_{qq'}^\text{A} + \frac{\tilde{Y}_{qq'}}{\omega^2 - {\Omega_D^\text{A}}^2} \, ,
\label{eq:OmegaDTDDFT}
\end{equation}
where
\begin{eqnarray}
\tilde{\Omega}_{qq'}^\text{A} &=& \Omega^\text{A}_{qq'} + \left[\left(\mathbf{A} -\mathbf{B}\right)^{1/2}\left(4\text{\textbf{X}}_0\right)\left(\mathbf{A} - \mathbf{B}\right)^{1/2}\right]_{qq'} \, , \\
    \tilde{Y}_{qq'} &=& \left[\left(\mathbf{A} -\mathbf{B}\right)^{1/2}\left(4\text{\textbf{Y}}_0\right)\left(\mathbf{A} - \mathbf{B}\right)^{1/2}\right]_{qq'} \, ,
\end{eqnarray}
with
\begin{eqnarray}
    Y_{0,qq'} &=& X_{0,qq'}(\Omega_q^\text{A} + \Omega_D^\text{A})(\Omega_{q'}^\text{A} + \Omega_D^\text{A}) \, , \label{eq:y0} \\
    X_{0,qq'} &=& \frac{H_{qD}H_{Dq'}}{4\sqrt{\nu_q\nu_{q'}}} \, . \label{eq:x0}
\end{eqnarray}
Finding the dressed eigenvalues and eigenvectors then requires diagonalizing Eq.~(\ref{eq:OmegaDTDDFT}), i.e., solving for $\det \left( \Omega_{qq'}^{\rm DTDDFT} -\omega^2\mathds{1} \right) = 0$ in the chosen truncated subspace, as described earlier. 
For the case of $N_q = 2$, this gives:

\begin{eqnarray}
\nonumber
-\tilde{\Omega}_{12}^\text{A}\tilde{\Omega}_{21}^\text{A} &+&\left(\omega^2 - \tilde{\Omega}_{11}^\text{A}\right)\left(\omega^2 - \tilde{\Omega}_{22}^\text{A}\right)   \\
\nonumber
&-& \frac{\tilde{Y}_{22}\left(\omega^2 - \tilde{\Omega}_{11}^\text{A}\right) + \tilde{Y}_{11}\left(\omega^2 - \tilde{\Omega}_{22}^\text{A}\right) + \tilde{\Omega}_{12}^\text{A} \tilde{Y}_{21} + \tilde{\Omega}_{21}^\text{A} \tilde{Y}_{12}}{\omega^2 - {\Omega_D^\text{A}}^2} \\
&+&\frac{\tilde{Y}_{11}\tilde{Y}_{22} - \tilde{Y}_{12}\tilde{Y}_{21}}{\left(\omega^2 - {\Omega_D^\text{A}}^2\right)^2} = 0 \, .
\label{eq:secular}
\end{eqnarray}
Noting that we should obtain 3 eigenvalues for $\omega^2$, given that we had originally started with 3 states, the last term in Eq.~(\ref{eq:secular}) must be zero, i.e., 
\ben
\tilde{Y}_{11}\tilde{Y}_{22} =\tilde{Y}_{12}\tilde{Y}_{21} \, .
\label{eq:exactcond}
\een
Equation~(\ref{eq:exactcond}) defines the exact condition for the $N_q = 2$ case for any proposed variant that has the structure of a term with a pole in $\omega^2$ added to an adiabatic matrix. 
For general $N_q$, there are $N_q - 1$ terms in Eq.~(\ref{eq:secular}), each with a denominator $\left(\omega^2 - {\Omega_D^\text{A}}^2\right)^{k}$ where $k = 2, 3, \dots, N_q$, and numerator that must also vanish such that only $N_q + 1$ eigenvalues are obtained.
Choosing a dyadic form for the residue of the second term in Eq.~(\ref{eq:OmegaDTDDFT}) will ensure the form is satisfied. This is the case for all the variants considered so far~\cite{DM23,BM25}. 
For DTDDFTa, for example, we may write ${\bf Y_0} = \mathbf{z z^T}$, where $z_q = \left(H_{qD}/2 \sqrt{\nu_{q}}\right)(\Omega^{\rm A}_{q} +\Omega^{\rm A}_D)$; similarly for the s-variant with $\Omega_q^{\rm A}$ replaced with $\nu_q$. 
But not all variants that might seem to be reasonable approximations necessarily satisfy this. For example, the general DTDDFT structure arose from first considering diagonalizing in the space of relevant KS singles plus the KS double excitation, resulting in a preliminary expression with a numerator involving $H_{qq'} - E_0$~\cite{DM25}, so a natural proposition could be to replace this with the corresponding adiabatic matrix element, yielding $Y_{0,qq'} = X_{0,qq'}(\Omega^{\rm A}_{qq'} + \Omega_D^{\rm A})^2$. However such a choice violates Eq.~(\ref{eq:exactcond}). In this way, Eq.~(\ref{eq:exactcond}) for the $N_q = 2$ case, and the dyadic form of $\mathbf{Y_0}$ for general $N_q$, serves as a condition for any proposed dressed approximation to satisfy, ensuring its connection to the physical Hamiltonian matrix diagonalization procedure from which the form was derived. 

Returning to Eq.~(\ref{eq:secular}), we now look for an `undownfolded' (udf) matrix of size $N_q +1$ whose eigenvalues are the same as those yielded by the secular equation. Keeping with the $2 \times 2$ KS single subspace for illustration, we write such a matrix as  

\begin{equation}
\boldsymbol{\Omega}_{\rm udf} =
    \begin{bmatrix}
      S_{11} & S_{12} & C_{1D} \\
      S_{21} & S_{22} & C_{2D} \\
      C_{1D} & C_{2D} & D  \\
    \end{bmatrix}
\end{equation}
Finding the roots of $\det(\Omega_{\rm udf} - \omega^2\mathds{1})  = 0$, and comparing with the solutions of Eq.~(\ref{eq:secular}) (with the last term there set to zero), we make the following identifications:
\begin{equation}
\boldsymbol{\Omega}^\text{DTDDFTa}_\text{udf} =
    \begin{bmatrix}
      \tilde{\Omega}_{11}^\text{A} & \tilde{\Omega}_{12}^\text{A} & \sqrt{\tilde{Y}_{11}} \\
      \tilde{\Omega}_{21}^\text{A} & \tilde{\Omega}_{22}^\text{A} & \frac{\tilde{Y}_{12}}{\sqrt{\tilde{Y}_{11}}} \\
      \sqrt{\tilde{Y}_{11}} & \frac{\tilde{Y}_{12}}{\sqrt{\tilde{Y}_{11}}} & {\Omega_D^\text{A}}^2  \\
    \end{bmatrix}
\end{equation}
Generalizing to the case of an $N_q$ KS single-excitation subspace:
\begin{equation}
    \boldsymbol{\Omega}^\text{DTDDFTa}_\text{udf} =
        \begin{bmatrix}
        \tilde{\Omega}_{11}^\text{A} & \tilde{\Omega}_{12}^\text{A} & \hdots & \tilde{\Omega}_{1N_q}^\text{A} & \sqrt{\tilde{Y}_{11}} \\
        \tilde{\Omega}_{21}^\text{A} & \tilde{\Omega}_{22}^\text{A} & \hdots & \tilde{\Omega}_{2N_q}^\text{A} & \frac{\tilde{Y}_{12}}{\sqrt{\tilde{Y}_{11}}} \\
        \vdots & \vdots & \ddots & \vdots & \vdots \\
        \tilde{\Omega}_{N_q1}^\text{A} & \tilde{\Omega}_{N2}^\text{A} & \hdots & \tilde{\Omega}_{N_qN_q}^\text{A} & \frac{\tilde{Y}_{1N}}{\sqrt{\tilde{Y}_{11}}}\\
        \sqrt{\tilde{Y}_{11}} & \frac{\tilde{Y}_{12}}{\sqrt{\tilde{Y}_{11}}} & \hdots & \frac{\tilde{Y}_{1N_q}}{\sqrt{\tilde{Y}_{11}}} & {\Omega_D^\text{A}}^2 \\
        \end{bmatrix} 
        \label{eq:DTDDFTa_udf}
\end{equation}
and replacements for DTDDFTs follow straightforwardly ($\Omega_D^{\rm A} \to \nu_D$, $\Omega_q^{\rm A} \to \nu_q$). We note that $H_{1D}$, which enters $\tilde{Y}_{11}$ (and $\tilde{Y}_{1q}$) through Eqs~\ref{eq:y0} and \ref{eq:x0}, is assumed to be non-zero; if not, the corresponding matrix elements in the last row and column in Eq.~\ref{eq:DTDDFTa_udf} should be defined in terms of another $\tilde{Y}_{ii}$ (and $\tilde{Y}_{iq}$).

Eq.~(\ref{eq:DTDDFTa_udf}) gives the same eigenvalues as would be obtained from the iterative solution for eigenvalues of the non-linear matrix Eq.~(\ref{eq:DTDDFT_gen}), but in a much simpler way, avoiding the question of which solution to iterate on, the question of whether the other eigenvalues produced in the iteration have meaning, and obtaining all the dressed excitation energies from one diagonalization (see earlier discussion in Question 7). Further, the $\vert G_I\vert^2$ for each eigenvalue can be read off from summing the squares of the first $N_q$ elements in the corresponding normalized eigenvector, and can be shown to agree with the $\vert G_I\vert^2$ obtained from Eq.~(\ref{eq:Casida_G_renorm}).

A similar undownfolding follows for the DTDA case, transforming the non-linear matrix of dimension $N_q$ [Eq.~(\ref{eq:DTDA_mat_ele})], into a linear one of dimension $N_q +1$. Since DTDA does not preserve the oscillator strength sum-rule, the $\vert G_I\vert^2$ obtained from the normalized eigenvector as described above (in the DTDDFT case) does not however correspond to the DTDA prediction correctly obtained from Eq.~(\ref{eq:Casida_G_renorm_TDA}).

\section{Results}
\label{sec:Results}

\subsection{Computational details}
\label{subsec:Comp_det}

Following Ref.~\citenum{DM25}, all DFT and ATDDFT (or ATDA) quantities used as input to the DTDDFT (or DTDA) problem were computed using the NWChem 7.2.3 program package~\cite{nwchem,nwchem20}.
The ground-state energy, KS orbital energy differences (i.e., KS excitation energies) and ATDDFT (or ATDA) excitation energies were extracted from the default NWChem output; adiabatic $(\mathbf{A} + \mathbf{B})$ and $(\mathbf{A} - \mathbf{B})^{1/2}$ Casida matrix elements were extracted after specifying the ‘debug’ print level; and two-electron (molecular orbital) integrals in the KS basis used to construct $H_{qD}$ matrix elements were extracted upon requesting FCIDUMP file generation.
The PBE~\cite{PBE96} (GGA), PBE0~\cite{PBE0} (hybrid GGA) and CAM-B3LYP~\cite{YTH04} (range-separated hybrid GGA) exchange-correlation functionals were used throughout in combination with the aug-cc-pVDZ (for acrolein and ethylene) or cc-pVDZ (for all other molecules) basis set~\cite{D89}.
Both the s- and a-variants of the dressing [Eqs.~\eqref{eq:DTDDFTs}/\eqref{eq:DTDAs} and \eqref{eq:DTDDFTa}/\eqref{eq:DTDAa}, respectively] are considered in the analysis of butadiene and hexatriene, however, for all other molecules, only the performance of the a-variant is investigated.
The reference wavefunction values (i.e., excitation energies and $\%T_1$ values) included in Tables~\ref{tab:octatetraene}-\ref{tab:cyclopentadienone} and Figs~\ref{fig:butadiene_curves} and \ref{fig:hexatriene_curves} were taken from the QUESTDB~\cite{VSCL21,KBLJ24,loos2025quest} or Ref.~\citenum{PSLP21}.
Table III in the SI collates the specific wavefunction method / basis set combinations used to calculate the reference excitation energies of each molecule and excited state studied in this work; all reference $\%T_1$ values were computed using CC3 and aug-cc-pVTZ, as reported in Ref.~\citenum{KBLJ24}.
Except for the butadiene and hexatriene curves (see Sec.\ref{subsec:polyenes}), all quantities (i.e., reference and those from this work) were calculated solely at the S$_0$ minimum (commonly denoted FC) geometry of each molecule, the latter taken from Ref.~\citenum{KBLJ24}, optimised at the CC3/aug-cc-pVTZ level of theory.

We note that the tables in this section are organized as follows: full ATDDFT excitation energies and DTDDFT/DTDA excitation energies are reported in Roman text, truncated ATDDFT excitation energies in italics and $\Delta$c-DTDDFT/$\Delta$c-DTDA excitation energies in bold. For states of genuine double-excitation character, where $\Delta$c-DTDDFTa / $\Delta$c-DTDAa results are not calculated, we also give the corresponding excitation energies in bold.
Note that the $\Delta$c correction applies only to the excitation energies;  $\vert G\vert^2$ values are unchanged from their dressed values.

\subsection{Polyene curves: butadiene and hexatriene}
\label{subsec:polyenes}

The low-lying excited states of short linear polyenes, such as butadiene and hexatriene, have been widely studied~\cite{HK72,SK72,SOK76,AH78,CD87,FLMSST98,CBBGF04,LM07,LM09,KOK13,B22,FPC26}, with the latter acting as models for natural chromophores critical to vision and light-harvesting, as well as technologically relevant molecular photoswitches and organic photovoltaics~\cite{OZZ91,AO67,B88_chapter,CPLDHC02,KMYWM05,HDRS98,W76,PKVCV02}.
The challenge for computational methods to capture these states stems from the partially doubly-excited nature of the dark 2A$_g$ ($\pi\rightarrow\pi^*$) state~\cite{OKS78,LC02,KMLJ24,LBSCJ19,PSLPJFC22}. 
This characterization of the state has been contested, e.g., in Ref.~\citenum{BAS25}, where adiabatic TDDFT out of a highly accurate ground-state KS potential yielded good energies for the state, but the designation of a state as having double-excitation character is certainly reference-dependent~\cite{M22}. In the present work, we find the state couples strongly to double excitations of the KS references computed with the ground-state DFA and basis set combinations chosen here, which suggests the designation of a state of double-excitation character.

\begin{figure*}[h!]
    \centering
    \includegraphics[width=\textwidth]{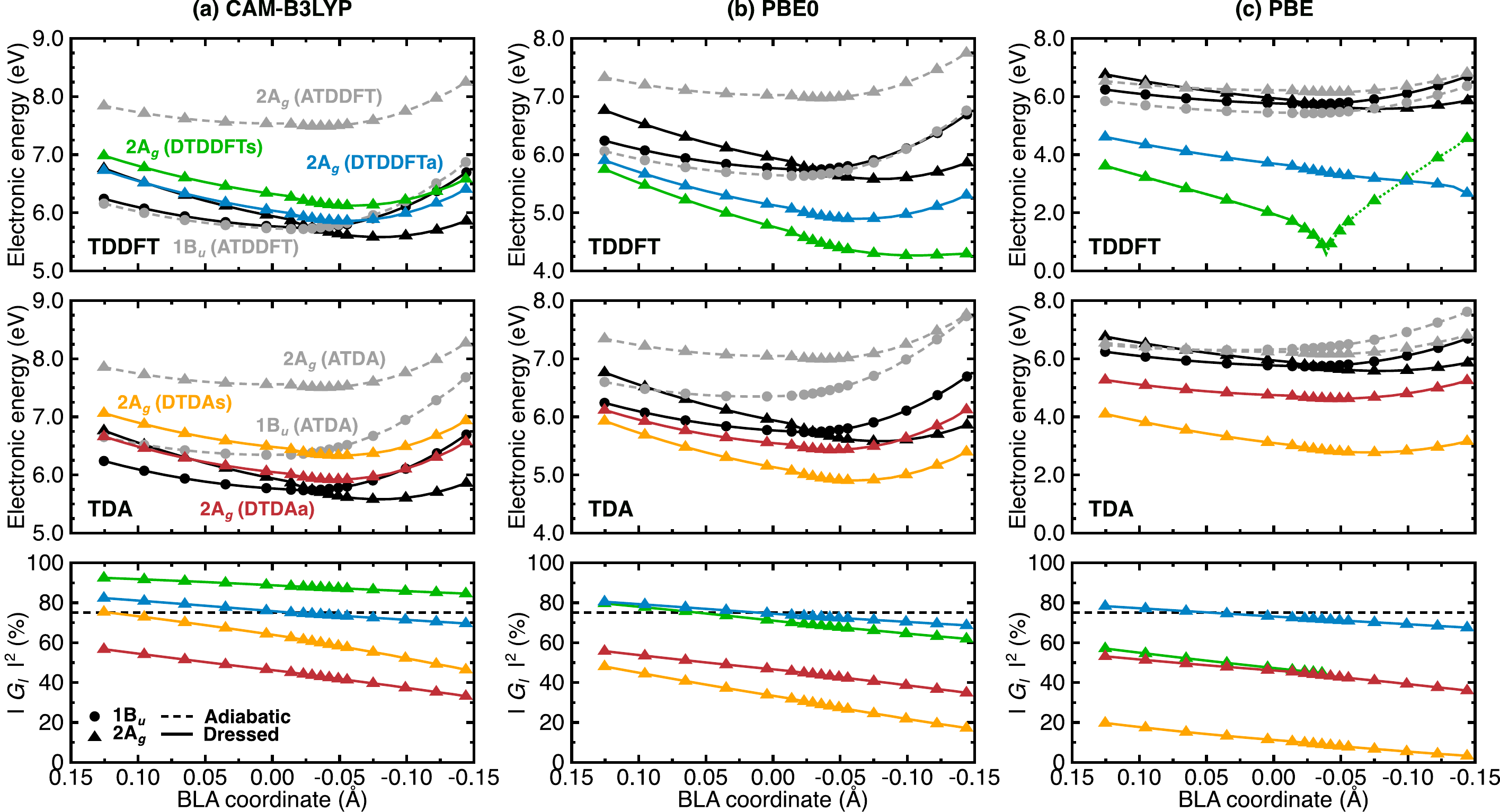}
    \caption{The 1B$_u$ (circles) and 2A$_g$ (triangles) electronic energies along the BLA coordinate in \textbf{butadiene} obtained with ATDDFT, DTDDFTs and DTDDFTa (top panel: grey, green, blue, respectively), as well as ATDA, DTDAs and DTDAa (middle panel: grey, yellow, red, respectively), using CAM-B3LYP (a), PBE0 (b) and PBE (c) computed within cc-pVDZ. 
    Dressed $|G_I|^2$ values corresponding to the 2A$_g$ state are shown in the bottom panel. 
    Reference data is given for $\Delta$-CR-EOMCC(2,3)/cc-pVTZ (electronic energies: black curves) and CC3/aug-cc-pVDZ ($\%T_1$ value for the BLA of $-0.030$~\AA: horizontal black dashed line). 
    Dotted lines indicate imaginary excitation energies: $i$ multiplied by the $y$-value.}
    \label{fig:butadiene_curves}
\end{figure*}

\begin{figure*}[h!]
    \centering
    \includegraphics[width=\textwidth]{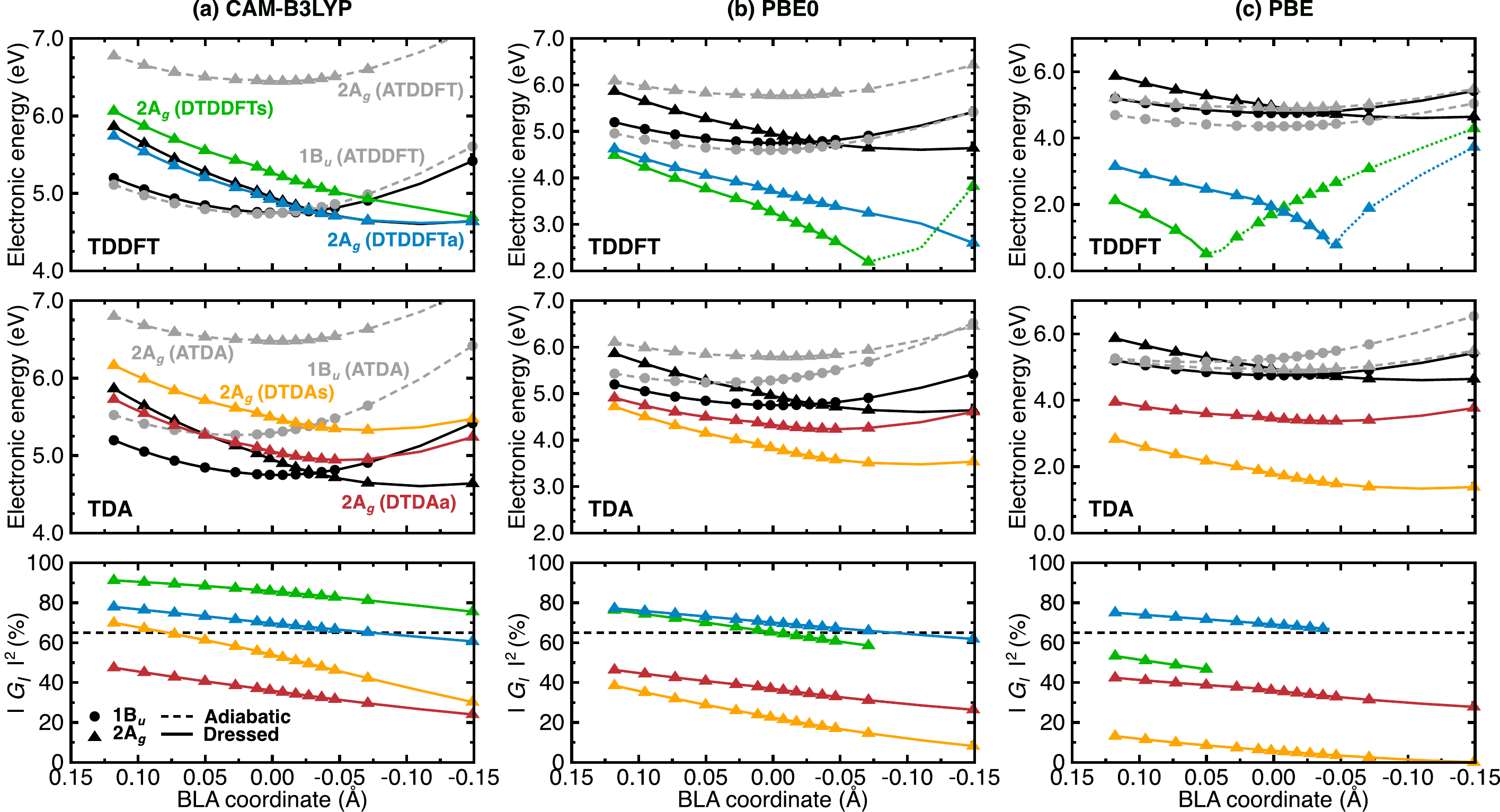}
    \caption{The 1B$_u$ (circles) and 2A$_g$ (triangles) electronic energies along the BLA coordinate in \textbf{hexatriene} obtained with ATDDFT, DTDDFTs and DTDDFTa (top panel: grey, green, blue, respectively), as well as ATDA, DTDAs and DTDAa (middle panel: grey, yellow, red, respectively), using CAM-B3LYP (a), PBE0 (b) and PBE (c) computed within cc-pVDZ. 
    Dressed $|G_I|^2$ values corresponding to the 2A$_g$ state are shown in the bottom panel. 
    Reference data is given for $\Delta$-CR-EOMCC(2,3)/cc-pVTZ (electronic energies: black curves) and CC3/aug-cc-pVDZ ($\%T_1$ value for the BLA of $-0.032$~\AA: horizontal black dashed line). 
    Dotted lines indicate imaginary excitation energies: $i$ multiplied by the $y$-value.}
    \label{fig:hexatriene_curves}
\end{figure*}

In the following, we compare the potential energy curves of the 1B$_u$ and 2A$_g$ singlet excited states in butadiene (Fig.~\ref{fig:butadiene_curves}) and hexatriene (Fig.~\ref{fig:hexatriene_curves}) along a particular cut through their conical intersection. 
Specifically, the geometries  were taken directly from Ref.~\citenum{PSLP21}, where they were generated from a geodesic interpolation between (EOM)CCSD/aug-cc-pVTZ optimised S$_0$(1A$_g$), S$_1$(1B$_u$) and S$_2$(2A$_g$) minimum geometries along the corresponding bond-alternation (BLA) coordinate, preserving molecular planarity.
The BLA coordinate is defined as the average length of the single bonds minus the average length of the double bonds, and approximates closely the tuning branching space coordinate of the accidental symmetry-allowed S$_2$(2A$_g$)/S$_1$(1B$_u$) conical intersection in short linear polyenes~\cite{KMHGD00,KOK13,KLDK17}.
Reference $\Delta$-CR-EOMCC(2,3)/cc-pVTZ 1B$_u$ and 2A$_g$ curves were reproduced from the corresponding curves in Figs 2 (butadiene) and 3 (hexatriene) in Ref.~\citenum{PSLP21}.
All excited-state energies  are given relative to the S$_0$ energy at the S$_0$ minimum.

The reference $\Delta$-CR-EOMCC(2,3) method~\cite{FLWPW11,PHA15,PSLP21} produces a crossing of the the 1B$_u$ and 2A$_g$ states at a BLA of $-0.030$~\AA~in butadiene and $-0.032$~\AA~in hexatriene, at energies 5.74~eV and 4.77~eV, respectively.
Following our working protocol in Section~\ref{sec:Operational}, two KS single excitations coupled to our chosen KS double excitation are considered (see Tables II, IV-VI and VI-IX in the SI for details).

{\underline{\it CAM-B3LYP}:} We first consider the performance of our chosen range-separated hybrid, CAM-B3LYP, for the underlying adiabatic DFA [Column (a) in Figs~\ref{fig:butadiene_curves} and \ref{fig:hexatriene_curves}, respectively].
ATDDFT (top panel, grey) with this functional approximation is remarkably accurate for the 1B$_u$ state, faithfully reproducing the energy and shape of the reference curve in both molecules.
However, the 2A$_g$ energy is significantly overestimated (i.e., by almost 2~eV), and, as a result no 1B$_u$/2A$_g$ curve-crossing appears.
Invoking the TDA (middle panel, grey) results in no appreciable change to the 2A$_g$ states, but significantly destabilises the 1B$_u$ state for both molecules, especially at larger negative BLAs.

Applying the dressing in the form of DTDDFTs (top panel, green) stabilises the 2A$_g$ curve, yielding a 1B$_u$/2A$_g$ curve-crossing in butadiene (hexatriene) but at a higher energy of 6.21~eV (4.96~eV) compared to the reference and a significantly shifted BLA of $-0.099$~\AA~($-0.071$~\AA).
DTDAs (middle panel, yellow), on the other hand, gives a curve-crossing at a BLA close to the reference (albeit over 0.5~eV higher in energy), however, this is not due to any better capturing of the 2A$_g$ state but rather is due to the raising of the 1B$_u$ state by ATDA relative to ATDDFT; in fact, the DTDAs 2A$_g$ energy is close to that for DTDDFTs, becoming worse at more negative BLAs.

With DTDDFTa (top panel, blue), the 2A$_g$ state is further stabilised beyond that of DTDDFTs, shifting the 1B$_u$/2A$_g$ curve-crossing remarkably close to that of the reference.
The 2A$_g$ state almost perfectly reproduces the $\Delta$-CR-EOMCC(2,3) curve at positive BLAs in butadiene, and for the entire range of BLAs in hexatriene, outperforming DTDDFTs in both cases.
DTDAa (middle panel, red) is also accurate for positive BLAs, but overestimates the 2A$_g$ energy at negative BLAs, where it follows more the curvature of the reference 1B$_u$ state than that of the 2A$_g$ state.
In either case, however, DTDAa still provides a significant improvement over ATDDFT or ATDA.
Due to the overestimation of the 1B$_u$ state by ATDA, the resulting curve-crossing is displaced significantly (i.e., 0.125 \AA~in butadiene and 0.050 \AA~ in hexatriene). 

The bottom panels in Figs~\ref{fig:butadiene_curves} and \ref{fig:hexatriene_curves} show the approximate measure of the single-excitation character, $|G_I|^2$, for the dressed 2A$_g$ state, which decreases upon going from positive to negative BLAs in all four dressed methods, particularly in DTDAs.
Of course, this approximate measure should be taken with a grain of salt, especially when directly comparing to the reference CC3 $\%T_1$ value, given the inherent difference in their meaning (Section~\ref{sec:Operational}).
Equally, the $\%T_1$ values apply only to the S$_0$ minimum geometry (i.e., BLA of 0.125~\AA~and 0.118~\AA~in butadiene and hexatriene, respectively).
That being said, we note that DTDDFTa, overall, gives $|G_I|^2$ values most consistent with the reference $\%T_1$ values. The DTDA values provide a less meaningful estimate of $|G_I|^2$, given their violation of the oscillator strength sum-rule.

{\underline{\it PBE0}:} Turning now to our chosen global hybrid DFA, PBE0, [Column (b) in Figs~\ref{fig:butadiene_curves} and \ref{fig:hexatriene_curves}], we find again that ATDDFT captures the 1B$_u$ state well, but overestimates the energy of the 2A$_g$ state significantly (albeit less so than for CAM-B3LYP), and no 1B$_u$/2A$_g$ curve-crossing occurs.
Applying the TDA again raises the 1B$_u$ states; this time, so much so that there is a fortuitous curve-crossing with the 2A$_g$ state; for both molecules, however, it is almost 2~eV higher than that of the reference and at a considerably shifted BLA.
Interestingly, for all three DFAs, the ATDDFT 1B$_u$ and 2A$_g$ state appear almost parallel to one another, whilst those in ATDA differ more in slope, the latter arguably allowing for a curve-crossing in some cases.

In DTDDFT (top panel), both variants overshoot the energy lowering of the 2A$_g$ state, particularly so for DTDDFTs, hinting at a curve-crossing with the 1B$_u$ state at a positive BLA a little beyond the range shown here.
The error in DTDDFTs is larger at larger negative BLAs; the reverse is true for DTDDFTa.
A similar behavior is seen in hexatriene, with DTDDFTs even beginning to yield imaginary frequencies at BLAs below $-0.047$~\AA. Using DTDA instead (middle panel) corrects the incorrect curvature and redeems positive frequencies.
Unlike, in the case of CAM-B3LYP, DTDA with PBE0 as the underling adiabatic DFA yields a smaller error than DTDDFT in both the s- and a-variant case for both molecules; for some BLAs, DTDAa is in fact very close to the reference, even if no 1B$_u$/2A$_g$ curve-crossing appears. 
However, DTDA predicts a larger double-excitation character (bottom panel) compared to the DTDDFT values; the latter are closer to the the CC3 $\%T_1$ values.
We note, the $|G_I|^2$ value ordering between the s- and a-variants of the dressing mirrors the corresponding energy order, consistent with the larger energy shift being produced by a larger component of double-excitation.

{\underline{\it PBE}:} Lastly, we consider the performance of our chosen GGA DFA, PBE [Column (c) in Figs~\ref{fig:butadiene_curves} and \ref{fig:hexatriene_curves}].
Unlike with the two hybrid DFAs, ATDDFT with PBE now underestimates the 1B$_u$ state slightly, and predicts the 2A$_g$ curve close in energy to the reference, albeit still with the wrong shape.
The ATDDFT 1B$_u$ and 2A$_g$ states still appear almost parallel to one another, as for the other DFAs.
Invoking the TDA again destabilises the 1B$_u$ state, not to the same extent as in PBE0 or CAM-B3LYP, but still enough for the 1B$_u$ state to become higher in energy than the 2A$_g$ state except at the fortuitous 1B$_u$/2A$_g$ curve-crossing, which now appears at large positive BLAs. 

Again, both variants of DTDDFT (top panel) overestimate the energy lowering of the 2A$_g$ state, more so than in PBE0.
No curve-crossing is observed, with now DTDDFTs (for both molecules) and DTDDFTa (for hexatriene) yielding 2A$_g$ curves with negative curvature and imaginary frequencies, doing so across a larger range of BLAs than PBE0.
Applying the TDA (middle panel), again, restores positive curvature and real frequencies, but still the 2A$_g$ state is overstabilised, resulting in no 1B$_u$/2A$_g$ curve-crossing.

The poor performance of DTDA energies when performed on top of a GGA functional had been previously observed in Ref.~\citenum{HIRC11}, which had advocated for using DTDA on top of a hybrid functional; here, we see that DTDDFT also, when performed on top of a GGA, is poor, and that it is susceptible to the imaginary frequency problem (Sec.~\ref{sec:Operational}).
Furthermore, DTDDFT (or DTDA) performed on top of the range-separated hybrid CAM-B3LYP gives better results than the local hybrid PBE0. 
Regarding the state character (bottom panel),  DTDDFTa (and DTDDFTs to a lesser extent) gives $|G_I|^2$ values most consistent with the reference $\%T_1$ values.
The other dressed approaches underestimate the single-excitation character of the 2A$_g$ state, especially DTDAs, which predicts almost entirely double-excitation character.

{\underline{\it Overall trends}}
Perhaps the most notable observations so far are that the results show a considerably improved description for the excitation energy and character of the 2A$_g$ state in butadiene and hexatriene over the range of BLA shown when DTDDFT is applied on top of CAM-B3LYP, compared with CAM-B3LYP itself, particularly with the a-variant but a poor performance when applied on top of PBE. Further, relating to the questions raised in Sec.~\ref{sec:Operational}, they demonstrate the possibility of imaginary frequencies in DTDDFT, but not in DTDA which otherwise generally gives larger errors in both the character and energies of the states. 
In line with the observations of Ref.~\citenum{HIRC11}, the dressing is also shown to have less of an effect (i.e., in terms of the energy shift and $|G_I|^2$ differing from their adiabatic value of 100$\%$) in the following order of underlying adiabatic DFA: PBE to PBE0 to CAM-B3LYP.
Generally, there is a  significant dependence on the underlying adiabatic DFA, dressing variant, and whether the TDA is invoked.
Isolating the effect of the dressing without showing the relation to the 1B$_u$ state, Fig. S1 in the SI shows the error in the 2A$_g$ energy [i.e., $E(\text{error}) = E(\text{dressed method}) - E(\text{reference})$] as function of BLA coordinate (top panels in Fig. S1), and as a function of the $\vert G_I\vert^2$ prediction of the method (bottom panels in Fig. S1). 
For a given BLA coordinate, the variation in the predictions of the four dressed methods increases upon going from CAM-B3LYP to PBE0 to PBE, and also in going from positive to negative BLAs. 
Considering instead the error in the energy as a function of $|G_I|^2$ value, we see instead that changing the underlying adiabatic DFA for the a-variant (in either DTDDFT or DTDA) does not correspondingly change the range of $|G_I|^2$ values obtained. 
The opposite is true for the s-variant: increasingly lower $|G_I|^2$ values are obtained as the DFA gets less sophisticated.

Given the poorer performance of the s-variant in general, and its higher propensity of yielding imaginary frequencies in DTDDFT (Question 6 of Sec.~\ref{sec:Operational}), at least for the systems shown here, we consider only the results of the a-variant for the remaining molecules we study here. 

Turning to the loss of dynamical correlation from the truncation within which DTDDFT is performed (Question 8 of Sec.~\ref{sec:Operational}), we note that  truncation has little effect on the ATDDFT 2A$_g$ state in butadiene using CAM-B3LYP, but that the truncated ATDDFT energies increase by roughly 0.2 eV in hexatriene compared to the full ATDDFT values. Thus, applying the $\Delta$c correction is assumed to have little effect on the DTDDFT 2A$_g$ state in butadiene, but would shift the equivalent state in hexatriene down by 0.2 eV, making the DTDDFTa curve slightly worse at some BLA and the DTDDFTs curve a little better.

The DTDDFT procedure should yield 3 states of A$_g$ character (Question 7 of Sec.~\ref{sec:Operational} and Sec.~\ref{sec:undownfolding}), but here we have only discussed the lowest. The other two resulting from the diagonalization of the 'un-downfolded' matrix lie much higher in the spectrum, embedded in higher-energy single-excitations, and so their meaning is limited without consider further mixing with those states. Both $\nu_D$ and $\Omega_D$ lie closer to the higher 3A$_g$ excitation, but are shifted less by the dressing (see Question 9 of Sec.~\ref{sec:Operational}). 

Finally, we note that the mixed-reference SF-TDDFT (MR-SF-TDDFT) approach of Choi and coworkers\cite{LFLC18,LKNLC19,PKLC23}, which alleviates some of the well-documented spin-contamination issues with conventional SF-TDDFT, has also been shown to provide accurate 1B$_u$ and 2A$_g$ curves in butadiene and hexatriene, as well as in octatetraene (one of the molecules we focus on next in Section~\ref{subsec:octaetraene_cyclopentadiene})\cite{PSLPJFC22,FPC26}. As mentioned in the introduction, (MR-)SF-TDDFT accesses certain KS double excitations through the response of   a mixed-reference state whilst remaining in the adiabatic approximation; in contrast, DTDDFT sticks with the ground-state determinant but accesses these states going beyond the adiabatic approximation. 

\subsection{Octatetraene and cyclopentadiene}
\label{subsec:octaetraene_cyclopentadiene}

\begin{table*}[]
\caption{Singlet 1B$_u$ and 2A$_g$ excitation energies, $\Omega_I$, in eV and corresponding $|G_I|^2$ values in $\%$ of {\bf octatetraene}, calculated with ATDDFT and DTDDFT, as well as their TDA equivalents, at the FC geometry using cc-pVDZ. Truncated ATDDFT / ATDA and $\Delta$c-DTDDFTa / DTDAa excitation energies are also provided in italics and bold, respectively. QUESTDB reference excitation energies and corresponding $\%T_1$ values (the latter given in parentheses) are taken from Ref.~\citenum{KBLJ24}. Note, CAM signifies CAM-B3LYP.}
\label{tab:octatetraene}
\centering
    \begin{tabular}{ll|ccc|ccc|ccc|ccc|c}
         & & \multicolumn{3}{c|}{ATDA} & \multicolumn{3}{c|}{ATDDFT} & \multicolumn{3}{c|}{DTDAa} & \multicolumn{3}{c|}{DTDDFTa} & QUESTDB \\
         & & PBE & PBE0 & CAM & PBE & PBE0 & CAM & PBE & PBE0 & CAM & PBE & PBE0 & CAM & \\
         \hline\hline
         1B$_u$ & $\Omega_I$ & 4.483 & 4.638 & 4.738 & 3.933 & 4.198 & 4.375 & & & & & & & 4.779 \\
         & $|G_I|^2$ & 100.0 & 100.0 & 100.0 & 100.0 & 100.0 & 100.0 & & & & & & & (91.5) \\
         \hline
         2A$_g$ & $\Omega_I$ & 4.282 & 5.154 & 5.933 & 4.273 & 5.140 & 5.906 & \textbf{3.048} & \textbf{3.970} & \textbf{4.739} & \textbf{2.072} & \textbf{3.626} & \textbf{4.709} & 4.680 \\
         & & \textit{4.298} & \textit{5.205} & \textit{6.124} & \textit{4.295} & \textit{5.200} & \textit{6.124} & 3.064 & 4.021 & 4.946 & 2.094 & 3.686 & 4.927 & \\
         & $|G_I|^2$ & 100.0 & 100.0 & 100.0 & 100.0 & 100.0 & 100.0 & 35.0 & 39.6 & 40.8 & 72.9 & 75.1 & 75.1 & (65.7) \\
         \hline\hline
    \end{tabular}
\end{table*}

We now continue along the polyene series to octatetraene, which we discuss together with cyclopentadiene, which serves as a cyclic polyene analogue.

Octatetraene is one of the smallest molecules to exhibit gas-phase dual-fluorescence\cite{GWMR78,CD06,BJPTFC90}, the origin of which was recently shown to likely come from the interplay between the 1B$_u$ and 2A$_g$ states\cite{PSLPJFC22}.
There is, however, a dispute in the literature as to the correct energy order of these two states at the S$_0$ minimum.
Some electronic-structure methods~\cite{PSLPJFC22,FPC26} predict the 1B$_u$ lower than the 2A$_g$, consistent with the shorter-chain polyenes at their S$_0$ minimum geometry, whereas others predict the reverse~\cite{KBLJ24,HK74}.
The QUESTDB reference method, CASPT3, falls into the latter group, giving the 2A$_g$ excitation energy as 4.680 eV ($\%T_1$ value of 65.7$\%$), with the 1B$_u$ excitation energy being slightly higher at 4.779 eV. While we use this as our reference value, we note that the reference method used for the  butadiene and hexatriene curves in Sec.~\ref{subsec:polyenes} gives instead the same  ordering for octatetraene as it did for those polyenes, i.e. the 1B$_u$ below the 2A$_g$ at the S$_0$ minimum geometry~\cite{PSLPJFC22} .

Similar to the 2A$_g$ state in the linear polyenes, our protocol suggests to consider coupling two KS single excitations to the chosen KS double for the low-lying 2A$_1$ ($\pi\rightarrow\pi^*$) excited state in cyclopentadiene for almost all functionals; 
there is one exception: when using PBE within DTDA, we instead include three KS single excitations.
See Tables II, X-XII  (octatetraene) and XIII-XV (cyclopentadiene) in the SI for the details. The QUESTDB reference method for cyclopentadiene is 
CCSDTQ, which predicts its 2A$_1$ state  to be 6.451 eV above the ground state with a $\%T_1$ value of 78.9$\%$.

Table~\ref{tab:octatetraene} shows the results for octatetraene. 
The ATDDFT 2A$_g$ state from PBE is already somewhat  close in energy to that of the reference, underestimated by about $0.4$ eV, but with inherently the wrong character (i.e., $|G_I|^2 = 1.0$); in fact PBE estimates the 2A$_g$ energy better than it does the predominantly  single 1B$_u$. The same ballpark errors and observation held for the butadiene and hexatriene molecules studied earlier. Like in those cases, introducing a fraction of HF exchange increases the excitation energies for the 1B$_u$ state and even more so for 2A$_g$, and the increase going from the local hybrid PBE0 to the range-separated hybrid CAM-B3LYP is almost as large as going from PBE to PBE0 for both cases. 
Making the TDA increases 2A$_g$ only very slightly while leading to a larger increase in 1B$_u$.
Noting that the dressing is applied to ATDDFT matrices that are truncated to the relevant single excitation subspace, we provide in italics the ATDDFT and ATDA values obtained from this truncation.  The truncation increases the predicted excitation energy by about 0.02 eV for PBE, 0.06 eV for PBE0, and a more significant 0.2 eV for CAM-B3LYP. 

Turning now to the dressed results, we first note that, similar to the smaller polyenes in Sec.~\ref{subsec:polyenes}, the second and third states from the DTDDFT procedure lie much higher in the spectrum; for the same reasons as given there we do not include them in Table~\ref{tab:octatetraene}.
Applying the dressing in the form of DTDDFTa acts to lower the 2A$_g$ state for all underlying adiabatic DFAs, with the extent of lowering decreasing in the order of PBE to PBE0 to CAM-B3LYP. 
As a result, DTDDFTa with PBE now significantly underestimates the reference value, with PBE0 reducing that underestimation and CAM-B3LYP providing very close energetic agreement to the reference. 
Despite the varied energetic agreement with the reference for all three underlying adiabatic DFAs, each yields a $|G_I|^2$ value indicating a partial double-excitation character consistent with the $\%T_1$ value of the reference.

Dressing within the TDA, unlike the adiabatic case, significantly increases the PBE and PBE0 energies compared with DTDDFTa, particularly for PBE.  For CAM-B3LYP, the DTDAa  is close to that of DTDDFTa for the energy, with a small increase similar to the adiabatic counterparts.  
For all functionals, the DTDAa $|G_I|^2$ values are significantly smaller, showing about twice as much double-excitation character than with the full DTDDFTa. 

These observations are consistent with those for butadiene and hexatriene in Section~\ref{subsec:polyenes}, with DTDDFTa/CAM-B3LYP yielding the best energetic agreement with the reference. In all cases, except for ATDA/PBE, the 2A$_g$ state is also predicted to be higher in energy than the 1B$_u$ state, which is inline with the behaviour in butadiene and hexatriene, but in contrast to the ordering of the reference CASPT3 energies. Interestingly, MR-SF-TDDFT also predicts the 2A$_g$ state to be above the 1B$_u$ at the S$_0$ minimum\cite{FPC26}.

\begin{table*}[h!]
\caption{Singlet 1B$_1$ and 2A$_1$ excitation energies, $\Omega_I$, in eV and corresponding $|G_I|^2$ values in $\%$ of {\bf cyclopentadiene}, calculated with ATDDFT and DTDDFTa, as well as their TDA equivalents, at the FC geometry using cc-pVDZ. Truncated ATDDFT / ATDA and $\Delta$c-DTDDFTa / $\Delta$c-DTDAa excitation energies are also provided in italics and bold, respectively. QUESTDB reference excitation energies and corresponding $\%T_1$ values (the latter given in parentheses) are taken from Ref.~\citenum{KBLJ24}. Note, CAM signifies CAM-B3LYP.}
\label{tab:cyclopentadiene}
\centering
    \begin{tabular}{ll|ccc|ccc|ccc|ccc|c}
         & & \multicolumn{3}{c|}{ATDA} & \multicolumn{3}{c|}{ATDDFT} & \multicolumn{3}{c|}{DTDAa} & \multicolumn{3}{c|}{DTDDFTa} & QUESTDB \\
         & & PBE & PBE0 & CAM & PBE & PBE0 & CAM & PBE & PBE0 & CAM & PBE & PBE0 & CAM & \\
         \hline\hline
         1B$_1$ & $\Omega_I$ & 5.512 & 5.643 & 5.701 & 5.081 & 5.249 & 5.309 & & & & & & & 6.407 \\
         & $|G_I|^2$ & 100.0 & 100.0 & 100.0 & 100.0 & 100.0 & 100.0 & & & & & & & (94.2) \\
         \hline
         2A$_1$ & $\Omega_I$ & 6.263 & 6.880 & 7.182 & 6.241 & 6.857 & 7.153 & \textbf{5.447} & \textbf{6.183} & \textbf{6.486} & \textbf{5.221} & \textbf{6.053} & \textbf{6.498} & 6.451 \\
         & & \textit{6.307} & \textit{6.934} & \textit{7.235} & \textit{6.322} & \textit{6.930} & \textit{7.235} & 5.491 & 6.237 & 6.547 & 5.302 & 6.126 & 6.580 & \\
         & $|G_I|^2$ & 100.0 & 100.0 & 100.0 & 100.0 & 100.0 & 100.0 & 56.5 & 62.4 & 62.6 & 82.6 & 84.3 & 85.9 & (78.9) \\
         \hline\hline
    \end{tabular}
\end{table*}

The same trends were observed for the 2A$_1$ state in cyclopentadiene (Table~\ref{tab:cyclopentadiene}), albeit to different degrees; notably, the DTDDFTa and DTDAa energies are less different with PBE and PBE0 in cyclopentadiene than in octatetraene.
Further, in both molecules, the ATDDFT-corrected HOMO-LUMO transition (i.e., $\Omega_{s_1}$) which affords the "adiabatic double" excitation (i.e., $\Omega_D = 2 \times \Omega_{s_1}$), that is, the 1B$_u$ state in octatetraene and 1B$_1$ state in cyclopentadiene, increases in accuracy from PBE to PBE0 to CAM-B3LYP, the same trend as in the accuracy in the DTDDFTa and DTDAa energies. 

For both octatetraene and cyclopentadiene, we note that when using CAM-B3LYP, the error from ATDDFT is notably worse for the single-excitation 1B$_u$/1B$_1$ than the error from DTDDFTa for the state of double-excitation character 2A$_g$/2A$_1$, once the effect of truncation is mitigated,  i.e. taking the $\Delta$c-DTDDFTa values (bold-type in the table). The $\Delta$c-DTDAa with CAM-B3LYP also gives notably less error for the energy of this state than the corresponding error of ATDA for the 1B$_u$/1B$_1$ excitation energy for the two molecules. We leave the question of why the ATDDFT and ATDA values are unusually lower than the reference for the 1B$_1$ state in cyclopentadiene for a future study. 

\subsection{Acrolein}
\label{subsec:acrolein}

Next, we consider the low-lying A' states of acrolein, a molecule commonly studied due to its role in radical generation in the atmosphere\cite{PMHHSGSC22}, owed to its release from natural combustion, as well as industrial sources\cite{MTMLMW02,G90}.
Unlike the low-lying states in the molecules discussed so far, one finds a relatively dense manifold within the ATDDFT spectrum between 6 and 8 eV, sharing similar KS single-excitation contributions.
We consider the 2A', 3A' and 4A' states, with the 4A' ($\pi\rightarrow\pi^*$) state having been categorized by CCSDTQ (Table~\ref{tab:acrolein} and Ref.~\citenum{KBLJ24}) as having partial double-excitation character~\cite{LBSCJ19, KBLJ24} with a $\%T_1$ value of $75.0\%$. 
Following the general procedure in Section~\ref{sec:Operational}, we consider four, three, and six KS single excitations coupling to our chosen KS double for PBE, PBE0 and CAM-B3LYP, respectively (see Tables II and XVI-XVIII in the SI for details).

For the molecules discussed so far, the dressing causes a significant energy shift to the full ATDDFT energy of interest. However, it will turn out that it is the truncation in acrolein that has a more significant effect than the dressing itself when DTDDFT or DTDA are performed (recall Question 8 of Sec.~\ref{sec:Operational}).  In this case we expect the $\Delta$c correction to be important but one might also wonder whether there is a significant correction beyond such a shift, by including more KS excitations in the truncation that couple directly to the KS double excitation. However, we find that including more (higher) KS excitations to the matrix does not bring the truncated ATDDFT values much closer to the full ATDDFT, and applying the dressing in the larger space appears only to lead to dressed higher states but not to affect the dressed energies of the lower ones that greatly. 

Adhering then to the truncated basis determined by our protocol, we first examine the adiabatic energies given in Table~\ref{tab:acrolein}. 
We report here only the lowest four A' states that result from the dressing in each case, even for the functionals where more states were included in the truncation, since the states not reported lie higher in energy. 
Within the full space, adiabatic PBE significantly underestimates all three A' states,  and PBE0 and CAM-B3LYP less so (and the CAM-B3LYP prediction for the 3A' state, which is $89.4\%$ single-excitation by the $\%T_1$ measure,   shows close agreement to the reference value).
Performing the truncation to the manifold in which the dressing will be applied, shifts the ATDDFT A' states higher in energy in all cases (although for 2A' and 3A' in PBE the shift is small). For PBE0, this truncation results in improving the energetic agreement to the reference for all three A' states;  for CAM-B3LYP the truncation overshoots the 2A' and 3A' but brings the 4A' prediction closer to the reference. 

DTDDFTa shifts the 4A' state higher in energy than its truncated adiabatic equivalent in the case of all three underlying adiabatic DFAs; the energies of the 2A' and 3A' states are essentially unaltered.
For PBE, this causes minimal improvement in the accuracy of the three A' states, with all still significantly underestimating the reference values.
For CAM-B3LYP, the dressing of the double excitation through DTDDFTa makes only a small correction to the truncated adiabatic energies for 2A' and 3A' ($-$0.02 eV and 0.03 eV  respectively), with a little larger correction of 0.08 eV for the 4A' state, bringing the DTDDFTa prediction a little closer to that of the reference  compared to the truncated adiabatic TDDFT. 
A similar behaviour is observed for PBE0.
If we were to compare with the full ATDDFT directly in this case, it might appear that DTDDFTa makes a significant improvement, however this is largely due to the truncation raising the ATDDFT value. In fact, reintroducing the missing dynamical correlation through the $\Delta$c correction makes the DTDDFTa 4A' prediction worse, and only a small improvement (0.08 eV) to the ATDDFT value. 
Despite the underwhelming effect of the dressing for the energetic performance among the different DFAs, it is interesting to note that all give reasonably consistent $|G_I|^2$ values to the reference $\%T_1$ values for all three A' states. 

Applying the TDA gives mixed outcomes both at the adiabatic and dressed level, depending on which underlying DFA is used.
We focus our attention here on the 4A' state.
 ATDA gives similar predictions to ATDDFT for most cases,  giving the largest difference (0.73 eV) for PBE0 when performed in the truncated space; differences of 0.2-0.3 eV are observed for PBE and CAM-B3LYP for this state. 
 For PBE0, the truncation yields a particularly large change from the full  ATDA value, increasing the 4A' energy by 1.4 eV (from full ATDA 7.025 eV to truncated value of 8.421 eV ), while for the other functionals, the truncation increases the energy much less, by about 0.4--0.5 eV, and for PBE0 with ATDDFT, the truncation increases the energy by 0.7 eV.
 
Turning to the dressing, DTDAa using PBE shifts the 4A' state lower than its truncated ATDA analogue, further underestimating the reference value, continuing the trend seen with the other molecules.
 With PBE0, DTDAa significantly overestimates the 4A' state (i.e., by over 0.5 eV), largely due to the significant overestimation by the underlying truncated ATDA 4A' state noted above.
 The DTDAa 4A' state using CAM-B3LYP, on the other hand, accurately reproduces the reference value, however this is mostly due to the truncation rather than the dressing:
 Like in the DTDDFTa case, applying the $\Delta$c correction worsens the DTDAa 4A' prediction with CAM-B3LYP and PBE0, again largely due to the truncated ATDDFT/ATDA values being fortuitously closer to those of the reference than the full ATDDFT/ATDA. 
 Finally, we note that the $|G_I|^2$ values by DTDAa are wide-ranging, depending on the functional, but we do not expect these to be meaningful, unlike for DTDDFTa.

Thus, the most salient aspect for acrolein is that truncation has the dominant effect, resulting in truncated ATDDFT energies for the partially-double 4A' state close to that of the reference even without dressing, particularly within the TDA. The dressing appears to give only a small improvement to the energy, while more significantly improving the character of the state.
Naturally, when $\Delta$c-DTDDFT is applied to bring back the correlation lost in the truncation, the results worsen since the full-ATDDFT value is farther from the reference. The extent of the truncation is determined by our protocol, but even going beyond this and including some more KS excitations that make up the relevant ATDDFT gives only a relatively small improvement. Finally, we note that the extra state coming from the dressing lies nested in the manifold, unlike for the previous molecules, and is predicted to have significant double-excitation character ($\vert G_I\vert^2$ around 35$\% $). We leave for future work to resolve the meaning of this excitation, as well as a deeper analysis of the truncation and basis-set dependence. 

\begin{table*}[]
\caption{Singlet 1A'' and three lowest A' excitation energies, $\Omega_I$, in eV and corresponding $|G_I|^2$ values in $\%$ of {\bf acrolein}, calculated with ATDDFT and DTDDFTa, as well as their TDA equivalents, at the FC geometry using aug-cc-pVDZ. Truncated ATDDFT / ATDA and $\Delta$c-DTDDFTa / $\Delta$c-DTDAa excitation energies are also provided in italics and bold, respectively. For states of genuine double-excitation character, where $\Delta$c-DTDDFTa / $\Delta$c-DTDAa results are not calculated, we also give the corresponding excitation energies in bold. QUESTDB reference excitation energies and corresponding $\%T_1$ values (the latter given in parentheses) are taken from Ref.~\citenum{KBLJ24}. Note, CAM signifies CAM-B3LYP. We also include one extra $\Delta$c-DTDDFTa / $\Delta$c-DTDAa A' state, denoted with a (?); see text for details.}
\label{tab:acrolein}
\centering
    \begin{tabular}{ll|ccc|ccc|ccc|ccc|c}
         & & \multicolumn{3}{c|}{ATDA} & \multicolumn{3}{c|}{ATDDFT} & \multicolumn{3}{c|}{DTDAa} & \multicolumn{3}{c|}{DTDDFTa} & QUESTDB \\
         & & PBE & PBE0 & CAM & PBE & PBE0 & CAM & PBE & PBE0 & CAM & PBE & PBE0 & CAM & \\
         \hline\hline
         1A'' & $\Omega_I$ & 3.145 & 3.619 & 3.791 & 3.128 & 3.593 & 3.761 & & & & & & & 3.723 \\
         & $|G_I|^2$ & 100.0 & 100.0 & 100.0 & 100.0 & 100.0 & 100.0 & & & & & & & (87.6) \\
         \hline
         2A' & $\Omega_I$ & 5.516 & 6.556 & 6.709 & 5.509 & 6.278 & 6.367 & \textbf{5.513} & \textbf{6.547} & \textbf{6.692} & \textbf{6.060} & \textbf{6.265} & \textbf{6.347} & 6.671 \\
         & & \textit{5.532} & \textit{6.734} & \textit{7.044} & \textit{5.531} & \textit{6.719} & \textit{7.007} & 5.529 & 6.723 & 7.026 & 5.526 & 6.707 & 6.988 & \\
         & $|G_I|^2$ & 100.0 & 100.0 & 100.0 & 100.0 & 100.0 & 100.0 & 97.7 & 92.9 & 88.1 & 99.3 & 97.3 & 96.2 & (91.2) \\
         \hline
         3A' & $\Omega_I$ & 5.861 & 6.869 & 7.074 & 5.841 & 6.739 & 7.018 & \textbf{5.854} & \textbf{6.860} & \textbf{7.080} & \textbf{5.499} & \textbf{6.722} & \textbf{7.045} & 7.108 \\
         & & \textit{5.888} & \textit{7.052} & \textit{7.639} & \textit{5.886} & \textit{7.045} & \textit{7.606} & 5.881 & 7.042 & 7.644 & 5.876 & 7.027 & 7.633 & \\
         & $|G_I|^2$ & 100.0 & 100.0 & 100.0 & 100.0 & 100.0 & 100.0 & 92.2 & 81.9 & 89.0 & 97.2 & 87.0 & 94.8 & (89.4) \\
         \hline
         (?)A' & $\Omega_I$ & & & & & & & & & & & & & \\
         & & & & & & & & \textbf{6.229} & \textbf{7.204} & \textbf{7.511} & \textbf{6.193} & \textbf{6.194} & \textbf{7.431} & \\
         & $|G_I|^2$ & & & & & & & 21.8 & 3.3 & 5.6 & 34.5 & 34.5 & 36.8 & \\
         \hline
         4A' & $\Omega_I$ & 6.232 & 7.025 & 7.418 & 6.065 & 7.010 & 7.411 & \textbf{6.196} & \textbf{7.078} & \textbf{7.421} & \textbf{5.871} & \textbf{7.076} & \textbf{7.493} & 7.928 \\
         & & \textit{6.265} & \textit{8.421} & \textit{7.946} & \textit{6.262} & \textit{7.694} & \textit{7.691} & 6.229 & 8.475 & 7.948 & 6.291 & 7.760 & 7.773 & \\
         & $|G_I|^2$ & 100.0 & 100.0 & 100.0 & 100.0 & 100.0 & 100.0 & 21.8 & 87.8 & 98.5 & 71.9 & 89.6 & 72.3 & (75.0) \\
         \hline\hline
    \end{tabular}
\end{table*}

\subsection{Naphthalene}
\label{subsec:naphthalene}

We next consider the low-lying states in naphthalene, the smallest polcyclic acene, with numerous domestic and industrial applications\cite{RSDSGKB26}.
In naphthalene, the 2A$_g$ excitation energy has previously been characterized by CCSDT (Table~\ref{tab:naphthalene} and Ref.~\citenum{KBLJ24}) as being of predominantly single-excitation character (5.902 eV with $\%T_1$ of 83.8$\%$), with the 3A$_g$ ($\pi\rightarrow\pi^*$) excitation energy of partial double-excitation character (6.748 eV with $\%T_1$ of 71.5$\%$).
For all three DFAs, we consider four KS single excitations coupling to our chosen double excitation (see Tables II and XIX-XXI in the SI for details).

From Table~\ref{tab:naphthalene}, we observe that ATDDFT using PBE gives a small (0.1 eV) overestimate to the reference 2A$_g$ value,  and about a 0.4 eV underestimate of 3A$_g$;  the truncation reduces the latter discrepancy to 0.26 eV. Again the hybrids raise the energies, particularly for the 3A$_g$ state, and more so for CAM-B3LYP than for PBE0. The lowest single excitation (1B$_{3u}$, the double of which composes the double-excitation that will be involved in the dressing), is most accurately predicted by CAM-B3LYP, and especially so within the TDA, but this functional yields a significant overestimate for both the 2A$_g$ and 3A$_g$ states, worse within both the TDA and within the truncation. 
The ATDA energies are similar to those of ATDDFT, particular for the A$_g$ states;  a small difference observed for the 3A$_g$ state increases from PBE to PBE0 to CAM-B3LYP.

Similar to the other molecules, DTDDFTa lowers the energies of the A$_g$ states. With our chosen truncation we produce five A' states, but we report only the lowest three, which lie in the energy manifold of interest; the other two are higher and, for the same reasons as discussed for the previous molecules, their energies are not meaningful. We find that the adiabatic 3A$_g$ is lowered by the dressing significantly more than the 2A$_g$ state, to the extent that it drops below it when the dressing is applied to the PBE or PBE0 functionals. That is, the adiabatic contribution to the second state of A$_g$ character in DTDDFTa/PBE and DTDDFTa/PBE0 is predominantly coming from the adiabatic 3A$_g$ state, and vice-versa. This can be seen from inspection of the DTDDFTa eigenvectors. Further, as reflected in the $\vert G_I|^2$ values, while the reference calculation predicts a larger double-excitation character for 3A$_g$, the reverse is true for DTDDFTa when computed on top of PBE and PBE0. Comparing then their 3A$_g$ predictions with the reference 2A$_g$ values, we see that they are only a little improved over their adiabatic predictions, even when the $\Delta$c correction is applied in approximately accounting for the  missing dynamical correlation from the truncation. This observation is consistent with the mixing with the double excitation being quite small, as reflected by their $\vert G_I\vert^2$ values. Instead, comparing  the DTDDFTa/PBE and DTDDFTa/PBE0 2A$_g$ predictions to the reference 3A$_g$, we see more mixing with the double excitation, with a $\vert G_I\vert^2$ comparable to the $\%T_1$, but an overcorrection to the energy lowering such that the dressing makes the PBE prediction worse than it was within the adiabatic approximation, while the error for the dressing performed with PBE0 is about the same as that for the adiabatic approximation, but as an underestimate instead of an overestimate. This is true within both the truncated approach as well as the $\Delta$c. 

When the dressing is applied to CAM-B3LYP, the adiabatic contributions to the 2A$_g$ and 3A$_g$ are not switched as they were for PBE and PBE0. The dressing again lowers the energy of both states, and the degrees of doubles character predicted for each are consistent with the reference $\%T_1$ values. The 2A$_g$ is lowered by about 0.17 eV compared to the adiabatic prediction, bringing it closer to the exact; with the $\Delta$c applied the error is about a $0.56$ eV overestimate. The 3A$_g$ is overestimated by about $0.36$ eV within $\Delta$c DTDDFTa/CAM-B3LYP, giving a significant improvement compared to its adiabatic error of about $1.07$ eV, together with the improved character of the state itself reflected in the $\vert G_I\vert^2$ value.

Finally, the DTDAa energy predictions do not improve the situation over DTDDFTa, and the $\vert G_I\vert^2$ are far less consistent with the reference $\%T_1$ as expected.

Overall, DTDDFTa/CAM-B3LYP again gives the best agreement with the reference for both states, in terms of energy and single-excitation contribution.

\begin{table*}[h!]
\caption{Singlet 1B$_{3u}$ and lowest two A$_g$ excitation energies, $\Omega_I$, in eV and corresponding $|G_I|^2$ values in $\%$ of {\bf naphthalene}, calculated with ATDDFT and DTDDFTa, as well as their TDA equivalents, at the FC geometry using cc-pVDZ. Truncated ATDDFT / ATDA and $\Delta$c-DTDDFTa / $\Delta$c-DTDAa excitation energies are also provided in italics and bold, respectively. QUESTDB reference excitation energies and corresponding $\%T_1$ values (the latter given in parentheses) are taken from Ref.~\citenum{KBLJ24}. Note, CAM signifies CAM-B3LYP.}
\label{tab:naphthalene}
\centering
    \begin{tabular}{ll|ccc|ccc|ccc|ccc|c}
         & & \multicolumn{3}{c|}{ATDA} & \multicolumn{3}{c|}{ATDDFT} & \multicolumn{3}{c|}{DTDAa} & \multicolumn{3}{c|}{DTDDFTa} & QUESTDB \\
         & & PBE & PBE0 & CAM & PBE & PBE0 & CAM & PBE & PBE0 & CAM & PBE & PBE0 & CAM & \\
         \hline\hline
         1B$_{3u}$ & $\Omega_I$ & 4.365 & 4.744 & 4.983 & 4.163 & 4.534 & 4.749 & & & & & & & 4.903 \\
         & $|G_I|^2$ & 100.0 & 100.0 & 100.0 & 100.0 & 100.0 & 100.0 & & & & & & & (90.3) \\
         \hline
         2A$_g$ & $\Omega_I$ & 6.013 & 6.487 & 6.664 & 5.997 & 6.464 & 6.630 & \textbf{5.426} & \textbf{6.259} & \textbf{6.491} & \textbf{5.218} & \textbf{6.143} & \textbf{6.459} & 5.902 \\
         & & \textit{6.056} & \textit{6.538} & \textit{6.731} & \textit{6.054} & \textit{6.537} & \textit{6.720} & 5.469 & 6.311 & 6.557 & 5.275 & 6.215 & 6.550 & \\
         & $|G_I|^2$ & 100.0 & 100.0 & 100.0 & 100.0 & 100.0 & 100.0 & 38.8 & 51.9 & 72.5 & 69.9 & 75.1 & 90.7 & (83.8) \\
         \hline
         3A$_g$ & $\Omega_I$ & 6.357 & 7.288 & 7.979 & 6.312 & 7.212 & 7.815 & \textbf{5.924} & \textbf{6.508} & \textbf{7.112} & \textbf{5.886} & \textbf{6.413} & \textbf{6.961} & 6.748 \\
         & & \textit{6.497} & \textit{7.452} & \textit{8.234} & \textit{6.487} & \textit{7.440} & \textit{8.206} & 6.064 & 6.672 & 7.368 & 6.062 & 6.641 & 7.352 & \\
         & $|G_I|^2$ & 100.0 & 100.0 & 100.0 & 100.0 & 100.0 & 100.0 & 99.3 & 74.0 & 51.8 & 99.9 & 93.9 & 75.8 & (71.5) \\
         \hline\hline
    \end{tabular}
\end{table*}

\subsection{Ethylene}
\label{subsec:ethylene}

In contrast to the previous molecules which have been characterized by Ref.~\citenum{KBLJ24} to have partial double-excitation character, we now consider a case characterized to have `genuine' double excitation character: the 2A$_g$ $(\pi,\pi\rightarrow\pi^*,\pi^*)$ state in ethylene, which in exFCI (Table~\ref{tab:ethylene}) is predicted to be at 12.899 eV above the ground state with a $\%T_1$ value of 20.0$\%$.
In the dressing problem, we consider one KS single excitation coupled to our chosen KS double excitation (see Tables II and XXII-XXIV in the SI for details). The symmetry of the HOMO-LUMO transition is B$_{2u}$ for the PBE and PBE0 functionals, and B$_{1u}$ for CAM-B3LYP; the corresponding double excitations have A$_g$ symmetry.

ATDDFT with PBE, PBE0, or CAM-B3LYP (Table~\ref{tab:ethylene}) all significantly underestimate the 2A$_g$ energy, with errors from almost 5 eV to 4.24 eV; ATDA is similar. In the truncated space, the values are raised by about 0.5 eV. 

Applying the dressing in the form of DTDDFTa, using PBE or PBE0, results in two dressed solutions, both reported in the table: a state which almost perfectly matches the energy of the truncated ATDDFT 2A$_g$ energy and a state at higher energy, which almost perfectly matches the energy of the chosen ``adiabatic double'' (i.e., 2$\times$ 1B$_{2u}$).
The former has a $|G_I|^2$ value of close to 1.0, whereas the latter a value close to 0.0, implying states of almost 100$\%$ single-excitation and double-excitation character, respectively. The mixing introduced by DTDDFT is relatively small: the $\vert H_{qD}\vert$ matrix element is about half the size of what it was for the polyenes (see SI), but also in this case  the energies of the KS single in the truncated space and the adiabatic double differ by over 5 eV for every functional, which perhaps is the main reason for the small mixing. 

With CAM-B3LYP, a similar negligible mixing occurs, but the adiabatic double is now twice the 1B$_{1u}$ energy, and is lower than that of the 1$B_{2u}$ energy (corresponding to the corrected HOMO-LUMO for PBE and PBE0), and is closer to the reference value.  Similar observations can be made for the TDA case, as well as the analysis above.

Since the reference method predicts the 2A$_g$ state to have a $\%T_1$ value of 20$\%$, we make a correspondence with this and the higher energy dressed solution, finding a better agreement with the energy than with ATDDFT, with an overestimate of around 2 eV.
DTDDFTa would predict the two lowest A$_g$ excitations as shown in the table; the literature appears to discuss the `lowest' A$_g$ excitation as either being around the value of the lower one here (e.g. Ref~\citenum{GARCIA1996}) or the higher one here (e.g. Ref.~\citenum{KBLJ24}). 
Whether in fact there are two A$_g$ states is difficult to determine by experiment due to their dark nature, the ground-state also being of A$_g$ character. Moreover, neither the reference calculation or our DTDDFTa treat continuum states adequately, and the higher A$_g$ state lies above the ionization continuum so is, in reality, a resonance, and is likely to be quite basis-set dependent. 
Nevertheless, here we simply compare with the TBE value given for the `genuine' excitation in Ref.~\citenum{KBLJ24}. 

Finally regarding the $\Delta$c-correction: this can be applied sensibly only to the lower dressed A$_g$ state, but gives a worse energy compared with the reference, stemming from the improved agreement with truncated-ATDDFT.

\begin{table*}[h!]
\caption{Singlet 1B$_{1u}$, 1B$_{2u}$ and 2A$_g$ excitation energies, $\Omega_I$, in eV and corresponding $|G_I|^2$ values in $\%$ of {\bf ethylene}, calculated with ATDDFT and DTDDFTa, as well as their TDA equivalents, at the FC geometry using cc-pVDZ. Truncated ATDDFT / ATDA and $\Delta$c-DTDDFTa / $\Delta$c-DTDAa excitation energies are also provided in italics and bold, respectively. For states of genuine double-excitation character, where $\Delta$c-DTDDFTa / $\Delta$c-DTDAa results are not calculated, we also give the corresponding excitation energies in bold. QUESTDB reference excitation energies and corresponding $\%T_1$ values (the latter given in parentheses) are taken from Ref.~\citenum{KBLJ24}. Note, CAM signifies CAM-B3LYP. We also include one extra $\Delta$c-DTDDFTa / $\Delta$c-DTDAa A$_g$ state, denoted with a (?); see text for details.}
\label{tab:ethylene}
\centering
    \begin{tabular}{ll|ccc|ccc|ccc|ccc|c}
         & & \multicolumn{3}{c|}{ATDA} & \multicolumn{3}{c|}{ATDDFT} & \multicolumn{3}{c|}{DTDAa} & \multicolumn{3}{c|}{DTDDFTa} & QUESTDB \\
         & & PBE & PBE0 & CAM & PBE & PBE0 & CAM & PBE & PBE0 & CAM & PBE & PBE0 & CAM & \\
         \hline\hline
         1B$_{1u}$ & $\Omega_I$ & & & 6.928 & & & 6.925 & & & & & & & 7.897 \\
         & $|G_I|^2$ & & & 100.0 & & & 100.0 & & & & & & & (95.8) \\
         \hline
         1B$_{2u}$ & $\Omega_I$ & 6.488 & 6.879 & & 7.415 & 7.523 & & & & & & & & \\
         & $|G_I|^2$ & 100.0 & 100.0 & & 100.0 & 100.0 & & & & & & & & \\
         \hline
         (?)A$_g$ & $\Omega_I$ & & & & & & & \textbf{8.634} & \textbf{8.939} & \textbf{9.000} & \textbf{8.639} & \textbf{8.935} & \textbf{8.983} & \\
         & & & & & & & & 8.274 & 8.647 & 8.698 & 8.258 & 8.616 & 8.657 & \\
         & $|G_I|^2$ & & & & & & & 90.8 & 97.0 & 99.4 & 99.3 & 99.7 & 99.9 & \\
         \hline
         2A$_g$ & $\Omega_I$ & 8.342 & 8.671 & 8.702 & 8.303 & 8.632 & 8.660 & & & & & & & 12.899 \\
         & & \textit{8.702} & \textit{8.963} & \textit{9.005} & \textit{8.685} & \textit{8.951} & \textit{8.987} & \textbf{13.045} & \textbf{13.783} & \textbf{13.861} & \textbf{14.866} & \textbf{15.059} & \textbf{13.853} & \\
         & $|G_I|^2$ & 100.0 & 100.0 & 100.0 & 100.0 & 100.0 & 100.0 & 0.0 & 0.0 & 0.0 & 0.7 & 0.3 & 0.1 & (20.0) \\
         \hline\hline
    \end{tabular}
\end{table*}

\subsection{Borole}
\label{subsec:borole}

Borole is considered to be the archetypal model of antiaromaticity, with relevance in photovoltaics and photoresponsive materials\cite{KRJ25}.
In the low-lying spectrum of borole, lies both a state of ``partial'' and a state of ``genuine'' double-excitation character.
CCSDTQ (Table~\ref{tab:borole} and Ref.~\citenum{KBLJ24}) predicts the 2A$_1$ ($\pi,\pi\rightarrow\pi^*,\pi^*$) excitation energy to be 4.708 eV and the 3A$_1$ ($\pi\rightarrow\pi^*$) excitation energy to be 6.484 eV, with $\%T_1$ values of 19.9$\%$ and 80.7$\%$, respectively.
For DTDDFTa with all DFAs, we consider coupling two KS single excitations that dominate the adiabatic excitations to our chosen KS double excitation. The same goes for DTDA, except for DTDAa/PBE, where we consider three.
Tables II and XXV-XXVII in the SI provide details of these choices.

Table~\ref{tab:borole} shows that ATDDFT significantly underestimates the HOMO-LUMO transition 1B$_1$, with the error decreasing in the order of PBE, PBE0, and CAM-B3LYP. There is a small improvement with ATDA, but it is still a large underestimate. On the other hand, the 2A$_1$ and 3A$_1$ energies are significantly overestimated by these functionals, with the error decreasing in the opposite order. Both the truncation and the ATDA makes the errors worse.  However, this may be an assignment problem of ATDDFT, as suggested by the DTDDFT analysis that we turn to next. 

Similar to the case of ethylene, DTDDFTa with all DFAs give dressed solutions that almost perfectly match the truncated ATDDFT 2A$_1$ and 3A$_1$ energies, as well as one dressed solution that almost perfectly matches the ``adiabatic double'' energy, i.e. twice the 1B$_1$ energy.
The latter is the lowest, so the 2A$_1$ predicted by DTDDFTa is predominantly a double excitation, consistent with the reference characterization of 'genuine double', and CAM-B3LYP in this framework approximates it very well. DTDDFTa predicts the 3A$_1$ state to be predominantly the single excitation, only slightly shifted lower from the corresponding ATDDFT 2A$_1$ root. In particular, the prediction from CAM-B3LYP is very close to the exact, but when dynamical correlation is reintroduced in the $\Delta$c approach it gets a little worse   (still within about 0.2 eV of the reference, which is a far smaller error than ATDDFT with this functional makes for the HOMO-LUMO transition). There is a significant discrepancy however between the DTDDFTa double-excitation character for these states: it is overestimated for 2A$_1$ and underestimated for 3A$_1$ compared to the $\%T_1$ reference.
DTDDFTa gives an extra state, the 4A$_1$ state of close to pure single-excitation character,  close in energy to the truncated ATDDFT 3A$_1$ state. Inspecting the table, it appears that ATDDFT's 2A$_1$ state is better interpreted as the true 3A$_1$ state, and ATDDFT's 3A$_1$ state maybe a fourth excitation of A$_1$ symmetry; that is, ATDDFT misses the 2A$_1$ state which DTDDFT retrieves, having a predominantly double excitation character. 
We note that the table reports all three roots from the dressed matrix for all cases; PBE/DTDAa gives a fourth one which is higher in energy and we do not report it here.

Similar conclusions can be found for DTDAa, with respect to the corresponding ATDA. Interestingly, in this case the double-excitation character suggested by $\vert G_I\vert^2$ is closer to the $\%T_1$ measure for DTDAa/PBE and DTDA/PBE0 than when DTDDFTa is applied on this functionals, however this is likely fortuitous since DTDAa does not satisfy the oscillator strength sum-rule. 

\begin{table*}[h!]
\caption{Singlet 1B$_1$ and lowest two A$_1$ excitation energies, $\Omega_I$, in eV and corresponding $|G_I|^2$ values in $\%$ of {\bf borole}, calculated with ATDDFT and DTDDFTa, as well as their TDA equivalents, at the FC geometry using cc-pVDZ. Truncated ATDDFT / ATDA and $\Delta$c-DTDDFTa / $\Delta$c-DTDAa excitation energies are also provided in italics and bold, respectively. For states of genuine double-excitation character, where $\Delta$c-DTDDFTa / $\Delta$c-DTDAa results are not calculated, we also give the corresponding excitation energies in bold. QUESTDB reference excitation energies and corresponding $\%T_1$ values (the latter given in parentheses) are taken from Ref.~\citenum{KBLJ24}. Note, CAM signifies CAM-B3LYP. We also include one extra $\Delta$c-DTDDFTa / $\Delta$c-DTDAa A$_1$ state, denoted with a (?); see text for details.}
\label{tab:borole}
\centering
    \begin{tabular}{ll|ccc|ccc|ccc|ccc|c}
         & & \multicolumn{3}{c|}{ATDA} & \multicolumn{3}{c|}{ATDDFT} & \multicolumn{3}{c|}{DTDAa} & \multicolumn{3}{c|}{DTDDFTa} & QUESTDB \\
         & & PBE & PBE0 & CAM & PBE & PBE0 & CAM & PBE & PBE0 & CAM & PBE & PBE0 & CAM & \\
         \hline\hline
         1B$_1$ & $\Omega_I$ & 2.281 & 2.499 & 2.580 & 2.093 & 2.320 & 2.397 & & & & & & & 4.863 \\
         & $|G_I|^2$ & 100.0 & 100.0 & 100.0 & 100.0 & 100.0 & 100.0 & & & & & & & (91.4) \\
         \hline
         2A$_1$ & $\Omega_I$ & 5.454 & 6.053 & 6.368 & 5.360 & 5.949 & 6.247 & & & & & & & 4.708 \\
         & & \textit{5.583} & \textit{6.195} & \textit{6.534} & \textit{5.527} & \textit{6.139} & \textit{6.471} & \textbf{4.523} & \textbf{4.951} & \textbf{5.129} & \textbf{4.150} & \textbf{4.605} & \textbf{4.774} & \\
         & $|G_I|^2$ & 100.0 & 100.0 & 100.0 & 100.0 & 100.0 & 100.0 & 0.5 & 0.6 & 0.4 & 1.9 & 1.8 & 0.9 & (19.9) \\
         \hline
         3A$_1$ & $\Omega_I$ & 8.450 & 8.792 & 8.903 & 7.913 & 8.216 & 8.308 & \textbf{5.486} & \textbf{6.094} & \textbf{6.395} & \textbf{5.387} & \textbf{5.977} & \textbf{6.262} & 6.484 \\
         & & \textit{9.404} & \textit{11.145} & \textit{11.054} & \textit{9.863} & \textit{9.889} & \textit{9.882} & 5.616 & 6.237 & 6.560 & 5.553 & 6.168 & 6.485 & \\
         & $|G_I|^2$ & 100.0 & 100.0 & 100.0 & 100.0 & 100.0 & 100.0 & 87.0 & 87.3 & 92.7 & 98.1 & 98.2 & 99.2 & (80.7) \\
         \hline
         (?)A$_1$ & $\Omega_I$ & & & & & & & \textbf{8.453} & \textbf{8.797} & \textbf{8.907} & \textbf{7.917} & \textbf{8.218} & \textbf{8.309} & \\
         & & & & & & & & 9.407 & 11.150 & 11.059 & 9.867 & 9.891 & 9.884 & \\
         & $|G_I|^2$ & & & & & & & 99.8 & 99.8 & 99.8 & 99.9 & 100.0 & 100.0 & \\
         \hline\hline
    \end{tabular}
\end{table*}

\subsection{Cyclopentadienone}
\label{subsec:cyclopentadienone}

Cyclopentadienone is unique amongst the nine molecules discussed in this work in that one of the states of interest  does not transform as the totally-symmetric representation of the molecular point group, and the KS double excitation that would contribute to it is no longer simply twice the HOMO to LUMO transition.
We consider the 2A$_1$ ($\pi\rightarrow\pi^*$), 3A$_1$ ($\pi,\pi\rightarrow\pi^*,\pi^*$) and 1B$_1$ ($n,\pi\rightarrow\pi^*,\pi^*$) states, with the former characterized as being of ``partial'' double-excitation character, and the latter two ``genuine'' double-excitation character.
In exFCI (Table~\ref{tab:cyclopentadienone} and Ref.~\citenum{KBLJ24}), the 1B$_1$ state is predicted to be at 5.009 eV ($\%T_1$ of 3.1$\%$) above the ground-state, whereas CC4 predicts the 1A$_1$ and 3A$_1$ states to be at 5.795 eV ($\%T_1$ of 73.6$\%$) and 6.714 eV ($\%T_1$ of 49.9$\%$), respectively.
Following our protocol, for the A$_1$ states, we consider three KS single excitations coupling to our chosen KS double excitation (taken as twice the HOMO to LUMO transition, like in all the other molecules). For the 1B$_1$ state, the protocol suggests only one KS single excitation is needed; this is coupled to the KS double excitation consisting of the HOMO to LUMO transition, plus HOMO-1 to LUMO transition.
See Tables II and XXVIII-XXXI in the SI for details.

Table~\ref{tab:cyclopentadienone} shows the performance of the hybrid DFAs, PBE0 and CAM-B3LYP only, because PBE using cc-pVDZ yields imaginary frequencies already at the adiabatic level at the equilibrium geometry, upon performing Davidson diagonalisation, thus, preventing us to build DTDDFTa on top of this.

The reference 1B$_1$ result indicates it is almost purely a double-excitation, and we expect that the ATDDFT and DTDDFT analysis follows closely to the corresponding  cases of ethylene and borole.
ATDDFT using both PBE0 and CAM-B3LYP overestimate the reference 1B$_1$ energy value by more than 1 eV with truncation raising this energy slightly. 
Performing DTDDFTa, results in two dressed solutions: as anticipated, the lower solution mirrors almost perfectly the sum of the full adiabatic 1B$_2$ and 1A$_2$ energies with a $\vert G_I\vert^2$ value close to the $\% T_1=3\%$ of the reference for both functionals. Instead, the upper B$_1$ solution almost perfectly matches the truncated ATDDFT 1B$_1$ state with very little mixing from the double excitation, as indicated  via the $\vert G_I\vert^2$ being close to 1. The lowest B$_1$ state predicted by ATDDFT B$_1$ state of the actual molecule, while DTDDFTa gets the character of the two lowest states well, although overestimating its energies, with DTDDFTa performed on top of PBE0 providing a closer estimate in this case. We report also the second root from DTDDFT, although we do not have an exact reference for it; we observe it is close to the uncorrected ATDDFT value, as expected.
Applying the TDA at either level has only a small effect on the energies.

Turning now to the A$_1$ states, ATDDFT using either of the hybrid functionals overestimates their energies, particularly for the 3A$_1$ state, with larger errors from CAM-B3LYP than PBE0. Truncation raises the energies further, even more so for the 3A$_1$ state.

DTDDFTa with these two functionals lowers both the 2A$_1$ and 3A$_1$ energies significantly, in comparison to the ATDDFT values. Comparing the $\vert G_I\vert^2$ and $\%T_1$ values of the two states suggest that the DTDDFTA-predicted 2A$_1$ state is more aligned with the reference 3A$_1$ and vice-versa, however the energies match better without switching them. We report the three lowest A$_1$ states resulting from DTDDFTa, although we do not have a reference value for the highest of the three, nor the fourth. Including the three lowest is helpful since it suggests that the ATDDFT's assignments should be shifted, given that the double-excitation interlopes in the manifold. 

Invoking the TDA at the adiabatic and dressed levels raises the energy of the 2A$_1$ and 3A$_1$ states in either case, giving a worse agreement to the reference values overall (with the exception of the DTDAa/PBE0 3A$_1$ state).
Overall, DTDDFTa/CAM-B3LYP gives the best energetic agreement to the reference for the 2A$_1$ and 3A$_1$ states, although the character of the states as indicated by $|G_I|^2$  are not so well-captured.

\begin{table*}[h!]
\caption{Singlet 1B$_2$, 1A$_2$, 1B$_1$ and two lowest A$_1$ excitation energies, $\Omega_I$, in eV and corresponding $|G_I|^2$ values in $\%$ of {\bf cyclopentadienone}, calculated with ATDDFT and DTDDFTa, as well as their TDA equivalents, at the FC geometry using cc-pVDZ. Truncated ATDDFT / ATDA and $\Delta$c-DTDDFTa / $\Delta$c-DTDAa excitation energies are also provided in italics and bold, respectively. For states of genuine double-excitation character, where $\Delta$c-DTDDFTa / $\Delta$c-DTDAa results are not calculated, we also give the corresponding excitation energies in bold. QUESTDB reference excitation energies and corresponding $\%T_1$ values (the latter given in parentheses) are taken from Ref.~\citenum{KBLJ24}. Note, CAM signifies CAM-B3LYP. We also include one extra $\Delta$c-DTDDFTa / $\Delta$c-DTDAa A$_1$ and B$_1$ state, denoted with a (?); see text for details.}
\label{tab:cyclopentadienone}
\centering
    \begin{tabular}{ll|cc|cc|cc|cc|c}
         & & \multicolumn{2}{c|}{ATDA} & \multicolumn{2}{c|}{ATDDFT} & \multicolumn{2}{c|}{DTDAa} & \multicolumn{2}{c|}{DTDDFTa} & QUESTDB \\
         & & PBE0 & CAM & PBE0 & CAM & PBE0 & CAM & PBE0 & CAM & \\
         \hline\hline
         1B$_2$ & $\Omega_I$ & 3.259 & 3.437 & 3.067 & 3.231 & & & & & 3.531 \\
         & $|G_I|^2$ & 100.0 & 100.0 & 100.0 & 100.0 & & & & & (91.2) \\
         \hline
         1A$_2$ & $\Omega_I$ & 2.734 & 2.927 & 2.702 & 2.891 & & & & & 2.951 \\
         & $|G_I|^2$ & 100.0 & 100.0 & 100.0 & 100.0 & & & & & (88.5) \\
         \hline
         2A$_1$ & $\Omega_I$ & 6.300 & 6.585 & 6.119 & 6.386 & & & & & 5.795 \\
         & & \textit{6.559} & \textit{6.888} & \textit{6.465} & \textit{6.794} & \textbf{6.262} & \textbf{6.559} & \textbf{5.623} & \textbf{6.006} & \\
         & $|G_I|^2$ & 100.0 & 100.0 & 100.0 & 100.0 & 9.3 & 12.0 & 34.1 & 33.2 & (73.6) \\
         \hline
         3A$_1$ & $\Omega_I$ & 8.384 & 8.677 & 8.019 & 8.247 & \textbf{6.419} & \textbf{6.754} & \textbf{6.533} & \textbf{6.762} & 6.714 \\
         & & \textit{10.251} & \textit{10.370} & \textit{9.418} & \textit{9.570} & 6.678 & 7.057 & 6.879 & 7.170 & \\
         & $|G_I|^2$ & 100.0 & 100.0 & 100.0 & 100.0 & 39.1 & 34.9 & 67.3 & 68.0 & (49.9) \\
         \hline
         (?)A$_1$ & $\Omega_I$ & & & & & \textbf{9.184} & \textbf{8.767} & \textbf{8.050} & \textbf{8.270} & \\
         & & & & & & 10.354 & 10.460 & 9.448 & 9.595 & \\
         & $|G_I|^2$ & & & & & 93.7 & 94.2 & 99.1 & 99.2 & \\
         \hline
         1B$_1$ & $\Omega_I$ & 6.493 & 6.776 & 6.474 & 6.753 & & & & & 5.009 \\
         & & \textit{6.550} & \textit{6.893} & \textit{6.541} & \textit{6.877} & \textbf{5.977} & \textbf{6.346} & \textbf{6.116} & \textbf{6.449} & \\
         & $|G_I|^2$ & 100.0 & 100.0 & 100.0 & 100.0 & 0.0 & 0.0 & 3.7 & 3.0 & (3.1) \\
         \hline
         (?)B$_1$ & $\Omega_I$ & & & & & \textbf{6.509} & \textbf{6.793} & \textbf{6.490} & \textbf{6.766} & \\
         & & & & & & 6.5663 & 6.910 & 6.558 & 6.890 & \\
         & $|G_I|^2$ & & & & & 80.3 & 79.5 & 96.3 & 97.0 & \\
         \hline\hline
    \end{tabular}
\end{table*}

\section{Discussion and Conclusions}
\label{sec:Concs}

\begin{figure}[h!]
    \centering
    \includegraphics[width=\linewidth]{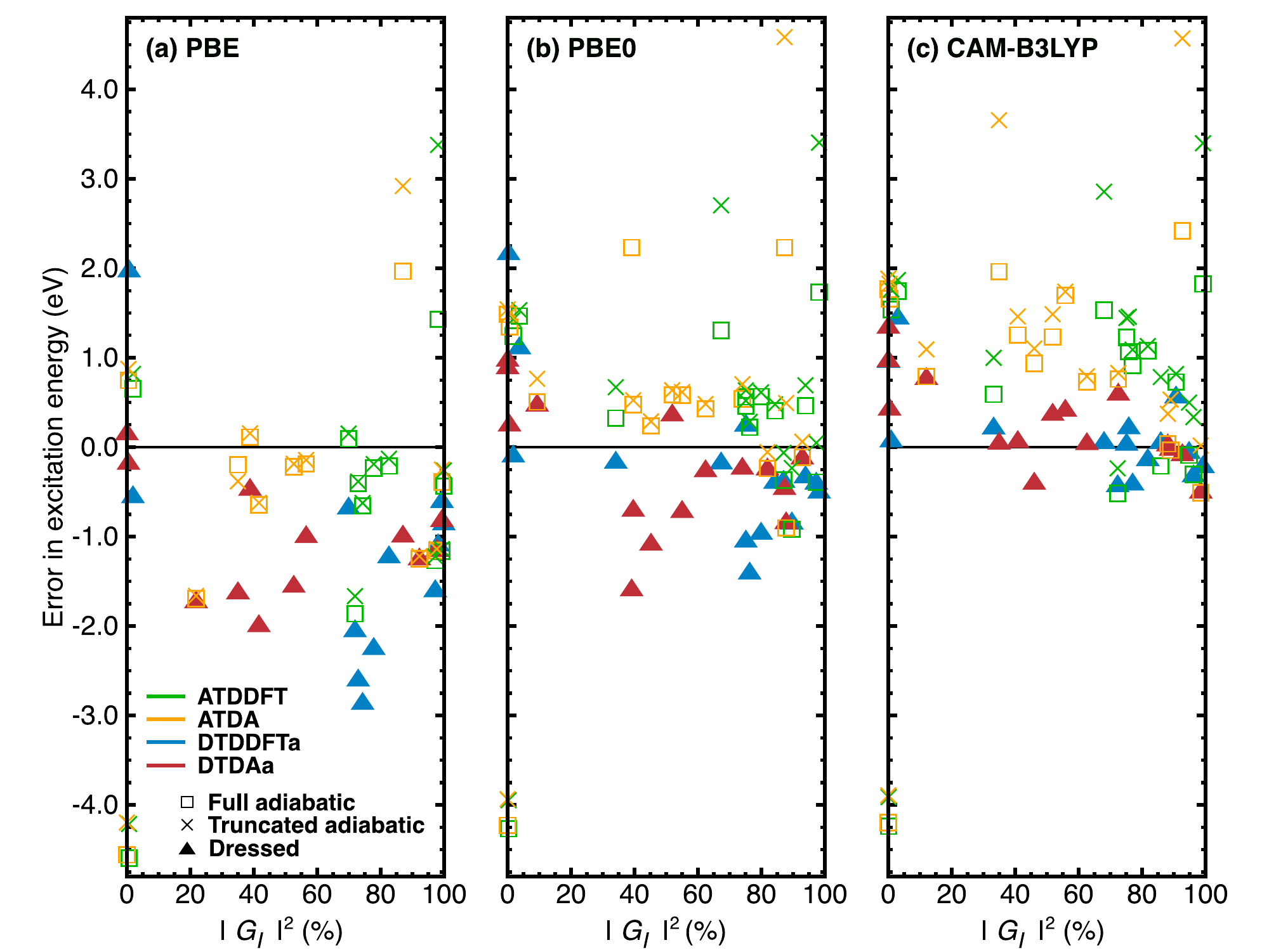}
    \caption{Error in the 15 excitation energies of interest across all nine molecules studied in this work as a function of $|G_I|^2$ value, calculated with full (un-filled square) and truncated (cross) ATDDFT (green) and ATDA (yellow), as well as their dressed (filled triangle) counterparts, DTDDFTa (blue) and DTDAa (red), for PBE (a), PBE0 (b) and CAM-B3LYP (c) using cc-pVDZ. Note, for states of single- and partial double-excitation character, we plot the error in the $\Delta$c-DTDDFTa/$\Delta$c-DTDAa values.}
    \label{fig:all_error}
\end{figure}

Before reviewing our findings, Fig.~\ref{fig:all_error} shows the error in the excitation energy as a function of $\vert G_I\vert^2$ for all molecules together, at their respective S$_0$ minimum geometries.
The error is defined in the same way as discussed in Section~\ref{subsec:polyenes}.
Several overall observations can be made.
First, regarding the effect of truncation on the adiabatic values themselves, we see that the truncated value is typically higher in energy than the full ATDDFT/ATDA counterpart, and this difference is least for PBE and more for the hybrids. 
As one changes the  adiabatic DFA from PBE to PBE0 to CAM-B3LYP, both truncated and full ATDDFT and ATDA go from underestimating the reference values to overestimating them. Most pertinent for the present study, the trend in using PBE to PBE0 to CAM-B3LYP as the underlying functional, is for DTDDFT and DTDA to go from being an underestimate to becoming relatively close to the reference values. 
Overall, out of all the methods, DTDDFTa/CAM-B3LYP is closest to the zero line, across the range of $\vert G_I\vert^2$, consistent with the discussions for each molecule in the previous sections.  
Interestingly, the range of $|G_I|^2$ values predicted by the three functionals is similar, despite the spread in the predicted energy error becoming smaller going from PBE to PBE0 to CAM-B3LYP. 

In providing benchmark data for the 15 excitations over 9 molecules, we have tested DTDDFT  across a range of states of double-excitation character of different natures, including from partial to genuine character and from relatively isolated to more densely embedded situations. Before this, we have delineated operational procedures and concerns arising with this relatively new method, and presented an `undownfolded' methodology to simplify the process in practise, addressing the complications discussed in Question 7 of Sec.~\ref{sec:Operational}. The aim is to progress DTDDFT towards a more well-defined, unambiguous approach for use in photochemical and photophysical applications, particularly since states of double-excitation character have a high propensity throughout excited potential energy landscapes of both simple and complex molecules. 

Embarking on applying DTDDFT, the questions touched on in Sec.~\ref{sec:Operational} naturally arise, and are partially addressed in this work.  The protocol described there (Question 1 in Sec.~\ref{sec:Operational}) is intended to provide a useful guide for how to proceed in practice. Future studies and experiences may modify its details. An immediate choice to be made is for the underlying adiabatic functional on top of which the dressing is to be built (Question 3 in Sec.~\ref{sec:Operational}), and our results using PBE, PBE0, and CAM-B3LYP suggest that hybrid functionals perform the best which is consistent with the conclusion  made in Ref.~\citenum{HIRC11} for the earlier DTDA approach. The earlier work considered only excitation energies and only within DTDA, but here we find hybrid functionals, particularly the range-separated CAM-B3LYP (which was also not studied in Ref.~\citenum{HIRC11}),  within DTDDFT, tends to be a good choice, for both the excitation energies as well as likely for the oscillator strengths as suggested by the single-vs double-excitation character of the state. This latter property of the excitation needs DTDDFT rather than DTDA to be meaningfully captured because of its satisfaction of the oscillator strength sum-rule; in fact we found that the DTDA energies were sometimes quite close to the DTDDFT ones, all else being equal, but not always, and DTDDFT is the preferred method due to the sum-rule conservation (Question 4 in Sec.~\ref{sec:Operational}). For the molecules studied here, we observed that the dressing has less of an effect as we go from PBE to PBE0 to CAM-B3LYP. While not explicitly discussed, or examined, in the present work, charge-transfer character in all the transitions involved in the dressing are important; that is, even if the state which we are aiming to correct does not have charge-transfer character, if the KS transition involved in the double-excitation it mixes with does (often the HOMO-LUMO transition), then a functional like CAM-B3LYP typically provides a more physical energy and orbitals for it, which are ingredients in the dressing for the double-excitation. 

Another choice to be made is that for the DTDDFT variant (Question 2 in Sec.~\ref{sec:Operational}). The results on the butadiene and hexatriene potential energy curves showed that DTDDFTa generally performed better than DTDDFTs, bearing out the expectation discussed in Question 2 of Sec.~\ref{sec:Operational}.

In some cases (e.g., octatetraene and cyclopentadiene),  we found that the errors of DTDDFTa applied on top of CAM-B3LYP often tend to be comparable to, or even better than, the error of CAM-B3LYP for single excitations of similar energies, which we found encouraging. However, given the range of performance across the whole molecule set, it would be too strong to say that DTDDFTa/CAM-B3LYP for states of double-excitation character could be used with as much confidence as CAM-B3LYP is used for single excitations. 

In some cases, the effect of truncation in the DTDDFT procedure dominated over the effect of the dressing (anticipated in Question 8 of Sec.~\ref{sec:Operational}). Sometimes applying the $\Delta$c DTDDFT to bring back the lost dynamical correlation then improved the results, giving a notable improvement over ATDDFT (e.g., octatetraene, cyclopentadiene, naphthalene); other times the $\Delta$c result was a little worse than the DTDDFTa but that still, using either DTDDFTa or $\Delta$c-DTDDFTa improved over ATDDFT (e.g., borole). In other cases, surprisingly, the truncated ATDDFT energies were closer to the reference than the full ATDDFT energies, and the dressing correction was negligible. This was seen most notably for acrolein; whether this is associated with a denser manifold of states with perhaps a large number of very small contributions from KS excitations constituting the ATDDFT excitations needs further investigation. 

One aspect that certainly needs more attention is basis-set dependence, which could affect both the raw ATDDFT ingredients, but especially the Hamiltonian matrix elements coupling the relevant states. Studies in this direction are underway. 

Another aspect is the extra state that is obtained in the dressing procedure. From considering the dimension of the state space one does expect this extra state, as discussed in Sec.~\ref{sec:Operational}, but one does not always find this extra state in the reference data (e.g., ethylene, borole, cyclopentadieone). In some cases we found its energy is high and so embedded in a denser manifold of higher-lying states with which it may interact and the DTDDFT protocol here would not be appropriate. But in other cases, its energy is predicted to fall close to the states listed in the usual references, but it is absent there (e.g., acrolein). One might surmise that since such a state has a predominantly double-excitation character (low $\vert G_I\vert^2$), the perhaps correspondingly low peak it would yield in linear response would mean that it simply does not show up. However, a closer investigation is needed. 

Relatedly, it should be borne in mind that some of the QUESTDB reference values have  been considered ``unsafe'' by their authors, [octatetraene (2A$_g$), naphthalene (1B$_{3u}$, 2A$_g$, 3A$_g$), borole (2A$_1$), cyclopentadienone (2A$_1$, 3A$_1$, 1B$_1$)] and in these cases, we should apply caution when comparing too closely the results of DTDDFT with them. 

Finally, we hope that this study is useful for both evaluating the performance of the DTDDFT method as well as pointing to possible future modifications. The exact condition derived in Sec.~\ref{sec:undownfolding} should be respected, and whether one can find a variant that avoids the possibility of imaginary frequencies that are sometimes encountered (Question 6 in Sec.~\ref{sec:Operational}), and how these are related to the recent insights of Ref.~\citenum{FC25}, would be an avenue for future work. We hope that the advances and analysis provided here have at least partially closed the Pandora's Box of double excitations in TDDFT.

\section*{Author contributions}
\textbf{Jack T. Taylor}: Methodology (equal); Investigation (lead); Visualisation (lead); Formal analysis (equal); Validation (lead); Writing -- original draft (lead); Write -- review and editing (equal); Data curation (lead). \textbf{Neepa T. Maitra}: Conceptualisation (lead); Methodology (equal); Formal analysis (equal); Write -- review and editing (equal); Supervision (lead); Funding acquisition (lead).

\section*{Conflicts of interest}
There are no conflicts to declare.

\section*{Data availability}
The data that support the findings of this study are available within the article and its supplementary material. 
The latter contains additional tables summarizing (i) two-electron integral orbital patterns, (ii) state characters of the excitation energies under investigation, along with the KS single and double excitations used in their dressing correction, (iii) details of the reference wavefunction methods, and (iv) KS contributions to the lowest four ATDDFT and ATDA states of each molecule.
An additional figure is also provided, highlighting the error in the 2A$_g$ energy for butadiene and hexatriene as a function of both BLA coordinate and $|G_I|^2$ value.

\section*{Acknowledgements}
We are grateful to Jun Shen and Piotr Piecuch for providing the butadiene and hexatriene BLA coordinate geometries, as well as the corresponding reference $\Delta$-CR-EOMCC(2,3)/cc-pVTZ $1B_u$ and $2A_g$ excited-state energies, and to Niri Govind for help with NWChem. 
Financial support from  the Department of Energy, Office of Basic Energy Sciences, Division of Chemical Sciences, Geosciences and Biosciences under Award No. DE‐SC0024496  is gratefully acknowledged. 
The authors acknowledge the Office of Advanced Research Computing (OARC) at Rutgers, The State University of New Jersey for providing access to the Amarel cluster and associated research computing resources that have contributed to the results reported here.

\bibliography{ref} 
\bibliographystyle{rsc} 

\end{document}


\title{Supplementary information for ``Dressed Time-Dependent Density Functional Theory: Opening and Closing a Pandora's Box''}
\author{Jack T. Taylor}
\affiliation{Department of Physics, Rutgers University, Newark 07102, New Jersey USA}
\author{Neepa T. Maitra}
\affiliation{Department of Physics, Rutgers University, Newark 07102, New Jersey USA}
\date{\today}

\maketitle

\section{Supplementary figures and tables}
\label{sec:supp_figs_tabs}

\begin{table*}[h!]
\caption{Summary of the possible two-electron integral orbital patterns and prefactors for constructing $H_{qD}$ matrix elements for a given KS double excitation in this work, considering an arbitrary occupied (unoccupied) orbital, $i$ ($a$), the HOMO$-$1 (H$-$1), HOMO (H) and/or LUMO (L).}
\label{tab:hqd_orb_pat}
\centering
    \begin{tabular}{llll}
    \hline\hline
    $D$ & $q$ & & $H_{qD}$ \\
    \hline\hline
    H$^2$ $\rightarrow$ L$^2$ & $i \rightarrow$ L & (for $i <$ H) & $\frac{-2}{\sqrt{2}}$ (L, H $|$ H, $i$)\\
    & H $\rightarrow$ $a$ & (for $a >$ L) & $\frac{2}{\sqrt{2}}$ ($a$, L $|$ L, H)\\
    \hline
    H, H$-$1 $\rightarrow$ L$^2$ & $i \rightarrow$ L & (for $i <$ H$-$1) & $-1 \times$ [(L, H $|$ H$-$1, $i$) + (L, H$-$1 $|$ H, $i$)] \\
    & H $\rightarrow$ $a$ & (for $a >$ L) & ($a$, L $|$ L, H$-$1)\\
    & H$-$1 $\rightarrow$ $a$ & (for $a >$ L) & ($a$, L $|$ L, H)\\
    \hline\hline
    \end{tabular}
\end{table*}

\begin{table*}[h!]
\caption{Summary of the molecules and electronic excited states under investigation in the present work, along with the KS double and single excitations ($D$ and $q$, respectively) considered in the corresponding dressed (full) TDDFT and TDA problems.}
\label{tab:summary}
\centering
    \begin{tabular}{lllllll}
    \hline\hline
    Molecule & Point & Excited & State & $D$ & $q$ & DTDDFT (functional$^\ddag$) \\
     & group & state(s) & character$^\dagger$ & & & \\
    \hline\hline
    butadiene & $C_{2h}$ & 2A$_g$ & pd & 2$\times$(15-16) & 14-16, 15-17 & full / TDA (P, P0) \\
    & & & & & 14-16, 15-18 & full / TDA (C) \\
    \hline
    hexatriene & $C_{2h}$ & 2A$_g$ & pd & 2$\times$(22-23) & 21-23, 22-24 & full / TDA (P, P0, C) \\
    \hline
    octatetraene & $C_{2h}$ & 2A$_g$ & pd & 2$\times$(29-30) & 28-30, 29-31 & full / TDA (P, P0, C) \\
    \hline
    cyclopentadiene & $C_{2v}$ & 2A$_1$ & pd & 2$\times$(18-19) & 17-19, 18-21 & full (P0, C), TDA (P0, C) \\
    & & & & & 17-19, 18-20 & full (P) \\
    & & & & & 17-19, 18-20, 12-19 & TDA (P) \\
    \hline
    acrolein & $C_s$ & 2A', 3A', 4A' & s, s, pd & 2$\times$(15-16) & 14-16, 15-17, 15-19, 15-20 & full / TDA (P) \\
    & & & & & 14-16, 15-17, 15-18 & full / TDA (P0) \\
    & & & & & 14-16, 15-17, 15-19, 15-20, 15-29, 15-31 & full / TDA (C) \\
    \hline
    naphthalene & $D_{2h}$ & 2A$_g$, 3A$_g$ & s, pd & 2$\times$(34-35) & 33-37, 32-36, 29-35, 34-39 & full / TDA (P) \\
    & & & & & 33-37, 32-36, 30-35, 34-41 & full / TDA (P0, C) \\
    \hline
    ethylene & $D_{2h}$ & 2A$_g$ & gd & 2$\times$(8-9) & 8-14 & full / TDA (P, P0, C)\\
    \hline
    borole & $C_{2v}$ & 2A$_1$, 3A$_1$ & gd, s & 2$\times$(17-18) & 14-18, 17-19 & full (P, P0), TDA (P0) \\
    & & & & & 14-18, 17-19, 14-20 & TDA (P) \\
    & & & & & 14-18, 17-21 & full / TDA (C) \\
    \hline
    cyclopentadienone & $C_{2v}$ & 2A$_1$, 3A$_1$ & pd, pd & 2$\times$(21-22) & 19-22, 21-23, 16-22 & full / TDA (P0, C) \\
     & & 1B$_1$ & gd & (21-22)+(20-22) & 18-22 & full / TDA (P0, C) \\
    \hline\hline
    \end{tabular}
    \begin{tablenotes}
        \item $^\dagger$Following Ref.~\cite{KBLJ24}, excited states are characterised as either single (s), partially double (pd) or genuinely double (gd).
        \item $^\ddag$For brevity, P = PBE, P0 = PBE0, C = CAM-B3LYP.
    \end{tablenotes}
\end{table*}

\begin{table*}[h!]
\caption{Summary of the wavefunction methods and basis sets used to generate reference excitation energies and $\%T_1$ values employed to compare to our ATDDFT/ATDA and DTDDFT/DTDA results from this work.}
\label{tab:ref_methods}
\centering
    \begin{threeparttable}
    \begin{tabular}{lll}
    \hline\hline
    Molecule & Excited states & Reference level of theory \\
    \hline\hline
    \textbf{Excitation energies} & & \\
    butadiene & 1B$_u$, 2A$_g$ & $\delta$-CR-EOMCC(2,3) / cc-pVTZ $^a$ \\
    hexatriene & 1B$_u$, 2A$_g$ & $\delta$-CR-EOMCC(2,3) / cc-pVTZ $^a$ \\
    octatetraene & 1B$_u$ & CCSDT / ``aug-cc-pVTZ'' $^{b,\dagger}$ \\
     & 2A$_g$ & CASPT3 / aug-cc-pVTZ $^b$ \\
    cyclopentadiene & 1B$_1$ & CC4 / "aug-cc-pVTZ" $^{b,\dagger}$ \\
     & 2A$_1$ & CCSDTQ / "aug-cc-pVTZ" $^{b,\dagger}$ \\
    acrolein & 1A", 2A', 3A', 4A' & CCSDTQ / ``aug-cc-pVTZ'' $^{b,\dagger}$ \\
    naphthalene & 1B$_{3u}$, 2A$_g$ & CCSDT / ``aug-cc-pVTZ'' $^{b,\dagger}$ \\
     & 3A$_g$ & CASPT3 / aug-cc-pVTZ $^b$ \\
    ethylene & 1B$_{1u}$ & CCSDTQP / ``aug-cc-pVTZ'' $^{b,\dagger}$ \\
     & 1B$_{2u}$ & $-$ \\
     & 2A$_g$ & exFCI / ``aug-cc-pVTZ'' $^{b,\dagger}$ \\
    borole & 1B$_1$, 2A$_1$, 3A$_1$ & CCSDTQ / ``aug-cc-pVTZ'' $^{b,\dagger}$ \\
    cyclopentadienone & 1B$_2$, 1A$_2$, 2A$_1$, 3A$_1$ & CC4 / ``aug-cc-pVTZ'' $^{b,\dagger}$ \\
     & 1B$_1$ & exFCI / ``aug-cc-pVTZ'' $^{b,\dagger}$ \\
     \hline
     \textbf{$\boldsymbol{\%T_1}$ values} & & \\
     All & All & CC3 / aug-cc-pVTZ$^b$ \\
    \hline\hline
    \end{tabular}
    \begin{tablenotes}[flushleft]
        \item $^a$ Ref.~\cite{PSLP21}
        \item $^b$ Ref.~\cite{KBLJ24}
        \item $^\dagger$The ``\dots'' denotes the use of a basis set correction (see Ref.~\cite{KBLJ24} for details).
    \end{tablenotes}
    \end{threeparttable}
\end{table*}

\begin{figure*}[h!]
    \centering
    \includegraphics[width=\textwidth]{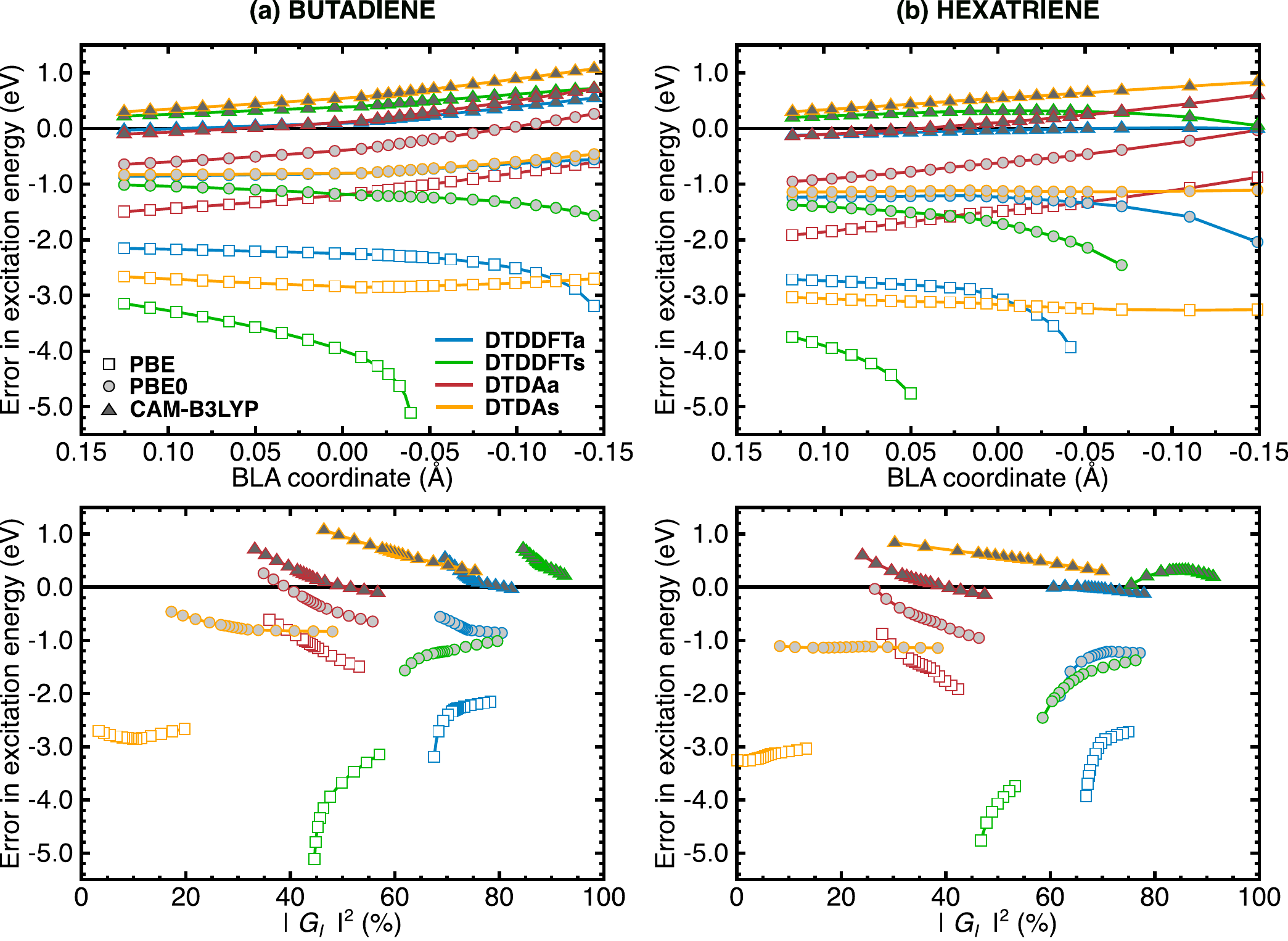}
    \caption{Error in the 2A$_g$ excitation energy in butadiene (a) and hexatriene (b) as function of BLA coordinate (top panel) or a function of $|G_I|^2$ value (bottom panel) for DTDDFTs (green), DTDDFTa (blue), DTDAs (yellow) and DTDAa (red) using PBE (square), PBE0 (circle) and CAM-B3LYP (triangle) in combination with cc-pVDZ.}
    \label{butadiene_hexatriene_error}
\end{figure*}

\clearpage
\subsection{Butadiene}

\begin{table*}[h!]
\caption{Quantities used to inform the decision of which KS single excitations, $q$, to consider in DTDDFT and DTDA along with the KS double excitation, $D = 2\times$(15-16), for an appropriate description of the lowest A$_g$ excitation energy in butadiene at the PBE/cc-pVDZ level of theory. We report quantities for the first four A$_g$ states of butadiene.}
\label{butadiene_pbe}
\centering
\begin{tabular}{cccc|cccc}
    \hline\hline
    ATDDFT root & $q$ & ($|X_q|^2 - |Y_q|^2) (\%)$ & $|H_{qD}| / 10^{-3} (\text{a.u.})$ & ATDA root & $q$ & $|X_q|^2 (\%)$ & $|H_{qD}| / 10^{-3} (\text{a.u.})$\\
    \hline\hline
    2A$_g$ &  14-16 &  53.4 &  90.040 & 2A$_g$ &  14-16 &  50.8 &  90.040 \\
    &  15-17 &  46.6 &  76.158 & &  15-17 &  49.1 &  76.158 \\
    \hline
    3A$_g$ &  15-17 &  50.8 &  76.158 & 3A$_g$ &  15-17 &  44.7 &  76.158 \\
    &  14-16 &  43.9 &  90.040 & &  14-16 &  42.2 &  90.040 \\
    & 10-22 & 0.4 & 0.000 & & 13-19 & 1.9 & 0.000 \\
    & 13-19 & 0.4 & 0.000 & & 12-18 & 1.2 & 0.000 \\
    & 10-19 & 0.4 & 0.000 & & 10-19 & 0.9 & 0.000 \\
    & 9-24 & 0.3 & 0.000 & & 10-22 & 0.9 & 0.000 \\
    & 11-25 & 0.3 & 0.000 & & 9-24 & 0.6 & 0.000 \\
    & 12-18 & 0.3 & 0.000 & & 11-25 & 0.5 & 0.000 \\
    & & & & & 15-35 & 0.4 & 14.315 \\
    & & & & & 12-23 & 0.4 & 0.000 \\
    & & & & & 7-19 & 0.4 & 0.000 \\
    & & & & & 8-23 & 0.4 & 0.000 \\
    & & & & & 15-32 & 0.4 & 53.084 \\
    & & & & & 13-25 & 0.4 & 0.000 \\
    & & & & & 13-28 & 0.4 & 0.000 \\
    & & & & & 9-26 & 0.3 & 0.000 \\ 
    & & & & & 14-36 & 0.3 & 0.000 \\
    & & & & & 11-34 & 0.3 & 0.000 \\
    & & & & & 12-27 & 0.3 & 0.000 \\
    & & & & & 13-21 & 0.3 & 0.000 \\
    \hline
    4A$_g$ & 13-19 & 98.2 & 0.000 & 4A$_g$ & 13-19 & 96.2 & 0.000 \\
    & 12-20 & 0.8 & 0.000 & & 12-20 & 1.0 & 0.000 \\
    & & & & &  15-17 &  0.8 &  76.158 \\
    & & & & &  14-16 &  0.6 &  90.040 \\
    & & & & & 13-22 & 0.3 & 0.000 \\
    \hline 
    5A$_g$ & 12-18 & 95.3 & 0.000 & 5A$_g$ & 12-18 & 93.0 & 0.000 \\
    & 13-21 & 3.4 & 0.000 & & 13-21 & 4.1 & 0.000 \\
    & & & & &  14-16 &  0.6 &  90.040 \\
    & & & & &  15-17 &  0.4 &  76.158 \\
    \hline
    \multicolumn{4}{c|}{Two KS single excitations considered:} & \multicolumn{4}{c}{Two KS single excitations considered:} \\
    \multicolumn{4}{c|}{14-16, 15-17} & \multicolumn{4}{c}{14-16, 15-17} \\
    \hline\hline
    \end{tabular}
\end{table*}

\begin{table*}[h!]
\caption{Quantities used to inform the decision of which KS single excitations, $q$, to consider in DTDDFT and DTDA along with the KS double excitation, $D = 2\times$(15-16), for an appropriate description of the lowest A$_g$ excitation energy in butadiene at the PBE0/cc-pVDZ level of theory. We report quantities for the first four A$_g$ states of butadiene.}
\label{butadiene_pbe0}
\centering
    \begin{tabular}{cccc|cccc}
    \hline\hline
    ATDDFT root & $q$ & $(|X_q|^2 - |Y_q|^2) (\%)$ & $|H_{qD}| / 10^{-3} (\text{a.u.})$ & ATDA root & $q$ & $|X_q|^2 (\%)$ & $|H_{qD}| / 10^{-3} (\text{a.u.})$\\
    \hline\hline
    2A$_g$ &  14-16 &  54.3 &  85.746 & 2A$_g$ &  14-16 &  50.6 &  85.746 \\
    &  15-17 &  45.7 &  71.132 & &  15-17 &  49.2 &  71.132 \\ 
    \hline
    3A$_g$ &  15-17 &  52.3 &  71.132 & 3A$_g$ &  15-17 &  46.0 &  71.132 \\
    &  14-16 &  43.4 &  85.746 & &  14-16 &  43.7 &  85.746 \\
    & 10-22 & 0.4 & 0.000 & & 13-19 & 0.9 & 0.000 \\
    & 10-19 & 0.4 & 0.000 & &  12-18 & 0.8 & 0.000 \\
    & 13-19 & 0.3 & 0.000 & & 10-19 & 0.7 & 0.000 \\ 
    & 9-24 & 0.3 & 0.000 & & 10-22 & 0.7 & 0.000 \\ 
    & 12-18 & 0.3 & 0.000 & & 9-24 & 0.5 & 0.000 \\
    & & & & & 7-19 & 0.4 & 0.000 \\ 
    & & & & & 8-23 & 0.4 & 0.000 \\
    & & & & & 11-25 & 0.4 & 0.000 \\
    & & & & & 13-28 & 0.4 & 0.000 \\
    & & & & & 11-34 & 0.3 & 0.000 \\
    & & & & & 13-25 & 0.4 & 0.000 \\
    & & & & & 9-26 & 0.3 & 0.000 \\
    & & & & & 12-23 & 0.3 & 0.000 \\
    & & & & & 12-27 & 0.3 & 0.000 \\
    \hline    
    4A$_g$ & 13-19 & 96.9 & 0.000 & 4A$_g$ & 13-19 & 96.4 & 0.000 \\
    & 12-20 & 2.2 & 0.000 & & 12-20 & 1.9 & 0.000 \\
    & & & & &  15-17 &  0.5 &  71.132 \\
    & & & & &  14-16 &  0.3 &  85.746 \\
    \hline
    5A$_g$ & 12-18 & 87.2 & 0.000 & 5A$_g$ & 12-18 & 86.4 & 0.000 \\
    & 11-19 & 5.5 & 0.000 & & 11-19 & 5.9 & 0.000 \\
    & 13-21 & 4.7 & 0.000 & & 13-21 & 4.2 & 0.000 \\
    & 12-20 & 1.5 & 0.000 & & 12-20 & 1.7 & 0.000 \\
    & & & & &  14-16 &  0.4 &  85.746 \\
    \hline
    \multicolumn{4}{c|}{Two KS single excitations considered:} & \multicolumn{4}{c}{Two KS single excitations considered:} \\
    \multicolumn{4}{c|}{14-16, 15-17} & \multicolumn{4}{c}{14-16, 15-17} \\
    \hline\hline
    \end{tabular}
\end{table*}

\begin{table*}[h!]
\caption{Quantities used to inform the decision of which KS single excitations, $q$, to consider in DTDDFT and DTDA along with the KS double excitation, $D = 2\times$(15-16), for an appropriate description of the lowest A$_g$ excitation energy in butadiene at the CAM-B3LYP/cc-pVDZ level of theory. We report quantities for the first four A$_g$ states of butadiene.}
\label{butadiene_cam-b3lyp}
\centering
    \begin{tabular}{cccc|cccc}
    \hline\hline
    ATDDFT root & $q$ & $(|X_q|^2 - |Y_q|^2) (\%)$ & $|H_{qD}| / 10^{-3} (\text{a.u.})$ & ATDA root & $q$ & $|X_q|^2 (\%)$ & $|H_{qD}| / 10^{-3} (\text{a.u.})$\\
    \hline\hline
    2A$_g$ &  14-16 &  53.6 &  82.930 & 2A$_g$ &  15-18 &  50.9 &  67.779 \\
    &  15-18 &  46.2 &  67.779 & &  14-16 &  48.9 &  82.930 \\ 
    \hline
    3A$_g$ &  15-18 &  51.7 &  67.779 & 3A$_g$ &  14-16 &  45.2 &  82.930 \\
    &  14-16 &  44.0 &  82.930 & &  15-18 &  44.3 &  67.779 \\ 
    & 10-22 & 0.4 & 0.000 & & 12-17 & 1.0 & 0.000 \\
    & 12-17 & 0.4 & 0.000 & & 13-19 & 1.0 & 0.000 \\
    & 13-19 & 0.4 & 0.000 & & 10-19 & 0.8 & 0.000 \\
\hline\hline
    \end{tabular}
\end{table*}

\clearpage
\subsection{Hexatriene}

\begin{table*}[h!]
\caption{Quantities used to inform the decision of which KS single excitations, $q$, to consider in DTDDFT and DTDA along with the KS double excitation, $D = 2\times$(22-23), for an appropriate description of the lowest A$_g$ excitation energy in hexatriene at the PBE/cc-pVDZ level of theory. We report quantities for the first four A$_g$ states of hexatriene.}
\label{hexatriene_pbe}
\centering
    \begin{tabular}{cccc|cccc}
    \hline\hline
    ATDDFT root & $q$ & $(|X_q|^2 - |Y_q|^2) (\%)$ & $|H_{qD}| / 10^{-3} (\text{a.u.})$ & ATDA root & $q$ & $|X_q|^2 (\%)$ & $|H_{qD}| / 10^{-3} (\text{a.u.})$ \\
    \hline\hline
    2A$_g$ &  21-23 &  56.9 &  79.763 & 2A$_g$ &  21-23 &  54.0 &  79.763 \\
    &  22-24 &  43.1 &  74.061 & &  22-24 &  45.9 &  74.061 \\
    \hline
    3A$_g$ &  22-24 &  53.2 &  74.061 & 3A$_g$ &  22-24 &  45.6 &  74.061 \\
    &  21-23 &  39.5 &  79.763 & &  21-23 &  37.2 &  79.763 \\
    & 19-24 & 2.1 & 0.000 & & 19-24 & 5.1 & 0.000 \\
    & 21-26 & 1.6 & 0.000 & & 21-26 & 3.7 & 0.000 \\
    & 16-30 & 0.4 & 0.000 & & 16-30 & 0.9 & 0.000 \\
    & 15-34 & 0.3 & 0.000  & & 15-34 & 0.5 & 0.000 \\
    & & & & & 17-28 & 0.5 & 0.000 \\
    & & & & & 14-25 & 0.4 & 0.000 \\
    & & & & & 18-36 & 0.4 & 0.000 \\
    & & & & & 14-31 & 0.4 & 0.000 \\
    & & & & & 17-40 & 0.3 & 0.000 \\
    & & & & & 14-27 & 0.3 & 0.000 \\
    & & & & & 16-28 & 0.3 & 0.000 \\
    & & & & & 22-41 & 0.3 & 5.584 \\
    \hline
    4A$_g$ & 21-26 & 56.8 & 0.000 & 4A$_g$ & 21-26 & 56.7 & 0.000 \\
    & 19-24 & 43.1 & 0.000 & & 19-24 & 43.2 & 0.000 \\
    \hline
    5A$_g$ & 19-24 & 52.2 & 0.000 & 5A$_g$ & 18-25 & 49.9 & 0.000 \\
    & 21-26 & 39.6 & 0.000 & & 19-24 & 21.7 & 0.000 \\
    &  21-23 &  1.8 &  79.763 & & 21-26 & 17.7 & 0.000 \\
    & 18-25 & 0.8 & 0.000 & &  21-23 &  1.8 &  79.763 \\
    & 20-30 & 0.6 & 0.000 & &  22-24 &  1.8 &  74.061 \\
    & & & & & 20-29 & 1.5 & 0.000 \\
    & & & & & 20-30 & 1.3 & 0.000 \\
    & & & & & 20-28 & 0.6 & 0.000 \\
    & & & & & 18-27 & 0.3 & 0.000 \\
    \hline
    \multicolumn{4}{c|}{Two KS single excitations considered:} & \multicolumn{4}{c}{Two KS single excitations considered:} \\
    \multicolumn{4}{c|}{21-23, 22-24} & \multicolumn{4}{c}{21-23, 22-24} \\
    \hline\hline
    \end{tabular}
\end{table*}

\begin{table*}[h!]
\caption{Quantities used to inform the decision of which KS single excitations, $q$, to consider in DTDDFT and DTDA along with the KS double excitation, $D = 2\times$(22-23), for an appropriate description of the lowest A$_g$ excitation energy in hexatriene at the PBE0/cc-pVDZ level of theory. We report quantities for the first four A$_g$ states of hexatriene.}
\label{hexatriene_pbe0}
\centering
    \begin{tabular}{cccc|cccc}
    \hline\hline
    ATDDFT root & $q$ & $(|X_q|^2 - |Y_q|^2) (\%)$ & $|H_{qD}| / 10^{-3} (\text{a.u.})$ & ATDA root & $q$ & $|X_q|^2 (\%)$ & $|H_{qD}| / 10^{-3} (\text{a.u.})$\\
    \hline\hline
    2A$_g$ &  21-23 &  59.8 &  74.666 & 2A$_g$ &  21-23 &  55.1 &  74.666 \\
    &  22-24 &  39.5 &  68.803 & &  22-24 &  44.0 &  68.803 \\
    & 21-26 & 0.5 & 0.000 & & 21-26 & 0.5 & 0.000 \\
    & 19-24 & 0.3 & 0.000 & & & & \\   
    \hline
    3A$_g$ &  22-24 &  58.1 &  68.803 & 3A$_g$ &  22-24 &  50.3 &  68.803 \\
    &  21-23 &  38.0 &  74.666 & &  21-23 &  39.2 &  74.666 \\
    & 19-24 & 0.5 & 0.000 & & 19-24 & 1.8 & 0.000 \\
    & 16-30 & 0.4 & 0.000 & &  21-26 & 1.1 & 0.000 \\
    & 18-36 & 0.3 & 0.000 & & 16-30 & 0.7 & 0.000 \\ 
    & & & & & 18-36 & 0.5 & 0.000 \\
    & & & & & 15-35 & 0.4 & 0.000 \\
    & & & & & 14-25 & 0.3 & 0.000 \\
    & & & & & 14-27 & 0.3 & 0.000 \\
    & & & & & 14-31 & 0.3 & 0.000 \\
    & & & & & 17-28 & 0.3 & 0.000 \\
    & & & & & 17-40 & 0.3 & 0.000 \\
    \hline
    4A$_g$ & 21-26 & 82.2 & 0.000 & 4A$_g$ & 21-26 & 70.8 & 0.000\\
    & 19-24 & 16.8 & 0.000 & & 19-24 & 28.2 & 0.000 \\
    &  21-23 &  0.5 &  74.666 & &  21-23 &  0.5 &  74.666 \\
    \hline
    5A$_g$ & 19-24 & 80.5 & 0.000 & 5A$_g$ & 19-24 & 65.3 & 0.000 \\
    & 21-26 & 15.8 & 0.000 & & 21-26 & 24.6 & 0.000 \\
    &  22-24 &  0.6 &  68.803 & &  22-24 &  1.3 &  68.803 \\
    & 18-25 & 0.4 & 0.000 & & 18-25 & 1.2 & 0.000 \\
    & & & & &  21-23 &  0.8 &  74.666 \\
    & & & & & 20-28 & 0.7 & 0.000 \\
    & & & & & 18-27 & 0.4 & 0.000 \\ 
    & & & & & 12-28 & 0.3 & 0.000 \\
    & & & & & 20-30 & 0.3 & 0.000 \\
    \hline
    \multicolumn{4}{c|}{Two KS single excitations considered:} & \multicolumn{4}{c}{Two KS single excitations considered:} \\
    \multicolumn{4}{c|}{21-23, 22-24} & \multicolumn{4}{c}{21-23, 22-24} \\
    \hline\hline
    \end{tabular}
\end{table*}

\begin{table*}[h!]
\caption{Quantities used to inform the decision of which KS single excitations, $q$, to consider in DTDDFT and DTDA along with the KS double excitation, $D = 2\times$(22-23), for an appropriate description of the lowest A$_g$ excitation energy in hexatriene at the CAM-B3LYP/cc-pVDZ level of theory. We report quantities for the first four A$_g$ states of hexatriene.}
\label{hexatriene_cam-b3lyp}
\centering
    \begin{tabular}{cccc|cccc}
    \hline\hline
    ATDDFT root & $q$ & $(|X_q|^2 - |Y_q|^2) (\%)$ & $|H_{qD}| / 10^{-3} (\text{a.u.})$ & ATDA root & $q$ & $|X_q|^2 (\%)$ & $|H_{qD}| / 10^{-3} (\text{a.u.})$\\
    \hline\hline
    2A$_g$ &  21-23 &  64.0 &  70.764 & 2A$_g$ &  21-23 &  54.7 &  70.764 \\
    &  22-24 &  32.8 &  64.807 & &  22-24 &  41.8 &  64.807 \\ 
    & 21-28 & 1.7 & 0.000 & & 21-28 & 1.9 & 0.000 \\ 
    & 19-24 & 1.5 & 0.000 & & 19-24 & 1.4 & 0.000 \\
    \hline
    3A$_g$ &  22-24 &  63.8 &  64.807 & 3A$_g$ &  22-24 &  52.1 &  64.807 \\
    &  21-23 &  32.9 &  70.764 & &  21-23 &  39.4 &  70.764 \\
    & 16-30 & 0.3 & 0.000 & & 16-30 & 0.7 & 0.000 \\ 
    & 18-36 & 0.3 & 0.000 & & 19-24 & 0.6 & 0.000 \\ 
    & & & & & 18-36 & 0.5 & 0.000 \\
    & & & & & 14-25 & 0.3 & 0.000 \\
    & & & & & 14-26 & 0.3 & 0.000 \\
    & & & & & 14-31 & 0.3 & 0.000 \\
    & & & & & 15-35 & 0.3 & 0.000 \\
    & & & & & 20-27 & 0.3 & 0.000 \\
    \hline
    4A$_g$ & 21-28 & 66.9 & 0.000 & 4A$_g$ & 21-28 & 73.2 & 0.000 \\
    & 19-24 & 29.6 & 0.000 & & 19-24 & 19.2 & 0.000 \\
    & 18-25 & 0.4 & 0.000 & & 18-25 & 0.9 & 0.000 \\ 
    & & & & & 20-27 & 0.7 & 0.000 \\ 
    & & & & &  21-23 &  0.7 &  70.764 \\ 
    & & & & & 18-26 & 0.5 & 0.000 \\
    \hline
    5A$_g$ & 19-24 & 65.4 & 0.000 & 5A$_g$ & 19-24 & 71.6 & 0.000 \\
    & 21-28 & 30.0 & 0.000 & & 21-28 & 22.0 & 0.000 \\ 
    &  22-24 &  1.7 &  64.807 & &  22-24 &  1.9 &  64.807 \\
    &  21-23 &  1.4 &  70.764 & & 18-25 & 1.7 & 0.000 \\
    & 18-25 & 0.8 & 0.000 & &  21-23 &  1.1 &  70.764 \\  
    & & & & & 20-27 & 0.4 & 0.000 \\
    \hline
    \multicolumn{4}{c|}{Two KS single excitations considered:} & \multicolumn{4}{c}{Two KS single excitations considered:} \\
    \multicolumn{4}{c|}{21-23, 22-24} & \multicolumn{4}{c}{21-23, 22-24} \\
    \hline\hline
    \end{tabular}
\end{table*}

\clearpage
\subsection{Octatetraene}

\begin{table*}[h!]
\caption{Quantities used to inform the decision of which KS single excitations, $q$, to consider in DTDDFT and DTDA along with the KS double excitation, $D = 2\times$(29-30), for an appropriate description of the lowest A$_g$ excitation energy in octatetraene at the PBE/cc-pVDZ level of theory. We report quantities for the first four A$_g$ states of octatetraene.}
\label{}
\centering
    \begin{tabular}{cccc|cccc}
    \hline\hline
    ATDDFT root & $q$ & $(|X_q|^2 - |Y_q|^2) (\%)$ & $|H_{qD}| / 10^{-3} (\text{a.u.})$ & ATDA root & $q$ & $|X_q|^2 (\%)$ & $|H_{qD}| / 10^{-3} (\text{a.u.})$\\
    \hline\hline
    2A$_g$ &  28-30 &  56.4 &  71.627 & 2A$_g$ &  28-30 &  53.9 &  71.627 \\
    &  29-31 &  43.6 &  68.603 & &  29-31 &  46.0 &  68.603 \\
    \hline
    3A$_g$ & 29-31 &  51.4 & 68.603 & 3A$_g$ & 24-30 &  53.2 & 5.307 \\
    &  28-30 &  38.9 &  71.627 & &  29-34 &  46.2 &  7.007 \\
    & 27-31 & 2.4 & 0.000 & & 27-31 & 0.3 & 0.000 \\
    & 29-34 & 2.1 & 7.007 & & & & \\
    & 28-32 & 1.3 & 0.000 & & & & \\
    & 24-30 & 1.0 & 5.307 & & & & \\  
    \hline
    4A$_g$ & 24-30 & 57.2 & 5.307 & 4A$_g$ &  29-31 &  35.8 &  68.603 \\
    &  29-34 &  42.1 &  7.007 & &  28-30 &  29.1 &  71.627 \\
    & 27-31 & 0.4 & 0.000 & & 27-31 & 9.8 & 0.000 \\
    & & & & & 29-34 & 8.4 & 7.007 \\
    & & & & & 24-30 & 7.3 & 5.307 \\
    & & & & & 28-32 & 4.4 & 0.000 \\
    & & & & & 23-38 & 0.4 & 0.000 \\    
    \hline
    5A$_g$ & 29-34 & 43.8 & 7.007 & 5A$_g$ & 29-34 & 29.0 & 7.007 \\
    & 24-30 & 27.0 & 5.307 & & 24-30 & 19.3 & 5.307 \\
    & 27-31 & 14.8 & 0.000 & & 27-31 & 17.0 & 0.000 \\
    & 28-32 & 4.6 & 0.000 & &  29-31 &  11.8 &  68.603 \\
    &  29-31 &  2.7 &  68.603 & &  28-30 &  10.7 &  71.627 \\
    &  28-30 &  2.5 &  71.627 & & 24-32 & 4.0 & 0.000 \\
    & 24-32 & 2.6 & 0.000 & & 27-34 & 4.0 & 0.000 \\
    & 27-34 & 2.1 & 0.000 & & 28-32 & 0.6 & 0.000 \\
    \hline
    \multicolumn{4}{c|}{ KS single excitations considered:} & \multicolumn{4}{c}{ KS single excitations considered:} \\
    \multicolumn{4}{c|}{28-30, 29-31} & \multicolumn{4}{c}{28-30, 29-31}  \\
    \hline\hline
    \end{tabular}
\end{table*}

\begin{table*}[h!]
\caption{Quantities used to inform the decision of which KS single excitations, $q$, to consider in DTDDFT and DTDA along with the KS double excitation, $D = 2\times$(29-30), for an appropriate description of the lowest A$_g$ excitation energy in octatetraene at the PBE0/cc-pVDZ level of theory. We report quantities for the first four A$_g$ states of octatetraene.}
\label{}
\centering
    \begin{tabular}{cccc|cccc}
    \hline\hline
    ATDDFT root & $q$ & $(|X_q|^2 - |Y_q|^2) (\%)$ & $|H_{qD}| / 10^{-3} (\text{a.u.})$ & ATDA root & $q$ & $|X_q|^2 (\%)$ & $|H_{qD}| / 10^{-3} (\text{a.u.})$\\
    \hline\hline
    2A$_g$ & 28-30 & 59.1 & 66.329 & 2A$_g$ & 28-30 & 55.0 & 66.329 \\
    & 29-31 & 39.9 & 63.259 & & 29-31 & 43.9 & 63.259 \\
    & 28-33 & 0.5 & 0.000 & & 28-33 & 0.5 & 0.000 \\
    & 27-31 & 0.4 & 0.000 & & 27-31 & 0.4 & 0.000 \\
    \hline
    3A$_g$ & 29-31 & 57.9 & 63.259 & 3A$_g$ & 28-30 & 39.1 & 66.329 \\
    & 28-30 & 38.9 & 66.329 & & 29-31 & 49.8 & 63.259 \\
    & & & & & 29-34 & 2.6 & 7.459 \\
    & & & & & 27-31 & 1.3 & 0.000 \\
    & & & & & 23-38 & 0.5 & 0.000 \\
    & & & & & 25-33 & 0.4 & 0.000 \\
    & & & & & 28-33 & 0.3 & 0.000 \\
    & & & & & 27-34 & 0.3 & 0.000 \\
    \hline
    4A$_g$ & 29-34 & 53.1 & 7.459 & 4A$_g$ & 29-34 & 52.0 & 7.459 \\
    & 25-30 & 45.8 & 6.098 & & 25-30 & 45.4 & 6.098 \\
    & 28-33 & 0.8 & 0.000 & & 28-33 & 0.9 & 0.000 \\
    & & & & & 28-30 & 0.7 & 66.329 \\
    & & & & & 29-31 & 0.5 & 63.259 \\
    \hline
    5A$_g$ & 25-30 & 50.3 &6.098  & 5A$_g$ & 25-30 & 44.5 & 6.098 \\
    & 29-34 & 44.0 & & & 29-34 & 38.6 & 7.459 \\
    & 27-31 & 2.7 & 0.000 & & 27-31 & 7.2 & 0.000 \\
    & 28-33 & 1.7 & 0.000 & & 28-33 & 4.3 & 0.000 \\
    & & & & & 29-31 & 0.8 & 63.259 \\
    & & & & & 28-30 & 0.7 & 66.329 \\
    & & & & & 27-34 & 0.6 & 0.000 \\
    & & & & & 25-33 & 0.5 & 0.000 \\
    \hline
    \multicolumn{4}{c|}{ KS single excitations considered:} & \multicolumn{4}{c}{ KS single excitations considered:} \\
    \multicolumn{4}{c|}{28-3, 29-31} & \multicolumn{4}{c}{28-30, 29-31}  \\
    \hline\hline
    \end{tabular}
\end{table*}

\begin{table*}[h!]
\caption{Quantities used to inform the decision of which KS single excitations, $q$, to consider in DTDDFT and DTDA along with the KS double excitation, $D = 2\times$(29-30), for an appropriate description of the lowest A$_g$ excitation energy in octatetraene at the CAM-B3LYP/cc-pVDZ level of theory. We report quantities for the first four A$_g$ states of octatetraene.}
\label{}
\centering
    \begin{tabular}{cccc|cccc}
    \hline\hline
    ATDDFT root & $q$ & $(|X_q|^2 - |Y_q|^2) (\%)$ & $|H_{qD}| / 10^{-3} (\text{a.u.})$ & ATDA root & $q$ & $|X_q|^2 (\%)$ & $|H_{qD}| / 10^{-3} (\text{a.u.})$\\
    \hline\hline
    2A$_g$ & 28-30 & 65.4 & 61.967 & 2A$_g$ & 28-30 & 55.1 & 1.967 \\
    & 29-31 & 30.0 & 58.859 & & 29-31 & 40.0 & 58.859 \\
    & 27-31 & 2.0 & 0.000 & & 28-33 & 2.0 & 0.000 \\
    & 28-33 & 1.8 & 0.000 & & 27-31 & 1.9 & 0.000 \\
    & 27-37 & 0.3 & 0.000 & & 27-37 & 0.4 & 0.000 \\
    \hline
    3A$_g$ & 29-31 & 66.0 & 58.859 & 3A$_g$ & 29-31 & 53.7 & 58.859 \\
    & 28-30 & 31.0 & 1.967 & & 28-30 & 39.0 & 1.967 \\
    & 28-33 & 0.4 & 0.000 & & 23-38 & 0.5 & 0.000 \\
    & & & & & 23-38 & 0.5 & 0.000 \\
    & & & & & 24-33 & 0.3 & 0.000 \\
    & & & & & 21-40 & 0.3 & 0.000 \\
    \hline
    4A$_g$ & 29-37 & 63.9 & & 4A$_g$ & 29-37 & 61.2 & \\
    & 24-30 & 33.1 & 6.530 & & 24-30 & 35.5 & 6.530 \\
    & 28-33 & 2.0 & 0.000 & & 28-33 & 1.9 & 0.000 \\
    & 27-31 & 0.6 & 0.000 & & 27-31 & 0.7 & 0.000 \\
    \hline
    5A$_g$ & 24-30 & 61.0 & 6.530 & 5A$_g$ & 24-30 & 58.9 & 6.530 \\
    & 29-37 & 30.8 & 7.689 & & 29-37 & 34.4 & 7.689 \\
    & 27-31 & 3.3 & 0.000 & & 21-32 & 0.5 & 0.000 \\
    & 28-33 & 1.6 & 0.000 & & 19-38 & 0.4 & 0.000 \\
    & 21-32 & 0.3 & 0.000 & & & & \\
    & 19-38 & 0.3 & 0.000 & & & & \\
    \hline
    \multicolumn{4}{c|}{ KS single excitations considered:} & \multicolumn{4}{c}{ KS single excitations considered:} \\
    \multicolumn{4}{c|}{28-30, 29-31} & \multicolumn{4}{c}{28-30, 29-31}  \\
    \hline\hline
    \end{tabular}
\end{table*}

\clearpage
\subsection{Cyclopentadiene}

\begin{table*}[h!]
\caption{Quantities used to inform the decision of which KS single excitations, $q$, to consider in DTDDFT and DTDA along with the KS double excitation, $D = 2\times$(18-19), for an appropriate description of the lowest A$_1$ excitation energy in cyclopentadiene at the PBE/cc-pVDZ level of theory. We report quantities for the first four A$_1$ states of cyclopentadiene.}
\label{cyclopentadiene_pbe}
\centering
    \begin{tabular}{cccc|cccc}
    \hline\hline
    ATDDFT root & $q$ & $(|X_q|^2 - |Y_q|^2) (\%)$ & $|H_{qD}| / 10^{-3} (\text{a.u.})$ & ATDA root & $q$ & $|X_q|^2 (\%)$ & $|H_{qD}| / 10^{-3} (\text{a.u.})$\\
    \hline\hline
    2A$_1$ &  17-19 &  57.1 &  46.885 & 2A$_1$ &  17-19 &  54.7 &  46.885 \\
    &  18-20 &  42.6 &  55.475 & &  18-20 &  44.7 &  55.475 \\
    & & & & &  12-19 &  0.3 &  51.242 \\
    \hline
    3A$_1$ &  18-20 &  50.3 &  55.475 & 3A$_1$ &  18-20 &  38.2 &  55.475 \\
    &  17-19 &  38.2 &  46.885 & &  17-19 &  33.0 &  46.885 \\
    & 12-19 & 4.9 & 51.242 & &  12-19 &  15.5 &  51.242 \\
    & 15-27 & 1.1 & 0.000 & & 16-21 & 2.7 & 0.000 \\
\hline\hline
    \end{tabular}
\end{table*}

\begin{table*}[h!]
\caption{Quantities used to inform the decision of which KS single excitations, $q$, to consider in DTDDFT and DTDA along with the KS double excitation, $D = 2\times$(18-19), for an appropriate description of the lowest A$_1$ excitation energy in cyclopentadiene at the PBE0/cc-pVDZ level of theory. We report quantities for the first four A$_1$ states of cyclopentadiene.}
\label{cyclopentadiene_pbe0}
\centering
    \begin{tabular}{cccc|cccc}
    \hline\hline
    ATDDFT root & $q$ & $(|X_q|^2 - |Y_q|^2) (\%)$ & $|H_{qD}| / 10^{-3} (\text{a.u.})$ & ATDA root & $q$ & $|X_q|^2 (\%)$ & $|H_{qD}| / 10^{-3} (\text{a.u.})$\\
    \hline\hline
    2A$_1$ &  17-19 &  59.2 &  45.340 & 2A$_1$ &  17-19 &  56.2 &  45.340 \\
    &  18-21 &  40.6 &  52.479 & &  18-21 &  43.4 &  52.479 \\
    \hline
    3A$_1$ &  18-21 &  56.0 &  52.479 & 3A$_1$ &  18-21 &  48.1 &  52.479 \\
    &  17-19 &  38.0 &  45.340 & &  17-19 &  36.8 &  45.340 \\
    & 14-28 & 1.0 & 0.000 & & 14-28 & 1.8 & 0.000 \\
    & 15-27 & 0.9 & 0.000 & & 12-19 & 3.8 & 48.553 \\
    & 12-19 & 0.8 & 48.553 & & 15-27 & 1.7 & 0.000 \\
    & 15-23 & 0.3 & 0.000 & & 15-23 & 0.7 & 0.000 \\
    & 16-20 & 0.3 & 0.000 & & 16-20 & 0.6 & 0.000 \\
    & & & & & 14-29 & 0.4 & 0.000 \\
    & & & & & 10-26 & 0.3 & 0.000 \\
    & & & & & 13-30 & 0.3 & 0.000 \\
    \hline
    4A$_1$ & 12-19 & 95.6 & 48.553 & 4A$_1$ & 12-19 & 88.6 & 48.553 \\
    & 17-25 & 1.5 & 0.000 & & 17-25 & 5.0 & 0.000 \\
    &  18-21 &  0.6 &  52.479 & &   18-21 &  1.8 &  52.479 \\
    & 15-27 & 0.5 & 0.000 & &  17-19 &  0.7 &  45.340 \\
    & & & & & 14-28 & 0.6 & 0.000 \\
    & & & & & 15-27 & 0.5 & 0.000 \\
    & & & & & 15-23 & 0.3 & 0.000 \\
    \hline
    5A$_1$ & 16-20 & 89.6 & 0.000 & 5A$_1$ & 16-20 & 88.3 & 0.000 \\
    & 17-25 & 6.2 & 0.000 & & 17-25 & 6.7 & 0.000 \\
    & 14-20 & 1.7 & 0.000 & & 14-20 & 1.8 & 0.000 \\
    & 15-23 & 0.9 & 0.000 & & 15-23 & 1.1 & 0.000 \\ 
    & 14-22 & 0.6 & 0.000 & & 14-22 & 0.6 & 0.000 \\
    & & & & &  18-21 &  0.3 &  52.479 \\ 
    \hline
    \multicolumn{4}{c|}{Two KS single excitations considered:} & \multicolumn{4}{c}{Two KS single excitations considered:} \\
    \multicolumn{4}{c|}{17-19, 18-21} & \multicolumn{4}{c}{17-19, 18-21} \\
    \hline\hline
    \end{tabular}
\end{table*}

\begin{table*}[h!]
\caption{Quantities used to inform the decision of which KS single excitations, $q$, to consider in DTDDFT and DTDA along with the KS double excitation, $D = 2\times$(18-19), for an appropriate description of the lowest A$_1$ excitation energy in cyclopentadiene at the  CAM-B3LYP/cc-pVDZ level of theory. We report quantities for the first four A$_1$ states of cyclopentadiene.}
\label{cyclopentadiene_cam-b3lyp}
\centering
    \begin{tabular}{cccc|cccc}
    \hline\hline
    ATDDFT root & $q$ & $(|X_q|^2 - |Y_q|^2) (\%)$ & $|H_{qD}| / 10^{-3} (\text{a.u.})$ & ATDA root & $q$ & $|X_q|^2 (\%)$ & $|H_{qD}| / 10^{-3} (\text{a.u.})$\\
    \hline\hline
    2A$_1$ &  17-19 &  61.5 &  44.504 & 2A$_1$ &  17-19 &  57.8 &  44.504 \\
    &  18-21 &  38.2 &  51.459 & &  18-21 &  41.8 &  51.459 \\
    \hline
    3A$_1$ &  18-21 &  59.1 &  51.459 & 3A$_1$ &  18-21 &  51.3 &  51.459 \\
    &  17-19 &  35.9 &  44.504 & &  17-19 &  36.2 &  44.504 \\
    & 14-28 & 1.0 & 0.000 & & 14-28 & 1.8 & 0.000 \\
    & 16-27 & 0.7 & 0.000 & & 12-19 & 1.6 & 47.609 \\
    & 14-29 & 0.3 & 0.000 & & 16-27 & 1.4 & 0.000 \\
\hline\hline
    \end{tabular}
\end{table*}

\clearpage
\subsection{Acrolein}

\begin{table*}[h!]
\caption{Quantities used to inform the decision of which KS single excitations, $q$, to consider in DTDDFT and DTDA along with the KS double excitation, $D = 2\times$(15-16), for an appropriate description of the lowest three A' excitation energies in acrolein at the PBE/aug-cc-pVDZ level of theory. We report quantities for the first four A' states of acrolein.}
\label{acrolein_pbe}
\centering
    \begin{tabular}{cccc|cccc}
    \hline\hline
    ATDDFT root & $q$ & $(|X_q|^2 - |Y_q|^2) (\%)$ & $|H_{qD}| / 10^{-3} (\text{a.u.})$ & ATDA root & $q$ & $|X_q|^2 (\%)$ & $|H_{qD}| / 10^{-3} (\text{a.u.})$\\
    \hline\hline
    2A' &  15-17 &  96.0 &  0.833 & 2A' &  15-17 &  98.0 &  0.833 \\
    &  14-16 &  3.8 &  10.576 & &  14-16 &  1.7 &  10.576 \\
    \hline
    3A' &  15-19 &  87.6 &  0.902 & 3A'&  15-19 &  97.4 &  0.902 \\
    &  14-16 &  10.6 &  10.576 & &  14-16 &  2.0 &  10.576 \\
    &  15-17 &  0.9 &  0.833 & & & & \\
    &  15-20 &  0.4 &  0.600 & & & & \\
    \hline
    4A' &  14-16 &  66.3 &  10.576 & 4A' &  15-20 &  91.2 &  0.600 \\
\hline\hline
    \end{tabular}
\end{table*}

\begin{table*}[h!]
\caption{Quantities used to inform the decision of which KS single excitations, $q$, to consider in DTDDFT and DTDA along with the KS double excitation, $D = 2\times$(15-16), for an appropriate description of the lowest three A' excitation energies in acrolein at the PBE0/aug-cc-pVDZ level of theory. We report quantities for the first four A' states of acrolein.}
\label{acrolein_pbe0}
\centering
    \begin{tabular}{cccc|cccc}
    \hline\hline
    ATDDFT root & $q$ & $(|X_q|^2 - |Y_q|^2) (\%)$ & $|H_{qD}| / 10^{-3} (\text{a.u.})$ & ATDA root & $q$ & $|X_q|^2 (\%)$ & $|H_{qD}| / 10^{-3} (\text{a.u.})$\\
    \hline\hline
    2A' &  14-16 &  85.5 &  9.778 & 2A' &  15-17 &  43.9 &  0.665 \\
    &  15-17 &  6.8 &  0.665 & &  14-16 &  40.2 &  9.778 \\
    &  15-18 &  1.8 &  0.734 & &  15-18 &  6.4 &  0.734 \\
    & 15-20 & 0.9 & 0.515 & & 15-20 & 3.9 & 0.515 \\
    & 14-19 & 0.8 & 0.000 & & 14-19 & 0.9 & 0.000 \\
    & 14-22 & 0.7 & 0.000 & & 14-22 & 0.8 & 0.000 \\
    & 15-23 & 0.6 & 0.506 & & 15-29 & 0.8 & 0.855 \\
\hline\hline
    \end{tabular}
\end{table*}

\begin{table*}[h!]
\caption{Quantities used to inform the decision of which KS single excitations, $q$, to consider in DTDDFT and DTDA along with the KS double excitation, $D = 2\times$(15-16), for an appropriate description of the lowest three A' excitation energies in acrolein at the CAM-B3LYP/aug-cc-pVDZ level of theory. We report quantities for the first four A' states of acrolein.}
\label{acrolein_cam-b3lyp}
\centering
    \begin{tabular}{cccc|cccc}
    \hline\hline
    ATDDFT root & $q$ & $(|X_q|^2 - |Y_q|^2) (\%)$ & $|H_{qD}| / 10^{-3} (\text{a.u.})$ & ATDA root & $q$ & $|X_q|^2 (\%)$ & $|H_{qD}| / 10^{-3} (\text{a.u.})$\\
    \hline\hline
    2A' &  14-16 &  91.6 &  9.503 & 2A' &  14-16 &  70.6 &  9.503 \\
    &  15-17 &  2.2 &  0.532 & &  15-17 &  11.5 &  0.532 \\
    &  15-18 &  1.0 &  0.575 & &  15-18 &  4.3 &  0.575 \\
    &  15-19 &  0.7 &  0.421 & &  15-19 &  4.1 &  0.421 \\
    & 14-23 & 0.7 & 0.000 & & 15-29 & 2.0 & 1.037 \\
    & 15-29 & 0.6 & 1.037 & & 14-23 & 1.5 & 0.000 \\
    & 15-22 & 0.4 & 0.410 & & 14-20 & 0.6 & 0.000 \\
\hline\hline
    \end{tabular}
\end{table*}

\clearpage
\subsection{Naphthalene}

\begin{table*}[h!]
\caption{Quantities used to inform the decision of which KS single excitations, $q$, to consider in DTDDFT and DTDA along with the KS double excitation, $D = 2\times$(34-35), for an appropriate description of the lowest two A$_g$ excitation energies in naphthalene at the PBE/cc-pVDZ level of theory. We report quantities for the first four A$_g$ states of naphthalene.}
\label{}
\centering
    \begin{tabular}{cccc|cccc}
    \hline\hline
    ATDDFT root & $q$ & $(|X_q|^2 - |Y_q|^2) (\%)$ & $|H_{qD}| / 10^{-3} (\text{a.u.})$ & ATDA root & $q$ & $|X_q|^2 (\%)$ & $|H_{qD}| / 10^{-3} (\text{a.u.})$\\
    \hline\hline
    2A$_g$ & 33-37 & 53.6 & 0.000 & 2A$_g$ & 33-37 & 53.1 & 0.000 \\
    & 32-36 & 45.7 & 0.000 & & 32-36 & 46.0 & 0.000 \\
    & 29-35 & 0.3 & 48.192 & & 29-35 & 0.4 & 48.192 \\
    & & & & & 34-39 & 0.3 & \\
    \hline
    3A$_g$ & 29-35 & 64.8 & 48.192 & 3A$_g$ & 29-35 & 68.6 & 48.192 \\
    & 32-36 & 20.2 & 0.000 & & 32-36 & 16.5 & 0.000 \\
    & 33-37 & 11.7 & 0.000 & & 33-37 & 8.3 & 0.000 \\
\hline\hline
    \end{tabular}
\end{table*}

\begin{table*}[h!]
\caption{Quantities used to inform the decision of which KS single excitations, $q$, to consider in DTDDFT and DTDA along with the KS double excitation, $D = 2\times$(34-35), for an appropriate description of the lowest two A$_g$ excitation energies in naphthalene at the PBE0/cc-pVDZ level of theory. We report quantities for the first four A$_g$ states of naphthalene.}
\label{}
\centering
    \begin{tabular}{cccc|cccc}
    \hline\hline
    ATDDFT root & $q$ & $(|X_q|^2 - |Y_q|^2) (\%)$ & $|H_{qD}| / 10^{-3} (\text{a.u.})$ & ATDA root & $q$ & $|X_q|^2 (\%)$ & $|H_{qD}| / 10^{-3} (\text{a.u.})$\\
    \hline\hline
    2A$_g$ & 33-37 & 49.8 & 0.000 & 2A$_g$ & 33-37 & 48.9 & 0.000 \\
    & 32-36 & 45.6 & 0.000 & & 32-36 & 46.6 & 0.000 \\
    & 30-35 & 3.2 & 47.6 & & 30-35 & 3.0 & 47.6 \\
    & 34-41 & 1.3 & & & 34-41 & 1.3 & \\
    \hline
    3A$_g$ & 30-35 & 47.7 & 47.6 & 3A$_g$ & 30-35 & 59.9 & 47.6 \\
    & 33-37 & 34.0 & 0.000 & & 33-37 & 27.3 & 0.000 \\
    & 32-36 & 17.7 & 0.000 & & 32-36 & 10.3 & 0.000 \\
    & & & & & 34-41 & 1.4 & 47.1 \\
    & & & & & 32-47 & 0.3 & 0.000 \\
    \hline06
    4A$_g$ & 34-41 & 35.6 & 47.1 & 4A$_g$ & 34-41 & 48.3 & 47.1 \\
    & 30-35 & 28.1 & 47.6 & & 32-36 & 28.0 & 0.000 \\
    & 32-36 & 27.9 & 0.000 & & 30-35 & 11.7 & 47.6 \\
    & 33-37 & 7.7 & 0.000 & & 33-37 & 10.3 & 0.000 \\
    & 27-37 & 0.4 & 0.000 & & 27-37 & 0.7 & 0.000 \\
    \hline
    5A$_g$ & 34-41 & 61.4 & 47.1 & 5A$_g$ & 34-41 & 45.5 & 47.1 \\
    & 30-35 & 19.7 & 47.6 & & 30-35 & 22.4 & 47.6 \\
    & 32-36 & 7.5 & 0.000 & & 32-36 & 11.9 & 0.000 \\
    & 33-37 & 7.5 & 0.000 & & 33-37 & 10.6 & 0.000 \\
    & 26-51 & 0.3 & 0.000 & & 31-50 & 1.1 & 0.000 \\
    & 31-50 & 0.5 & 0.000 & & 26-51 & 0.7 & 0.000 \\
    & 21-49 & 0.3 & 0.000 & & 21-49 & 0.6 & 0.000 \\
    & & & & & 22-48 & 0.5 & 0.000 \\
    & & & & & 25-53 & 0.5 & 0.000 \\
    & & & & & 22-40 & 0.3 & 0.000 \\
    & & & & & 23-42 & 0.3 & 0.000 \\
    & & & & & 29-52 & 0.3 & 0.000 \\
    & & & & & 20-43 & 0.3 & 0.000 \\
    \hline
    \multicolumn{4}{c|}{ KS single excitations considered:} & \multicolumn{4}{c}{ KS single excitations considered:} \\
    \multicolumn{4}{c|}{ 30-35, 34-41 (33-37, 32-36)} & \multicolumn{4}{c}{30-35, 34-41 (33-37, 32-36)}  \\
    \hline\hline
    \end{tabular}
\end{table*}

\begin{table*}[h!]
\caption{Quantities used to inform the decision of which KS single excitations, $q$, to consider in DTDDFT and DTDA along with the KS double excitation, $D = 2\times$(34-35), for an appropriate description of the lowest two A$_g$ excitation energies in naphthalene at the CAM-B3LYP/cc-pVDZ level of theory. We report quantities for the first four A$_g$ states of naphthalene.}
\label{}
\centering
    \begin{tabular}{cccc|cccc}
    \hline\hline
    ATDDFT root & $q$ & $(|X_q|^2 - |Y_q|^2) (\%)$ & $|H_{qD}| / 10^{-3} (\text{a.u.})$ & ATDA root & $q$ & $|X_q|^2 (\%)$ & $|H_{qD}| / 10^{-3} (\text{a.u.})$\\
    \hline\hline
    2A$_g$ & 33 37 & 49.7 & 0.000 & 2A$_g$ & 33-37 & 48.1 & 0.000 \\
    & 32-36 & 41.0 & 0.000 & & 32-36 & 42.6 & 0.000 \\
    & 30-35 & 6.1 & 47.7 & & 30-35 & 5.8 & 47.7 \\
    & 34-41 & 3.0 & 47.2 & & 34-41 & 3.1 & 47.2 \\ 
    \hline
    3A$_g$ & 33-37 & 46.6 & 0.000 & 3A$_g$ & 30-35 & 42.7 & 47.7 \\
    & 32-36 & 35.5 & 0.000 & & 33-37 & 38.5 & 0.000 \\
    & 30-35 & 17.1 & 47.7 & & 32-36 & 17.1 & 0.000 \\
    \hline
    4A$_g$ & 30-35 & 65.9 & 47.7 & 4A$_g$ & 34-41 & 34.0 & 47.2 \\
    & 32-36 & 22.4 & 0.000 & & 32-36 & 31.5 & 0.000 \\
    & 34-41 & 9.3 & 47.2 & & 30-35 & 27.8 & 47.7 \\
    & 27-37 & 1.0 & 0.000 & & 33-37 & 3.9 & 0.000 \\
    & 33-37 & 0.5 & 0.000 & & 27-37 & 1.0 & 0.000 \\
    & & & & & 32-47 & 0.3 & 0.000 \\
    \hline
    5A$_g$ & 34-41 & 85.9 & 47.2 & 5A$_g$ & 34-41 & 58.9 & 47.2 \\
    & 30-35 & 8.8 & 47.7 & & 30-35 & 20.0 & 47.7 \\
    & 33-37 & 2.5 & 0.000 & & 33-37 & 6.6 & 0.000 \\
    & 32-47 & 0.5 & 0.000 & & 32-36 & 5.5 & 0.000 \\
    & 31-50 & 0.3 & 0.000 & & 31-50 & 0.8 & 0.000 \\
    & & & & & 26-51 & 0.6 & 0.000 \\
    & & & & & 21-49 & 0.5 & 0.000 \\
    & & & & & 25-53 & 0.5 & 0.000 \\
    & & & & & 22-48 & 0.4 & 0.000 \\
    & & & & & 22-40 & 0.3 & 0.000 \\
    & & & & & 20-43 & 0.3 & 0.000 \\
    & & & & & 32-47 & 0.3 & 0.000 \\
    & & & & & 23-42 & 0.3 & 0.000 \\
    \hline
    \multicolumn{4}{c|}{ KS single excitations considered:} & \multicolumn{4}{c}{ KS single excitations considered:} \\
    \multicolumn{4}{c|}{ 30-35, 34-41 (33-37, 32-36)} & \multicolumn{4}{c}{30-35, 34-41 (33-37, 32-36)}  \\
    \hline\hline
    \end{tabular}
\end{table*}

\clearpage
\subsection{Ethylene}

\begin{table*}[h!]
\caption{Quantities used to inform the decision of which KS single excitations, $q$, to consider in DTDDFT and DTDA along with the KS double excitation, $D = 2\times$(8-9), for an appropriate description of the lowest A$_g$ excitation energy in ethylene at the PBE/aug-cc-pVDZ level of theory. We report quantities for the first four A$_g$ states of ethylene.}
\label{}
\centering
    \begin{tabular}{cccc|cccc}
    \hline\hline
    ATDDFT root & $q$ & $(|X_q|^2 - |Y_q|^2) (\%)$ & $|H_{qD}| / 10^{-3} (\text{a.u.})$ & ATDA root & $q$ & $|X_q|^2 (\%)$ & $|H_{qD}| / 10^{-3} (\text{a.u.})$\\
    \hline\hline
    2A$_g$ & 8-14 & 97.6 & 19.978 & 2A$_g$ & 8-14 & 96.6 & 19.978 \\
    & 7-13 & 0.6 & 0.000 & & 7-13 & 0.8 & 0.000 \\
    & 6-10 & 0.5 & 0.000 & & 6-10 & 0.7 & 0.000 \\
    & 5-12 & 0.3 & 0.000 & & 5-12 & 0.4 & 0.000 \\
    & & & & & 7-21 & 0.3 & 0.000 \\
    \hline
    3A$_g$ & 7-13 & 87.2 & 0.000 & 3A$_g$ & 7-13 & 85.2 & 0.000 \\
    & 6-10 & 12.4 & 0.000 & & 6-10 & 14.3 & 0.000 \\
    \hline
    4A$_g$ & 6-10 & 86.1 & 0.000 & 4A$_g$ & 6-10 & 83.6 & 0.000 \\
    & 7-13 & 11.6 & 0.000 & & 7-13 & 13.1 & 0.000 \\
    & 8-14 & 0.7 & 19.978 & & 8-14 & 0.9 & 19.978 \\
    & 7-21 & 0.5 & 0.000 & & 7-21 & 0.6 & 0.000 \\
    & 5-12 & 0.4 & 0.000 & & 5-12 & 0.5 & 0.000 \\
    \hline
    5A$_g$ & 5-12 & 62.6 & 0.000 & 5A$_g$ & 5-12 & 60.2 & 0.000 \\
    & 6-15 & 35.5 & 0.000 & & 6-15 & 37.6 & 0.000 \\
    & 7-21 & 1.3 & 0.000 & & 7-21 & 1.5 & 0.000 \\
    \hline
    \multicolumn{4}{c|}{ KS single excitations considered:} & \multicolumn{4}{c}{ KS single excitations considered:} \\
    \multicolumn{4}{c|}{8-14} & \multicolumn{4}{c}{8-14}  \\
    \hline\hline
    \end{tabular}
\end{table*}

\begin{table*}[h!]
\caption{Quantities used to inform the decision of which KS single excitations, $q$, to consider in DTDDFT and DTDA along with the KS double excitation, $D = 2\times$(8-9), for an appropriate description of the lowest A$_g$ excitation energy in ethylene at the PBE0/aug-cc-pVDZ level of theory. We report quantities for the first four A$_g$ states of ethylene.}
\label{}
\centering
    \begin{tabular}{cccc|cccc}
    \hline\hline
    ATDDFT root & $q$ & $(|X_q|^2 - |Y_q|^2) (\%)$ & $|H_{qD}| / 10^{-3} (\text{a.u.})$ & ATDA root & $q$ & $|X_q|^2 (\%)$ & $|H_{qD}| / 10^{-3} (\text{a.u.})$\\
    \hline\hline
    2A$_g$ & 8-14 & 98.1 & 12.584 & 2A$_g$ & 8-14 & 97.3 & 12.584 \\
    & 7-13 & 0.4 & 0.000 & & 7-13 & 0.5 & 0.000 \\
    & 6-10 & 0.3 & 0.000 & & 6-10 & 0.5 & 0.000 \\
    & 7-21 & 0.2 & 0.000 & & 7-21 & 0.4 & 0.000 \\
    & & & & & 5-12 & 0.3 & 0.000 \\
    \hline
    3A$_g$ & 7-13 & 93.7 & 0.000 & 3A$_g$ & 7-13 & 91.7 & 0.000 \\
    & 6-10 & 5.4 & 0.000 & & 6-10 & 7.3 & 0.000 \\
    & 7-21 & 0.3 & 0.000 & & 7-21 & 0.3 & 0.000 \\
    \hline
    4A$_g$ & 6-10 & 93.0 & 0.000 & 4A$_g$ & 6-10 & 90.8 & 0.000 \\
    & 7-13 & 5.4 & 0.000 & & 7-13 & 7.2 & 0.000 \\
    & 8-14 & 0.4 & 12.584 & & 8-14 & 0.5 & 12.584 \\
    & 6-19 & 0.4 & 0.000 & & 6-15 & 0.3 & 0.000 \\
    & 6-15 & 0.3 & 0.000 & & 6-19 & 0.3 & 0.000 \\
    \hline
    5A$_g$ & 6-15 & 73.2 & & 5A$_g$ & 6-15 & 73.0 & 0.000 \\
    & 5-12 & 24.7 & 0.000 & & 5-12 & 25.0 & 0.000 \\
    & 7-21 & 0.7 & 0.000 & & 7-21 & 0.6 & 0.000 \\
    & 5-18 & 0.6 & 0.000 & & 5-18 & 0.6 & 0.000 \\
    \hline
    \multicolumn{4}{c|}{ KS single excitations considered:} & \multicolumn{4}{c}{ KS single excitations considered:} \\
    \multicolumn{4}{c|}{8-14} & \multicolumn{4}{c}{8-14}  \\
    \hline\hline
    \end{tabular}
\end{table*}

\begin{table*}[h!]
\caption{Quantities used to inform the decision of which KS single excitations, $q$, to consider in DTDDFT and DTDA along with the KS double excitation, $D = 2\times$(8-9), for an appropriate description of the lowest A$_g$ excitation energy in ethylene at the CAM-B3LYP/aug-cc-pVDZ level of theory. We report quantities for the first four A$_1$ states of borole.}
\label{}
\centering
    \begin{tabular}{cccc|cccc}
    \hline\hline
    ATDDFT root & $q$ & $(|X_q|^2 - |Y_q|^2) (\%)$ & $|H_{qD}| / 10^{-3} (\text{a.u.})$ & ATDA root & $q$ & $|X_q|^2 (\%)$ & $|H_{qD}| / 10^{-3} (\text{a.u.})$\\
    \hline\hline
    2A$_g$ & 8-14 & 97.8 & 5.486 & 2A$_g$ & 8-14 & 96.9 & 5.486 \\
    & 6-9 & 0.4 & 1.311 & & 6-9 & 0.5 & 1.311 \\
    & 7-13 & 0.3 & 0.000 & & 7-13 & 0.5 & 0.000 \\
    & 7-20 & 0.3 & 0.000 & & 7-20 & 0.5 & 0.000 \\
    & & & & & 5-11 & 0.3 & 0.000 \\
    \hline
    3A$_g$ & 7-13 & 79.4 & 0.000 & 3A$_g$ & 7-13 & 75.5 & 0.000 \\
    & 6-9 & 16.4 & 1.311 & & 6-9 & 20.2 & 1.311 \\
    & 7-20 & 2.3 & 0.000 & & 7-20 & 2.2 & 0.000 \\
    & 6-17 & 1.0 & 0.000 & & 6-17 & 1.2 & 0.000 \\
    & 5-11 & 0.4 & 0.000 & & 5-11 & 0.3 & 0.000 \\
    \hline
    4A$_g$ & 6-9 & 78.0 & 1.311 & 4A$_g$ & 6-9 & 74.0 & 1.311 \\
    & 7-13 & 17.1 & 0.000 & & 7-13 & 21.0 & 0.000 \\
    & 6-17 & 2.4 & 0.000 & & 6-17 & 2.1 & 0.000 \\
    & 6-15 & 0.9 & 0.000 & & 6-15 & 1.0 & 0.000 \\
    & 8-14 & 0.7 & 5.486 & & 8-14 & 0.8 & 5.486 \\
    \hline
    5A$_g$ & 6-15 & 78.0 & 0.000 & 5A$_g$ & 6-15 & 77.5 & 0.000 \\
    & 5-11 & 14.5 & 0.000 & & 5-11 & 15.1 & 0.000 \\
    & 5-18 & 1.9 & 0.000 & & 5-18 & 1.9 & 0.000 \\
    & 7-20 & 1.6 & 0.000 & & 7-20 & 1.6 & 0.000 \\
    & 6-26 & 0.9 & 0.000 & & 6-26 & 0.8 & 0.000 \\
    & 6-9 & 0.7 & 1.311 & & 6-9 & 0.7 & 1.311 \\
    & 4-10 & 0.6 & 0.000 & & 4-10 & 0.6 & 0.000 \\
    & 6-21 & 0.6 & 0.000 & & 6-21 & 0.5 & 0.000 \\
    & 7-13 & 0.5 & 0.000 & & 7-13 & 0.5 & 0.000 \\
    & 4-16 & 0.3 & 0.000 & & 4-16 & 0.3 & 0.000 \\
    \hline
    \multicolumn{4}{c|}{ KS single excitations considered:} & \multicolumn{4}{c}{ KS single excitations considered:} \\
    \multicolumn{4}{c|}{8-14} & \multicolumn{4}{c}{8-14}  \\
    \hline\hline
    \end{tabular}
\end{table*}

\clearpage
\subsection{Borole}

\begin{table*}[h!]
\caption{Quantities used to inform the decision of which KS single excitations, $q$, to consider in DTDDFT and DTDA along with the KS double excitation, $D = 2\times$(17-18), for an appropriate description of the lowest two A$_1$ excitation energies in borole at the PBE/cc-pVDZ level of theory. We report quantities for the first four A$_1$ states of borole.}
\label{borole_pbe}
\centering
    \begin{tabular}{cccc|cccc}
    \hline\hline
    ATDDFT root & $q$ & $(|X_q|^2 - |Y_q|^2) (\%)$ & $|H_{qD}| / 10^{-3} (\text{a.u.})$ & ATDA root & $q$ & $|X_q|^2 (\%)$ & $|H_{qD}| / 10^{-3} (\text{a.u.})$\\
    \hline\hline
    2A$_1$ &  14-18 &  77.0 &  9.937 & 2A$_1$ &  14-18 &  68.9 & 9.937 \\
    &  17-19 &  22.6 &  1.643 & &  17-19 &  30.2 &  0.827\\    
    \hline
    3A$_1$ &  17-19 &  67.5 &  1.643 & 3A$_1$ &  17-19 &  48.3 &  1.643 \\
    &  14-18 &  18.1 &  9.937 & &  14-20 &  21.2 &  0.000 \\
    & 14-20 & 8.9 & 0.000 & &  14-18 &  18.3 &  9.937 \\
    & 15-21 & 1.4 & 0.000 & & 15-21 & 4.0 & 0.000 \\
    & 16-26 & 0.8 & 0.000 & & 16-26 & 1.6 & 0.000 \\
\hline\hline
    \end{tabular}
\end{table*}

\begin{table*}[h!]
\caption{Quantities used to inform the decision of which KS single excitations, $q$, to consider in DTDDFT and DTDA along with the KS double excitation, $D = 2\times$(17-18), for an appropriate description of the lowest two A$_1$ excitation energies in borole at the PBE0/cc-pVDZ level of theory. We report quantities for the first four A$_1$ states of borole.}
\label{borole_pbe0}
\centering
    \begin{tabular}{cccc|cccc}
    \hline\hline
    ATDDFT root & $q$ & $(|X_q|^2 - |Y_q|^2) (\%)$ & $|H_{qD}| / 10^{-3} (\text{a.u.})$ & ATDA root & $q$ & $|X_q|^2 (\%)$ & $|H_{qD}| / 10^{-3} (\text{a.u.})$\\
    \hline\hline
    2A$_1$ &  14-18 &  82.6 &  10.444 & 2A$_1$ &  14-18 &  73.7 &  10.444 \\
    &  17-19 &  17.2 &  0.191 & &  17-19 &  25.3 &  0.191 \\ 
    \hline 
    3A$_1$ &  17-19 &  78.5 &  0.191 & 3A$_1$ &  17-19 &  63.3 &  0.191 \\
    &  14-18 &  14.8 &  10.444 & &  14-18 &  18.5 &  10.444 \\
    & 14-20 & 2.5 & 0.000 & & 14-20 & 8.6 & 0.000 \\
    & 16-28 & 0.6 & 0.000 & & 16-28 & 1.2 & 0.000 \\
    & 16-26 & 0.5 & 0.000 & & 16-26 & 1.1 & 0.000 \\
\hline\hline
    \end{tabular}
\end{table*}

\begin{table*}[h!]
\caption{Quantities used to inform the decision of which KS single excitations, $q$, to consider in DTDDFT and DTDA along with the KS double excitation, $D = 2\times$(17-18), for an appropriate description of the lowest two A$_1$ excitation energies in borole at the CAM-B3LYP/cc-pVDZ level of theory. We report quantities for the first four A$_1$ states of borole.}
\label{borole_cam-b3lyp}
\centering
    \begin{tabular}{cccc|cccc}
    \hline\hline
    ATDDFT root & $q$ & $(|X_q|^2 - |Y_q|^2) (\%)$ & $|H_{qD}| / 10^{-3} (\text{a.u.})$ & ATDA root & $q$ & $|X_q|^2 (\%)$ & $|H_{qD}| / 10^{-3} (\text{a.u.})$\\
    \hline\hline
    2A$_1$ &  14-18 &  86.4 &  9.225 & 2A$_1$ &  14-18 &  76.8 &  9.225 \\
    &  17-21 &  13.0 &  0.830 & &  17-21 &  22.0 &  0.830 \\
    \hline
    3A$_1$ &  17-21 &  83.9 &  0.830 & 3A$_1$ &  17-21 &  69.5 &  0.830 \\
    &  14-18 &  11.5 &  9.225 & &  14-18 &  16.9 &  9.225 \\
    & 14-20 & 0.7 & 0.000 & & 14-20 & 4.1 & 0.000 \\
    & 16-28 & 0.6 & 0.000 & & 16-28 & 1.4 & 0.000 \\
    & 16-26 & 0.3 & 0.000 & & 16-26 & 0.7 & 0.000 \\
\hline\hline
    \end{tabular}
\end{table*}

\clearpage
\subsection{Cyclopentadienone}

\begin{table*}[h!]
\caption{Quantities used to inform the decision of which KS single excitations, $q$, to consider in DTDDFT and DTDA along with the KS double excitation, $D = 2\times$(21-22), for an appropriate description of the lowest three A$_1$ excitation energies in cyclopentadienone at the PBE0/cc-pVDZ level of theory. We report quantities for the first four A$_1$ states of cyclopentadienone.}
\label{}
\centering
    \begin{tabular}{cccc|cccc}
    \hline\hline
    ATDDFT root & $q$ & $(|X_q|^2 - |Y_q|^2) (\%)$ & $|H_{qD}| / 10^{-3} (\text{a.u.})$ & ATDA root & $q$ & $|X_q|^2 (\%)$ & $|H_{qD}| / 10^{-3} (\text{a.u.})$\\
    \hline\hline
    2A$_1$ & 19-22 & 80.3 & 12.250 & 2A$_1$ & 19-22 & 67.0 & 12.250 \\
    & 21-23 & 16.6 & 10.372 & & 21-23 & 26.7 & 10.372 \\
    & 16-22 & 2.0 & 23.583 & & 16-22 & 4.2 & 23.583 \\
    & 20-30 & 0.6 & 0.000 & & 20-30 & 0.8 & 0.000 \\
    \hline
    3A$_1$ & 21-23 & 41.9 & 10.372 & 3A$_1$ & 16-22 & 59.8 & 23.583 \\
    & 16-22 & 40.8 & 23.583 & & 21-23 & 15.0 & 10.372 \\
    & 19-22 & 13.6 & 12.250 & & 19-22 & 14.8 & 12.250 \\
    & 20-30 & 1.8 & 0.000 & & 20-30 & 3.1 & 0.000 \\
    & 20-33 & 0.3 & 0.000 & & 19-25 & 2.6 & 0.000 \\
    & 16-25 & 0.3 & 0.000 & & 16-25 & 1.1 & 0.000 \\
    & & & & & 20-33 & 0.6 & 0.000 \\
    & & & & & 20-35 & 0.4 & 0.000 \\
    \hline
    4A$_1$ & 16-22 & 50.9 & 23.583 & 4A$_1$ & 21-23 & 46.6 & 10.372 \\
    & 21-23 & 37.2 & 10.372 & & 16-22 & 22.4 & 23.583 \\
    & 19-25 & 6.4 & 0.000 & & 19-25 & 13.4 & 0.000 \\
    & 19-22 & 2.7 & 12.250 & & 19-22 & 9.0 & 12.250 \\
    & 18-29 & 0.4 & 0.000 & & 20-30 & 2.2 & 0.000 \\
    & & & & & 18-29 & 0.9 & 0.000 \\
    & & & & & 18-31 & 0.4 & 0.000 \\
    & & & & & 15-30 & 0.3 & 0.000 \\
    & & & & & 20-26 & 0.3 & 0.000 \\
    \hline
    5A$_1$ & 20-26 & 87.5 & 0.000 & 5A$_1$ & 20-26 & 87.4 & 0.000 \\
    & 20-28 & 12.2 & 0.000 & & 20-28 & 12.0 & 0.000 \\
    \hline
    \multicolumn{4}{c|}{ KS single excitations considered:} & \multicolumn{4}{c}{ KS single excitations considered:} \\
    \multicolumn{4}{c|}{19-22, 21-23, 16-22} & \multicolumn{4}{c}{19-22, 21-23, 16-22}  \\
    \hline\hline
    \end{tabular}
\end{table*}

\begin{table*}[h!]
\caption{Quantities used to inform the decision of which KS single excitations, $q$, to consider in DTDDFT and DTDA along with the KS double excitation, $D = 2\times$(21-22), for an appropriate description of the lowest three A$_1$ excitation energies in cyclopentadienone at the CAM-B3LYP/cc-pVDZ level of theory. We report quantities for the first four A$_1$ states of cyclopentadienone.}
\label{}
\centering
    \begin{tabular}{cccc|cccc}
    \hline\hline
    ATDDFT root & $q$ & $(|X_q|^2 - |Y_q|^2) (\%)$ & $|H_{qD}| / 10^{-3} (\text{a.u.})$ & ATDA root & $q$ & $|X_q|^2 (\%)$ & $|H_{qD}| / 10^{-3} (\text{a.u.})$\\
    \hline\hline
    2A$_1$ & 19-22 & 84.7 & 13.047 & 2A$_1$ & 19-22 & 71.6 & 13.047 \\
    & 21-23 & 12.7 & 11.332 & & 21-23 & 22.9 & 11.332 \\
    & 16-22 & 1.4 & 23.753 & & 16-22 & 3.1 & 23.753 \\
    & 20-30 & 0.6 & 0.000 & & 20-30 & 0.8 & 0.000 \\
    & 21-23 & 0.3 & 0.000 & & & & \\
    \hline
    3A$_1$ & 21-23 & 64.8 & 11.332 & 3A$_1$ & 16-22 & 41.1 & 23.753 \\
    & 16-22 & 19.7 & 23.753 & & 21-23 & 33.4 & 11.332 \\
    & 19-22 & 11.8 & 13.047 & & 19-22 & 16.9 & 13.047 \\
    & 20-30 & 1.4 & 0.000 & & 20-30 & 3.1 & 0.000 \\
    & 20-33 & 0.3 & 0.000 & & 20-33 & 0.7 & 0.000 \\
    & & & & & 20-35 & 0.4 & 0.000 \\
    & & & & & 7-45 & 0.3 & 0.000 \\
    & & & & & 16-25 & 0.3 & 0.000 \\
    & & & & & 19-25 & 0.3 & 0.000 \\
    \hline
    4A$_1$ & 16-22 & 75.3 & 23.753 & 4A$_1$ & 16-22 & 45.6 & 23.753 \\
    & 21-23 & 19.4 & 11.332 & & 21-23 & 34.7 & 11.332 \\
    & 19-25 & 2.4 & 0.000 & & 19-25 & 9.4 & 0.000 \\
    & 19-22 & 0.4 & 13.047 & & 19-22 & 3.5 & 13.047 \\
    & 18-29 & 0.3 & 0.000 & & 18-29 & 0.8 & 0.000 \\
    & & & & & 20-30 & 0.6 & 0.000 \\
    & & & & & 15-30 & 0.4 & 0.000 \\
    & & & & & 18-31 & 0.3 & 0.000 \\
    & & & & & 14-30 & 0.3 & 0.000 \\
    & & & & & 12-27 & 0.3 & 0.000 \\
    \hline
    5A$_1$ & 20-26 & 67.6 & 0.000 & 5A$_1$ & 20-26 & 67.5 & 0.000 \\
    & 20-28 & 31.1 & 0.000 & & 20-28 & 30.9 & 0.000 \\
    & & & & & 20-30 & 0.3 & 0.000 \\
    \hline
    \multicolumn{4}{c|}{ KS single excitations considered:} & \multicolumn{4}{c}{ KS single excitations considered:} \\
    \multicolumn{4}{c|}{19-22, 21-23, 16-22} & \multicolumn{4}{c}{19-22, 21-23, 16-22}  \\
    \hline\hline
    \end{tabular}
\end{table*}

\begin{table*}[h!]
\caption{Quantities used to inform the decision of which KS single excitations, $q$, to consider in DTDDFT and DTDA along with the KS double excitation, $D$ = (21-22)+(20-22), for an appropriate description of the lowest B$_1$ excitation energy in cyclopentadienone at the PBE0/cc-pVDZ level of theory. We report quantities for the first four B$_1$ states of cyclopentadienone.}
\label{}
\centering
    \begin{tabular}{cccc|cccc}
    \hline\hline
    ATDDFT root & $q$ & $(|X_q|^2 - |Y_q|^2) (\%)$ & $|H_{qD}| / 10^{-3} (\text{a.u.})$ & ATDA root & $q$ & $|X_q|^2 (\%)$ & $|H_{qD}| / 10^{-3} (\text{a.u.})$\\
    \hline\hline
    2B$_1$ & 18-22 & 99.4 & 3.521 & 2B$_1$ & 18-22 & 99.3 & 3.521 \\
    \hline
    3B$_1$ & 17-22 & 95.6 & 9.966 & 3B$_1$ & 17-22 & 95.8 & 9.966 \\
    & 20-23 & 3.4 & 0.007 & 0.000 & 20-23 & 3.1 & 0.007 \\
    \hline
    4B$_1$ & 20-23 & 95.8 & 0.007 & 4B$_1$ & 20-23 & 96.1 & 0.007 \\
    & 17-22 & 3.5 & 9.966 & 0.000 & 17-22 & 3.2 & 9.966 \\
    \hline
    5B$_1$ & 21-26 & 96.5 & 0.000 & 5B$_1$ & 21-26 & 96.5 & 0.000 \\
    & 19-24 & 2.0 & 0.000 & & 19-24 & 2.0 & 0.000 \\
    & 21-28 & 0.9 & 0.000 & & 21-28 & 0.9 & 0.000 \\
    \hline
    \multicolumn{4}{c|}{ KS single excitations considered:} & \multicolumn{4}{c}{ KS single excitations considered:} \\
    \multicolumn{4}{c|}{18-22} & \multicolumn{4}{c}{18-22}  \\
    \hline\hline
    \end{tabular}
\end{table*}

\begin{table*}[h!]
\caption{Quantities used to inform the decision of which KS single excitations, $q$, to consider in DTDDFT and DTDA along with the KS double excitation, $D$ = (21-22)+(20-22), for an appropriate description of the lowest B$_1$ excitation energy in cyclopentadienone at the CAM-B3LYP/cc-pVDZ level of theory. We report quantities for the first four B$_1$ states of cyclopentadienone.}
\label{}
\centering
    \begin{tabular}{cccc|cccc}
    \hline\hline
    ATDDFT root & $q$ & $(|X_q|^2 - |Y_q|^2) (\%)$ & $|H_{qD}| / 10^{-3} (\text{a.u.})$ & ATDA root & $q$ & $|X_q|^2 (\%)$ & $|H_{qD}| / 10^{-3} (\text{a.u.})$\\
    \hline\hline
    2B$_1$ & 18-22 & 98.6 & 3.521 & 2B$_1$ & 18-22 & 98.5 & 3.521 \\
    & 18-25 & 0.4 & 0.000 & & 18-25 & 0.4 & 0.000 \\
    \hline
    3B$_1$ & 17-22 & 93.6 & 9.966 & 3B$_1$ & 17-22 & 93.6 & 9.966 \\
    & 20-23 & 3.7 & 0.007 & & 20-23 & 3.6 & 0.007 \\
    & 17-25 & 0.6 & 0.000 & & 13-25 & 0.6 & 0.000 \\
    & 13-25 & 0.6 & 0.000 & & 14-23 & 0.6 & 0.000 \\
    & 14-23 & 0.6 & 0.000 & & 17-25 & 0.6 & 0.000 \\
    \hline
    4B$_1$ & 20-23 & 93.1 & 0.007 & 4B$_1$ & 20-23 & 93.3 & 0.007 \\
    & 17-22 & 3.9 & 9.966 & & 17-22 & 3.8 & 9.966 \\
    & 14-23 & 0.8 & 0.000 & & 14-23 & 0.7 & 0.000 \\
    & 13-22 & 0.6 & 3.261 & & 13-22 & 0.5 & 3.261 \\
    & 17-25 & 0.5 & 0.000 & & 15-23 & 0.4 & 0.000 \\
    & 15-23 & 0.4 & 0.000 & & 17-25 & 0.4 & 0.000 \\
    & 12-22 & 0.3 & 8.238 & & 12-22 & 0.3 & 8.238 \\
    \hline
    5B$_1$ & 21-26 & 90.5 & 0.000 & 5B$_1$ & 21-26 & 90.5 & 0.000 \\
    & 19-24 & 6.6 & 0.000 & & 19-24 & 6.6 & 0.000 \\
    & 21-28 & 1.2 & 0.000 & & 21-28 & 1.2 & 0.000 \\
    & 16-24 & 0.8 & 0.000 & & 16-24 & 0.8 & 0.000 \\
    \hline
    \multicolumn{4}{c|}{ KS single excitations considered:} & \multicolumn{4}{c}{ KS single excitations considered:} \\
    \multicolumn{4}{c|}{18-22} & \multicolumn{4}{c}{18-22}  \\
    \hline\hline
    \end{tabular}
\end{table*}

\bibliography{ref}
\bibliographystyle{rsc}